\documentclass[prd,amssymb,amsmath,amsfonts,superscriptaddress,nofootinbib,reprint,showpacs, longbibliography]{revtex4-1}

\usepackage{graphicx}
\usepackage{lmodern}
\usepackage{amsmath,amssymb}
\usepackage{mathrsfs}
\usepackage{amsfonts}
\usepackage[utf8]{inputenc}
\usepackage{url}
\usepackage[colorlinks]{hyperref}
\usepackage{xcolor}
\usepackage{multirow}
\usepackage[normalem]{ulem}
\usepackage{orcidlink}
\usepackage{placeins}
\usepackage{mathtools}

\definecolor{dodgerblue}{HTML}{1E90FF}
\definecolor{viennared}{HTML}{DA0A14}
\definecolor{ctorange}{HTML}{FF6C0C}
\definecolor{wales}{HTML}{ff0038}
\definecolor{benettongreen}{HTML}{009421}
\definecolor{ferrarired}{HTML}{ff2800}
\definecolor{austriawienpurple}{HTML}{441678}

\hypersetup{
     colorlinks=true,
     linkcolor=dodgerblue,
     filecolor=dodgerblue,
     citecolor = dodgerblue,      
     urlcolor=dodgerblue,
     }

\newcommand{\Birmingham}{School of Physics and Astronomy and Institute for Gravitational Wave Astronomy, University of Birmingham, Edgbaston, Birmingham, B15 2TT, United Kingdom}

\newcommand{\keras}{\texttt{Keras}{ }}
\newcommand{\tensorflow}{\texttt{Tensorflow}{ }}
\newcommand{\numpy}{\texttt{NumPy}{ }}

\newcommand{\minmax}{\texttt{MinMax}{ }}
\newcommand{\standard}{\texttt{Standard}{ }}
\newcommand{\hyperopt}{\texttt{Hyperopt}{ }}
\newcommand{\scikitlearn}{\texttt{scikit-learn}{ }}

\newcommand{\Y}{\textbf{Y}{ }}
\newcommand{\X}{\textbf{X}{ }}
\newcommand{\SEOB}{\texttt{SEOBNRv4PHM}}
\newcommand{\SEOBNN}{\texttt{SEOBNN\textunderscore v4PHM\textunderscore4dq2}}

\DeclarePairedDelimiterX{\norm}[1]{\lVert}{\rVert}{#1}

\begin{document}

\title{Accelerating multimodal gravitational waveforms from precessing compact binaries with artificial neural networks}

\author{Lucy M. Thomas \orcidlink{0000-0003-3271-6436}}
\email{lthomas@star.sr.bham.ac.uk}
\affiliation{\Birmingham}
\author{Geraint Pratten \orcidlink{0000-0003-4984-0775}}
\email{G.Pratten@bham.ac.uk}
\affiliation{\Birmingham}
\author{Patricia Schmidt \orcidlink{0000-0003-1542-1791}}
\email{P.Schmidt@bham.ac.uk}
\affiliation{\Birmingham}

\date{\today}

\begin{abstract}
Gravitational waves from the coalescences of black hole and neutron stars afford us the unique opportunity to determine the sources' properties, such as their masses and spins, with unprecedented accuracy. To do so, however, theoretical models of the emitted signal that are i) extremely accurate and ii) computationally highly efficient are necessary. The inclusion of more detailed physics such as higher-order multipoles and relativistic spin-induced orbital precession increases the complexity and hence also computational cost of waveform models, which presents a severe bottleneck to the parameter inference problem. A popular method to generate waveforms more efficiently is to build a fast surrogate model of a slower one.  
In this paper, we show that traditional surrogate modelling methods combined with 
artificial neural networks can be used to build a computationally highly efficient while still accurate emulation of multipolar time-domain waveform models of precessing binary black holes. We apply this method to the state-of-the-art waveform model \SEOB{} and find significant computational improvements: On a traditional CPU, the typical generation of a single waveform using our neural network surrogate \SEOBNN{} takes $18$ms for a binary black hole with a total mass of $44M_\odot$ when generated from $20$Hz. In comparison to \SEOB{} itself, this amounts to an improvement in computational efficiency by two orders of magnitude.
Utilising additional graphics processing units (GPU) acceleration, we find that this speed-up can be increased further with the generation of batches of waveforms simultaneously. 
Even without additional GPU acceleration, this dramatic decrease in waveform generation cost can reduce the inference timescale from weeks to hours. 
\end{abstract}

\maketitle 

\section{Introduction}
\label{sec:Introduction}
Since the first observing run of the currently operating ground-based gravitational-wave (GW) detector network consisting of Advanced LIGO~\cite{LIGOScientific:2014pky} and Advanced Virgo~\cite{VIRGO:2014yos}, and soon also KAGRA~\cite{KAGRA:2020tym}, detections of GWs from more than 90 compact binary inspirals and mergers consisting of black holes and neutron stars~\cite{LIGOScientific:2018mvr, LIGOScientific:2020ibl, LIGOScientific:2021usb, LIGOScientific:2021djp} have been announced. These observations have a transformative impact on our understanding of the properties of black holes and neutron stars, allowing us to determine their mass and spin distributions~\cite{LIGOScientific:2021psn} and to put them into their astrophysical context. 

The inference of the source properties is predicated on the availability of accurate theoretical models of the emitted GW signal through inspiral, merger and ringdown (IMR). Recent years have seen much progress in the improvement of such waveform models through the inclusion of higher-order multipoles and spin-induced precession of the orbital plane. These improvements come, however, at the expense of computational efficiency. Fast model evaluation speeds are a necessary requirement for key analysis such as Bayesian inference where on average $10^6-10^8$ such model evaluations are needed to obtain well-sampled posterior probability distributions. Different strategies are employed to make waveform models computationally more efficient, some prominent ones include: Phenomenological ansatzes~\cite{Pratten:2020fqn}; reduced-order surrogate modelling~\cite{Field:2011mf,Field:2013cfa,Purrer:2014fza,Blackman:2015pia,Blackman:2017dfb,Galley:2016mvy,Varma:2019csw}; dimensional reduction~\cite{Schmidt:2014iyl,Thomas:2020uqj}; and most recently the incorporation of machine learning into reduced-order models~\cite{Khan:2020fso}. Alternatively, routes to either accelerate the likelihood evaluation directly~\cite{Cornish:2010kf,Canizares:2013ywa,Canizares:2014fya,Smith:2016qas,Zackay:2018qdy,Cornish:2021lje,Leslie:2021ssu,Morisaki:2021ngj} or to perform likelihood-free inference~\cite{Chua:2019wwt,Dax:2021tsq} have also been developed.

In this work we combine reduced-order modelling with the power of artificial neural networks (ANNs) to build a computationally vastly more efficient surrogate model of the state-of-the-art inspiral-merger-ringdown (IMR) waveform model \SEOB{} \cite{Ossokine:2020kjp} that includes both spin-induced orbital precession~\cite{Apostolatos:1994mx} and higher-order modes beyond the quadrupole emission. While the efficacy of this approach has previously been demonstrated for the quadrupole ($(2,2)$-) mode of aligned-spin binary black holes (BBHs)~\cite{Khan:2020fso,Fragkouli:2022lpt}, here we demonstrate its feasibility for the multimodal, precessing case. To achieve this, we decompose the \SEOB{} waveform model into eight components that describe the modes in a non-inertial, co-prcessing coordinate frame and three components that encode the precession dynamics. Using a combination of traditional surrogate modelling steps and neural networks to produce parameter fits, we build a fast surrogate model for each component. Using extensive optimisation we determine an optimal network for each component, which allows us to speed up the model evaluation by a factor of a few hundred on average on a CPU and even further on a graphics processing units (GPU), demonstrating the efficacy of this approach for state-of-art multimodal waveforms with precession.

The paper is organised as follows: First in Sec.~\ref{sec:Methodology} we introduce the methodology behind this work, including a brief overview of surrogate waveform modelling in Sec.~\ref{subsec:SurrogateModelling}, details of precessing waveform decomposition in Sec.~\ref{subsec:WaveformDecomposition}, and details of the mismatch metrics we use to assess the accuracy of our model in Sec.~\ref{subsec:Mismatch}. We then detail the construction of the model in Sec.~\ref{sec:Model}, first describing the training data upon which the model is built in Sec.~\ref{subsec:TrainingAndValidationData}, the reduced basis and empirical interpolant construction in Sec.~\ref{subsec:ReducedBasisAndEmpiricalInterpolant}, and the neural networks for each of the model components in Sec.~\ref{subsec:ANNs}, both coprecessing modes and Euler angles, putting these together to describe the full surrogate model construction in Sec.~\ref{subsec:CompleteSurrogateModel}. We then discuss the completed model in Sec.~\ref{sec:ModelEvaluation}, both the accuracy of the waveforms generated in Sec.~\ref{subsec:WaveformAccuracy} and the timing of the model evaluation in Sec.~\ref{subsec:Timing}. Finally, we summarise the model construction and results in Sec.~\ref{sec:Discussion}, also discussing caveats and further work. 

Throughout this paper we use geometric units, $G=c=1$, unless stated otherwise.

\section{Methodology}
\label{sec:Methodology}
\subsection{Surrogate Modelling}
\label{subsec:SurrogateModelling}
Surrogate models are fast, accurate approximations to an underlying (slower) model, over a chosen parameter space region. Therefore, their range of validity in parameter space is limited to the region over which they are constructed, the training space, plus an extrapolation region over which the model has been tested and shown to be accurate to within some tolerance. Recent examples of surrogate models for waveforms from coalescing compact binaries include Numerical Relativity (NR) and NR-hybrid surrogate models~\cite{Blackman:2015pia, Blackman:2017dfb, Varma:2018mmi,Varma:2019csw, Islam:2021mha, Islam:2022laz}, surrogates for the aligned-spin effective-one-body (EOB) model \texttt{SEOBNRv4}~\cite{Bohe:2016gbl} using artificial neural networks \cite{Khan:2020fso,Fragkouli:2022lpt} and a machine learning emulation of a different EOB model, TEOBResumS~\cite{Nagar:2018zoe, Schmidt:2020yuu}. 

In the following, we will provide a brief outline of the main steps for building a surrogate model. For a more complete explanation we refer the reader to e.g. \cite{Field:2013cfa}. 

The process of building a surrogate model may begin with building a \emph{reduced basis}, which enables us to represent any arbitrary function, e.g. a time-domain waveform $h(t,\vec{\lambda})$ with intrinsic parameters $\vec{\lambda}$, within the discrete training space $\mathcal{T}_M = \{ \vec{\lambda}_i\}_{i=1}^{M} \subseteq \mathcal{T} = \{ \vec{\lambda}_i\}_{i=1}^{\infty}$ as a linear combination of an $n$-dimensional orthonormal basis $\{ \hat{e}_i(t)\}_{i=1}^{n}$ and projection coefficients $\{c_n (\vec{\lambda})\}_{i=1}^{n}$,
\begin{equation}
    h(t, \vec{\lambda}) \approx \sum_{i=1}^{n} c_{i}(\vec{\lambda})\hat{e}_{i}(t),
\end{equation}
with $n \leq M$. 
The reduced basis is constructed recursively using a greedy algorithm~\cite{Field:2011mf, Field:2013cfa} until all waveforms in the training space $\mathcal{T}_M$ are represented by the basis to within a certain tolerance $\sigma$, which is related to the representation error $\epsilon$ by
\begin{equation}
    \max{\epsilon} = \max \norm[\Big]{ h(t; \vec{\lambda}) - \sum_{i=1}^n c_{i}(\vec{\lambda})\hat{e}_{i}(t) }^2 \leq \sigma,
    \label{eq:tol}
\end{equation}
where $||*||$ denotes the $L^2$-norm, which we compute via the Chebyshev-Gauss quadrature rule.
To achieve this, at each step the waveform with the largest representation error using the current basis is chosen, orthogonalised with respect to all current basis elements, and normalised, before being added to the basis as the next basis element. The greedy algorithm stops once Eq~\eqref{eq:tol} is fulfilled or if the waveform with the largest representation error is already a basis element. The latter is an indication that the training space $\mathcal{T}_M$ is sampled too coarsely to achieve the desired accuracy $\sigma$. 
If the discrete training space is sampled sufficiently densely, then the reduced basis representation allows us to approximate \emph{any} waveform in the entire parameter space $\mathcal{T}$.

After the basis has been constructed, we proceed to build an \emph{empirical interpolant} (EI) using the empirical interpolation method~\cite{BARRAULT2004667, Maday:2009}, which allows us to reconstruct each waveform $h(t;\vec{\lambda})$ for $\vec{\lambda} \in \mathcal{T}_M$ to within a high accuracy, only using information at certain (sparse) time nodes $\{T_i\}^{n}_{i=1}$. These carefully selected empirical times or nodes are determined exclusively by the reduced basis waveforms, and the number of time nodes will be equal to the number of waveforms within the reduced basis:
\begin{align}
    {\rm EI}[h](t; \vec{\lambda}) &= \sum_{j=1}^n B_j(t) h(T_j; \vec{\lambda}),\\
    &= \sum_{i=1}^n \sum_{j=1}^n \hat{e}_i(t)(V^{-1})_{ij} h(T_j; \vec{\lambda}),
\end{align}
where $(V)_{ij} = (\hat{e}_i(T_j))$ is the interpolation matrix.

The final step for building a surrogate model is to perform a parameter space fit which allows us to predict waveforms at the empirical times $\{T_i\}_{i=1}^n$ for arbitrary parameters $\vec{\lambda} \in \mathcal{T}$ based on the greedy points $\{\vec{\lambda}_i\}_{i=1}^n$ selected to construct the reduced basis. This requires us to fit $h(t;\vec{\lambda})$ across the parameter space at each empirical node such that
\begin{equation}
\label{eq:hfit}
    h(T_i; \vec{\lambda}) \approx A_i (\vec{\lambda}) e^{i \phi_i(\vec{\lambda})},
\end{equation}
where $A_i$ and $\phi_i$ are the amplitude and phase at the $i$-th empirical node. 
The $2n$-functions that determine the parameter space fits can be determined by different means, for example via traditional fitting functions such as splines or polynomials~\cite{Field:2011mf, Blackman:2015pia, Varma:2019csw} or by using machine learning algorithms such artificial neural networks~\cite{Khan:2020fso, Fragkouli:2022lpt} or Gaussian processes~\cite{Varma:2018mmi,Williams:2019vub, Islam:2021mha, Yoo:2022erv}. In this work, we will follow Ref.~\cite{Khan:2020fso} and use ANNs to determine the fitting coefficients $A_i(\vec{\lambda})$ and $\phi_i(\vec{\lambda})$.
The final surrogate model for a waveform $h(t; \vec{\lambda})$ is then given by 
\begin{equation}
    h^S(t; \vec{\lambda}) \equiv \sum_{i=1}^n \sum_{j=1}^n (V^{-1})_{ji} \hat{e}_j(t) A_i(\vec{\lambda})e^{-i\phi_i(\vec{\lambda})}.
    \label{eq:sur}
\end{equation}
We note that this prescription applies to generic functions up to the parameter space fits Eq.~\eqref{eq:hfit}, whose RHS decomposition depends on the function that is being modelled. In subsequent sections, we will follow this approach for individual waveform modes decomposed into amplitude and phase as well as angle functions.

To construct the reduced bases and empirical interpolants we use the publicly available \texttt{Python} package \texttt{RomPy}~\cite{Field:2013cfa, rompy}.

\subsection{Waveform Decomposition}
\label{subsec:WaveformDecomposition}
Binary black holes on quasi-spherical orbits span a seven-dimensional (intrinsic) parameter space characterised by the mass ratio $q=m_1/m_2 \geq 1$ and the (dimensionless) spin angular momenta $\vec{\chi}_1$ and $\vec{\chi}_2$.  
If the spin angular momenta are misaligned with the direction of the instantaneous orbital angular momentum $\hat{L}(t)$, then spin-induced precession occurs~\cite{Apostolatos:1994mx,Kidder:1995zr}. This causes the orbital plane to change its spatial orientation as the binary inspirals due to GW emission. This more complex two-body dynamics leads to amplitude and phase modulations of the emitted GW signal $h(t; \vec{\lambda})$ and is also a source of the excitation of higher-order multipoles, $h_{\ell m}$, in the radiation field, which must be included to accurately describe the GW signal:
\begin{equation}
    h(t; \vec{\lambda}; \theta, \varphi) = \sum_{\ell=2}^\infty \sum_{m=-\ell}^\ell h_{\ell m}(t; \vec{\lambda}) {}^{-2}Y_{\ell m}(\theta, \varphi),
    \label{eq:strain}
\end{equation}
where $(\theta, \varphi)$ denote the angles on the unit sphere. Due to the increased complexity, modelling the signal from precessing BBHs is a challenging task but is accomplished as follows~\cite{Apostolatos:1994mx, Buonanno:2002fy, Schmidt:2010it, Schmidt:2012rh}: 
The GW modes from precessing binaries, $h^P_{\ell m}(t;\vec{\lambda})$, can be conveniently decomposed into a simpler carrier signal corresponding to a non-inertial coprecessing observer, $h^{\rm co-prec}_{\ell m}(t;\vec{\lambda})$, and a time-dependent rotation operator $\mathbf{R}$ which encodes the orbital precession dynamics, i.e.
\begin{equation}
    h^P_{\ell m}(t; \vec{\lambda}) = \sum_{m'= -\ell}^{\ell} \mathbf{R}_{\ell m m'}(t; \vec{\lambda}) h^{\rm co-prec}_{\ell m'}(t; \vec{\lambda}),
\label{eq:modes}
\end{equation}
where $\vec{\lambda}$ denotes the binary's intrinsic parameters. 

As a first approximation, the coprecessing waveform modes can be approximated by aligned-spin modes~\cite{Schmidt:2010it,Schmidt:2012rh}. This simplifying approximation is made in many of the state-of-the-art waveform models~\cite{Hannam:2013oca,Pratten:2020ceb,Ossokine:2020kjp} and is a known source of modelling errors~\cite{Ramos-Buades:2020noq}. Importantly, this approximation assumes a conjugate symmetry between the $+m$ and $-m$ modes, which no longer holds in the case of precessing binaries~\cite{Schmidt:2012rh,Boyle:2014ioa}. The waveform model we emulate here, \SEOB{}~\cite{Cotesta:2018fcv, Ossokine:2020kjp}, contains the $(2,\pm 2), (2, \pm 1), (3,\pm 3), (4, \pm 4)$ and $(5,\pm 5)$ coprecessing modes defined in a time-dependent coordinate frame that tracks $\hat{L}(t)$ ($L$-frame), and assumes conjugate mode symmetry, i.e.,
\begin{equation}
    h^{\rm co-prec}_{\ell,-m}(t; \vec{\lambda}) = (-1)^\ell h^{\rm co-prec *}_{\ell m}(t; \vec{\lambda}).
\label{eq:conj}
\end{equation}
Therefore, we only model the positive $m-$modes in the coprecessing frame and obtain the $-m$-modes via Eq. \eqref{eq:conj}. 
The coprecessing waveform modes are then further decomposed into amplitude and phase, 
\begin{equation}
    h^{\rm co-prec}_{\ell m}(t; \vec{\lambda})=A_{\ell m}(t; \vec{\lambda}) e^{i\phi_{\ell m}(t; \vec{\lambda})}.
\end{equation}

\begin{figure}[t!]
     \centering
     \includegraphics[width=0.85\columnwidth]{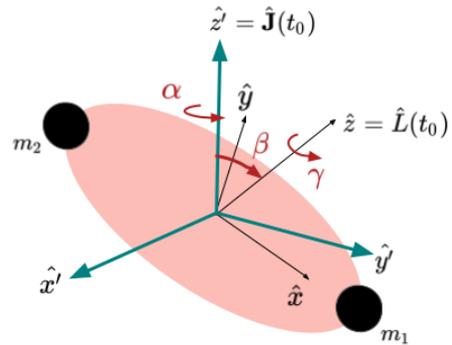}
     \caption{Definition of the inertial $J$-frame and the Euler angles. The three Euler angles $\alpha$, $\beta$ and $\gamma$ define the rotation from the $L$-frame where the $z$-component is parallel to the orbital angular momentum of the binary $\hat{z}=\hat{L}(t_0)$, to the $J$-frame where it is parallel to the total angular momentum $\hat{z'}=\hat{\mathbf{J}}(t_0)$ at the start time of the waveform $t_0$. The angle $\gamma$ is defined by the minimal rotation condition, $\dot{\gamma}=-\dot{\alpha} \cos{\beta}$~\cite{Boyle:2013nka, Buonanno:2002fy}.}
     \label{fig:Jframe}
 \end{figure}

For the rotation operator we will use its $\rm SO(3)$ representation and model the three Euler angles $\alpha(t; \vec{\lambda}), \beta(t; \vec{\lambda})$ and $\gamma(t; \vec{\lambda})$ in an inertial Cartesian coordinate frame that is aligned with the total angular momentum at the the initial time $t_0$, i.e. $\mathbf{J}(t_0) = \hat{z}$, as shown in Fig.~\ref{fig:Jframe}, henceforth referred to as the $J$-frame. 

Utilising this decomposition, we build (i) ANN surrogates for the amplitude and phase of each coprecessing positive $m-$mode contained in \texttt{SEONBNRv4PHM} (see Sec.~\ref{subsec:CoprecessingModes}) and (ii) ANN surrogates for the three Euler angles (see Sec.~\ref{subsec:EulerAngles}).

\subsection{Mismatch}
\label{subsec:Mismatch}
To determine the agreement between the input waveform model \SEOB{} and its ANN emulation, we will use the mismatch as the metric to assess the accuracy of the various elements in the surrogate modelling process.

To quantify the agreement between two coprecessing modes/Euler angles/polarisations, we will employ the (frequency-domain, noise-weighted) inner product or match $\mathcal{M}_{f}$ optimised over a time and phase shift, given as
\begin{equation}
    \label{eq:Match}
    \mathcal{M}_{f}(h_1,h_2) = \max_{t_{c}, \phi_{0}}\frac{\langle h_1, h_2\rangle}{\sqrt{\langle h_1, h_1\rangle \langle h_2, h_2\rangle}},
\end{equation}
where the inner product is defined as
\begin{equation}
    \label{eq:InnerProduct}
    \langle h_1,h_2 \rangle = 4\Re \displaystyle\int^{f_{\rm{max}}}_{f_{\rm{min}}} \frac{\tilde{h}_1(f)\tilde{h}_{2}^*(f)}{S_n (|f|)} \, df,
\end{equation}
with $S_n(f)$ the one-sided power spectral density (PSD) of the detector noise, $\tilde{h}$ indicates the Fourier transform of $h$, and $\tilde{h}^*$ the complex conjugate. The \textit{mismatch} can now be defined as
\begin{equation}
    \bar{\mathcal{M}}_{f} (h_1, h_2) \equiv 1 - \mathcal{M}_{f} (h_1, h_2). 
\end{equation}
\newline 
We will also find it convenient to introduce a normalized waveform $\hat{h} = h / \sqrt{\langle h , h \rangle}$. 

When using a white-noise PSD, i.e. independent of frequency, it is convenient to define a time-domain overlap
\begin{equation}
    \label{eq:InnerProductTD}
    \langle h_1 , h_2 \rangle_t = \Re \displaystyle\int^{t_{\rm max}}_{t_{\rm min}} h_1 (t) h^{\ast}_2 (t) \, dt ,
\end{equation}
\newline 
with an inherited norm $\| h \|^2_t = \langle h, h \rangle_t$. We can then define an analogous time domain match as 
\begin{equation}
    \label{eq:MismatchTD}
    \mathcal{M}_t (h_1,h_2) = \frac{\langle h_1, h_2 \rangle_t}{\| h_1 \|_t \; \| h_2 \|_t},
\end{equation}
\newline 
and the associated \textit{mismatch} as $\bar{\mathcal{M}}_t = 1 - \mathcal{M}_t$. 

\begin{table*}[t!]
    \centering
    \begin{tabular}{c|c|c|c|c|c|c}
    \hline
    \hline
    \multicolumn{2}{c|}{Training data subset} & Number of binaries & $q$ & $\left|\chi_1\right|$ & $\theta_1$ [rad] & $\phi_1$ [rad]\\
    \hline
    \parbox[t]{8mm}{\multirow{6}{*}{\rotatebox[origin=c]{90}{\parbox{2cm}{Systematically\\ sampled}}}} & Nonspinning & $6$ & $\left[1, 1.2, 1.4, 1.6, 1.8, 2\right]$ & $\left[0\right]$ & - & - \\
    & & & & & & \\
    & Spin-aligned & $48$ & $\left[1, 1.2, 1.4, 1.6, 1.8, 2\right]$ & $\left[0.2, 0.4, 0.6, 0.8\right]$ & $\left[0, \pi\right]$ & - \\
    & & & & & & \\
    & \multirow{2}{*}{Precessing} & \multirow{2}{*}{$720$} & \multirow{2}{*}{$\left[1, 1.2, 1.4, 1.6, 1.8, 2\right]$} & \multirow{2}{*}{$\left[0.2, 0.4, 0.6, 0.8\right]$} & $[\pi/6, \pi/3, \pi/2,$ & $[0, \pi/3, 2\pi/3,$\\
    & & & & & $2\pi/3, 5\pi/6]$ &  $\pi, 4\pi/3, 5\pi/3]$ \\
    & & & & & & \\
    \hline
    \multicolumn{2}{c|}{\multirow{2}{*}{Randomly sampled}} & \multirow{2}{*}{$199,226$} & \multirow{2}{*}{$U\left[1,2\right]$} & \multirow{2}{*}{$U\left[0,0.8\right]$} & \multirow{2}{*}{$U\left[0,\pi\right]$} & \multirow{2}{*}{$U\left[0,2\pi\right]$} \\
    \multicolumn{2}{c|}{} & & & & & \\
    \hline
    \hline
    \end{tabular}
    \label{tab:binaryparams}
    \caption{Parameters of the $200,000$ binaries which span our training dataset. The training space is split into two sectors: (i) a systematically sampled subset which is included to ensure coverage of the parameter space boundaries; (ii) a random but uniformly sampled subset. All spin parameters are quoted at a reference frequency of $4$Hz for the $(2,2)$-mode for a binary with a total mass of $60 M_\odot$}
\end{table*}

\section{Model}
\label{sec:Model}
In this section we describe the construction of our surrogate model. As described in Sec.~\ref{subsec:WaveformDecomposition}, we model the coprecessing modes and Euler angles separately. We detail the training dataset upon which the model is built in Sec.~\ref{subsec:TrainingAndValidationData}, as well as the validation and test datasets. In Sec.~\ref{subsec:ReducedBasisAndEmpiricalInterpolant} we describe the construction of the reduced bases and empirical interpolants for each component, and in Secs.~\ref{subsec:CoprecessingModes} and \ref{subsec:EulerAngles} we describe the neutral network architecture and training for the coprecessing modes and Euler angles respectively, assessing the accuracy of each component. We will then consider the accuracy and speed of the complete model to produce a fully precessing signal in Sec.~\ref{sec:ModelEvaluation}.

\subsection{Training, Validation and Testing Data}
\label{subsec:TrainingAndValidationData}
\subsubsection{Waveforms}
Our waveform training dataset consists of $2 \times 10^5$ multipolar \SEOB{} waveforms with mass ratios $q \in [1,2]$ and arbitrarily oriented spin on the primary black hole with magnitude $|\chi_1| \leq 0.8$\footnote{We note that the spin orientation is defined relative to the orbital angular momentum $\hat{L}(t_0)$ at the initial time. Decomposed in Cartesian coordinates the spin vector is given by $\{\chi_{1x}, \chi_{1y}, \chi_{1z}\}$, where $\chi_{1z} = \vec{\chi}_1 \cdot \hat{L}(t_0)$.}; the secondary black hole is nonspinning.
Both the coprecessing waveform modes and the time-dependent Euler angles are obtained directly from the \SEOB{} implementation in the public LIGO Algorithm Library \texttt{LAL}~\cite{lalsuite}, mitigating the need to perform any additional post-processing.

We first randomly sample $199,226$ binaries from the reduced parameter space, drawing the parameters from distributions uniform in mass ratio $q$, uniform in spin magnitude $|\chi_1|$ and isotropic in spin orientation $(\theta_1, \phi_1)$. We supplement these binaries with an additional $774$ systematically chosen points to accurately sample the boundaries. 
The parameters of these systematically chosen binaries are listed in Tab. I and a visualisation of the training set can be found in Fig.~\ref{fig:spins} in Appendix~\ref{sec:app}. 

We initially generate waveforms such that the $(2,2)$-mode of a binary with a total mass of $60\, M_\odot$ starts from an initial frequency of $4$Hz. In geometrised units this corresponds to an approximate length of $\sim 2\times10^5 M$ before merger, though the duration varies due to mass ratio and inspiral spin~\cite{Racine:2008qv, Harry:2016ijz}. 
The modes are then aligned such that the peak of the quadrature of all modes occurs at $t=0M$. 
Each waveform is first generated on a uniform time grid with a time spacing $\Delta t =0.1M,$ and then reinterpolated onto a non-uniform grid which is $20$ times coarser in the early inspiral, but retains the $0.1M$ spacing in the later-inspiral, merger and ringdown. 
The waveforms are of varying length pre- and post-merger, and so have different time grids of the same resolution, but are required to be of equal length and evaluated upon the same times in order to build the reduced bases. 
As a data preprocessing step we choose the waveform with the shortest length pre-merger, and reinterpolate all waveforms onto this common time grid. 
We then choose the waveform with the shortest post-merger signal and truncate all waveforms such that the final time matches that of the shortest. We note that this truncation is less than $10 M$ for all waveforms, and contains a negligible amount of the ringdown signal in all cases. 
For computational reasons, we truncate all waveforms to be of length $10^4M$ pre-merger. 
Therefore all waveforms start at $10^4M$ before the peak, and include $110M$ of post-merger signal. We note that due to this truncation, the spin parameters are specified at the initial time $t\sim -2 \times 10^5M$ and not at the start of the waveforms. Since the spins in precessing binaries evolve with time, it is necessary to define the reference time or frequency at which they are defined. Being able to do this for some arbitrary time/frequency requires either code (see e.g.~\cite{Johnson-McDaniel:2021rvv}) or additional NNs that track the spin evolution. We leave building neural networks for the spin evolution for future work.
The truncated waveforms are then re-interpolated onto a uniform grid with spacing $\Delta t = 1M$ in order to build the reduced bases and empirical interpolants, as we found that the finer $0.1 M$ spacing was not required. 

We note that when constructing the models for the coprecessing odd-$m$ mode amplitudes and phases, not all of the $2\times10^5$ training waveforms are used. We first remove training points where there is very little spin or mass asymmetry in the system, as we expect the odd-$m$ amplitudes to be small and therefore noisy in the true \SEOB{} data. We impose a cut of $q>1.01$, $\chi_{1z}>10^{-2}$, which removes $109$ points from the training set. Next, we remove any training data which show signs of (unphysical) discontinuities in the phase, possibly due to next-to-quasi-circular corrections in the \SEOB{} data. For the coprecessing $(2,1)$-mode, this amounts to $11,091$ points, and $198$ for the $(3,3)$-mode. Therefore, for the $(2,1)$-mode amplitude and phase, the total training dataset is $188,800$ waveforms, whereas for the $(3,3)$-mode it is $199,693$.

To validate our neural networks as they train, we also produce a validation dataset of $10^4$ waveforms which covers the same parameter space as the training set. We sample this validation set uniformly is mass ratio, primary spin magnitude, spin tilt and azimuthal angles. All preprocessing steps for the validation data are the same as for the training data: we interpolate these waveforms onto the same common time grid with an equivalent spacing. For the coprecessing odd-$m$ modes, we remove $4$ points with little asymmetry, $553$ which show signs of phase discontinuity in the $(2,1)$-mode phase, and $9$ in the $(3,3)$-mode. This equates to a validation set size of $9,443$ for the coprecessing $(2,1)$-mode, and $9,987$ for the $(3,3)$-mode.

Lastly, we also produce a separate test dataset of $10^4$ waveforms in exactly the same way as the validation set, which is completely independent and unseen by the neural networks. Of this dataset, $3$ points are removed for the coprecessing odd-$m$ modes due to little symmetry, $672$ due to discontinuities in the $(2,1)$-phase, and $13$ due to the $(3,3)$-phase. Therefore for the $(2,1)$-mode amplitude and phase, the test set is of size $9325$, and for the $(3,3)$-mode it is $9,984$.

\subsubsection{Euler Angles}
For the Euler angles, we use the same dataset of $2 \times 10^5$ waveforms as described above. However, as the Euler angles become ill defined in the non-precessing limit, we restrict our training data to only those binaries with an initial in-plane spin magnitude $|\chi_{1,\perp}|=\sqrt{\chi_{1x}^2 + \chi_{1y}^2} > 10^{-3}$. In contrast to above, we decompose this initial dataset into a training dataset of $1.8 \times 10^5$ binaries and a validation dataset of $18,634$ binaries. As no hyperparameter optimization was performed on the Euler angle networks, the validation dataset is never used to train the network or to inform the network hyperparameters. We therefore treat the validation dataset as being effectively independent. The data conditioning is identical to the procedure described above for the waveform modes, with the Euler angles being evaluated on a uniform grid with spacing $\Delta t = 1 M$ and a length of $10^4 M$. 

\subsection{Reduced Basis and Empirical Interpolant}
\label{subsec:ReducedBasisAndEmpiricalInterpolant}
We construct our reduced bases and empirical interpolants following the algorithm described in Sec.\ref{subsec:SurrogateModelling}. We separate each coprecessing mode into its constituent amplitude and phase, and construct a reduced basis, empirical interpolant and neural network for each component. We also construct a reduced basis and empirical interpolant for each Euler angle separately, leading to a total of $11$ different components to make up the full precessing signal\footnote{The odd-$m$ modes are obtained via conjugation and hence do not need to be modelled separately but are included in the full precessing signal.}. When discussing the construction and evaluation of these models, we use the following terminology: 
\X describes the input parameters of the model, i.e. the four intrinsic parameters of the binary \X$=\vec{\lambda} = \{q, \chi_{1x}, \chi_{1y}, \chi_{1z}\}$; \Y is an $n$-dimensional vector that denotes the fitting coefficients, for example the mode amplitudes in Eq.~\eqref{eq:sur}. 

We choose to condition the data before building our reduced bases as we found this to be beneficial for the neural network performance: For the coprecessing modes we use a \scikitlearn\cite{scikit-learn} \standard scaler on the \X data and a \minmax scaler on the \Y data for the phases as we found that without scaling the greedy algorithm for the coprecessing $(2,1)$- and $(3,3)$-mode phases was unable to converge and produce a reduced basis to within the greedy tolerance accuracy. We also remove the initial phase at time $t=-10,000M$, such that all phase data begin at zero. We note that we do not explicitly model these initial phases, and leave this to future work.

In contrast, we find no major benefit to scaling the \X data for the Euler angles and the amplitude \Y data for the coprecessing modes. For $\alpha$ and $\gamma$, we apply a \minmax scaler to them$\Y$ data but we do not apply any preprocessing to the $\Y$ data for $\beta$. A summary of the data conditioning can be found in Table III. 

To build the reduced bases, we use an absolute greedy error tolerance of $\sigma = 10^{-6}$ for all components of the coprecessing modes, except for the phases of the $(2,1)$ and $(3,3)$-modes. For the $(2,1)$-mode, we decreased the greedy tolerance to $10^{-8}$ as we found a significant tail of poor mismatches against the reduced basis representation with a tolerance of $10^{-6}$. Conversely, for the $(3,3)$-mode, we reduced the tolerance to $10^{-3}$ in order to achieve a reduced basis of manageable size. The tolerances and the sizes of the resulting reduced bases (and therefore the number of empirical interpolation nodes) are given in the fourth and fifth column of Table II.

\begin{table*}[t!]
    \centering
    \begin{tabular}{c|c|c|c|c|c|c|c|c}
    \hline
    \hline
    \multirow{2}{*}{$(\ell,m)$} & \multirow{2}{*}{Component} & Training & Greedy & Basis & \multirow{2}{*}{$\bar{\mathcal{M}}_f^{\rm max,train}$} & \multirow{2}{*}{$\bar{\mathcal{M}}_f^{\rm median,train}$}& \multirow{2}{*}{$\bar{\mathcal{M}}_f^{\rm max,val}$} & \multirow{2}{*}{$\bar{\mathcal{M}}_f^{\rm median,val}$}\\
     & & Set Size & Tolerance & Size & & & &\\
    \hline
    $(2,2)$ & Amplitude & \multirow{2}{*}{$200,000$} & $10^{-6}$ & $23$ & \multirow{2}{*}{$1.1\times10^{-5}$} & \multirow{2}{*}{$5.0\times10^{-7}$} & \multirow{2}{*}{$1.4\times10^{-5}$} & \multirow{2}{*}{$6.0\times10^{-7}$} \\
    $(2,2)$ & Phase & & $10^{-6}$ & $29$ & & & &\\
    \hline
    $(2,1)$ & Amplitude & \multirow{2}{*}{$188,800$} & $10^{-6}$ & $26$ & \multirow{2}{*}{$3.2\times10^{-3}$} & \multirow{2}{*}{$1.8\times10^{-6}$}& \multirow{2}{*}{$3.4\times10^{-4}$} & \multirow{2}{*}{$1.7\times10^{-6}$}\\
    $(2,1)$ & Phase & & $10^{-8}$ & $40$ & & & &\\
    \hline
    $(3,3)$ & Amplitude & \multirow{2}{*}{$199,693$} & $10^{-6}$ & $26$ & \multirow{2}{*}{$1.8\times10^{-2}$} & \multirow{2}{*}{$3.0\times10^{-4}$} & \multirow{2}{*}{$2.7\times10^{-2}$} & \multirow{2}{*}{$3.1\times10^{-4}$}\\
    $(3,3)$ & Phase & & $10^{-3}$ & $46$ & & & &\\
    \hline
    $(4,4)$ & Amplitude  & \multirow{2}{*}{$200,000$} & $10^{-6}$ & $4$ & \multirow{2}{*}{$1.6\times10^{-4}$} & \multirow{2}{*}{$4.0\times10^{-5}$} & \multirow{2}{*}{$1.9\times10^{-4}$} & \multirow{2}{*}{$4.1\times10^{-5}$}\\
    $(4,4)$ & Phase  & & $10^{-6}$ & $29$ & & & &\\
    \hline
    \hline
    \multicolumn{2}{c|}{Euler Angle} & \multicolumn{3}{|c|}{} & $\bar{\mathcal{M}}_t^{\rm max,train}$ & $\bar{\mathcal{M}}_t^{\rm median,train}$ & $\bar{\mathcal{M}}_t^{\rm max,val}$ & $\bar{\mathcal{M}}_t^{\rm median,val}$\\
    \hline
    \multicolumn{2}{c|}{$\alpha$} & 180,000 & $7 \times 10^{-9}$ & 18 & $3.0 \times 10^{-5}$ &$1.4 \times 10^{-9}$ & $2.6 \times 10^{-6}$ & $1.4 \times 10^{-9}$\\
    \multicolumn{2}{c|}{$\beta$} & 180,000 & $6 \times 10^{-7}$ & 19 & $4.1 \times 10^{-6}$ & $1.3 \times 10^{-6}$ & $4.4 \times 10^{-6}$ & $1.3 \times 10^{-7}$\\
    \multicolumn{2}{c|}{$\gamma$} & 180,000 & $7 \times 10^{-9}$ & 18 & $4.2 \times 10^{-5}$ & $1.5 \times 10^{-9}$ & $1.0\times 10^{-6}$ & $1.5 \times 10^{-9}$\\
    \hline
    \hline
    \end{tabular}
    \label{tab:ReducedBases}
    \caption{Greedy tolerances, reduced basis sizes and the maximum and median training space mismatches for the amplitude and phase of each mode.  For phases, MinMax scaling was used on the Y data. For both amplitudes and phases, standard scaling was used on the X data. For the Euler angles, only MinMax scaling was used on the Y-data for $\alpha$ and $\gamma$ and we use the time-domain mismatch $\bar{\mathcal{M}}_t$ as our metric.}
\end{table*}

To assess the accuracy of the coprecessing $(\ell, m)$-modes reconstructed from their reduced basis representations in amplitude and phase, we compute frequency-domain white noise mode-by-mode mismatches $\bar{\mathcal{M}}_f$, defined by Eq.~\eqref{eq:Match} against the original \SEOB{} data. Columns 6-9 of Table II show the maximum and median mismatch across the full training and validation datasets for each coprecessing mode, noting that the validation data is not used in the construction of the reduced bases. Generally, we find that the odd-$m$ modes are less accurately represented than the even-$m$ modes and that that their bases sizes are larger. This is perhaps not too surprising as the odd-$m$ modes are (i) subdominant and (ii) contain more structure, therefore requiring more basis elements to achieve the same representation accuracy~\cite{Galley:2016}.

Similarly, we compute time-domain mismatches $\bar{\mathcal{M}}_t$, defined by Eq.~\eqref{eq:MismatchTD} between the original \SEOB{} data Euler angles, and those reconstructed from the reduced basis projections. We do this across both the training and validation datasets, and state the median and maximum values for each dataset in columns 5-8 in the bottom half of Table II. We see that for both the coprecessing modes and the Euler angles, the median mismatch across both datasets is comparable to the greedy tolerance used to create the reduced basis (for the mode mismatches, it is limited by whichever greedy tolerance is larger, amplitude or phase).

Lastly, the similar mismatches for the coprecessing modes across both the training dataset, which was used to construct the bases, and the validation dataset, which was previously unseen, suggests that the reduced bases are large enough to accurately represent waveforms across our chosen parameter space. We note that for the Euler angles, the mismatches (both median and maximum) over the validation dataset can be up to an order of magnitude smaller than over the training dataset. This suggests that the validation dataset is not large enough to accurately represent the full distribution over the entire parameter space, especially for the $\beta$ angle which is typically much flatter than either $\alpha$ or $\gamma$.

\subsection{Parameter space fits with ANNs}
\label{subsec:ANNs}
\begin{figure}
    \centering
    \includegraphics[width=0.8\columnwidth]{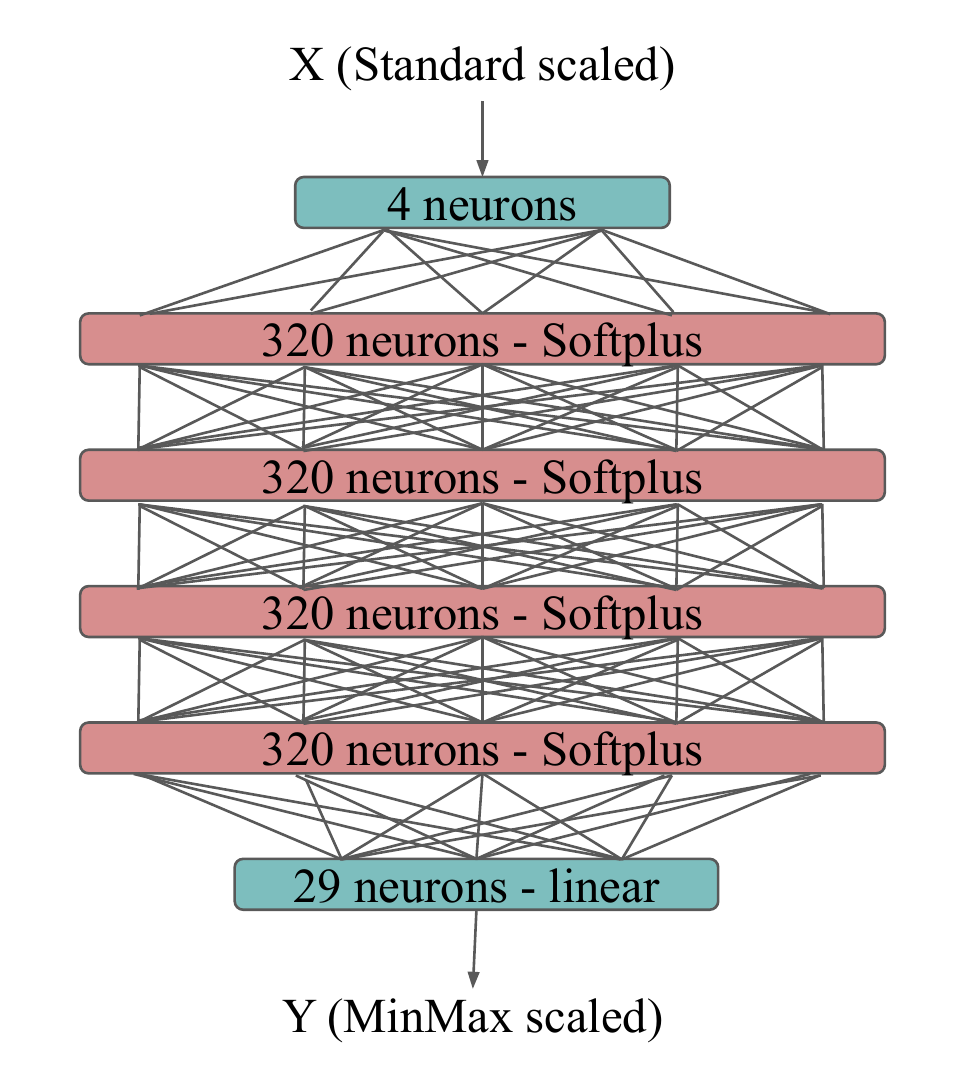}
    \caption{Graphical representation of the ANN architecture for the coprecessing $(2,2)$-mode phase $\phi_{22}(t;\vec{\lambda})$, as an example. This neural network takes in the \standard scaled intrinsic binary parameters \X as input, and outputs the \minmax scaled \Y, a prediction of the coprecessing $(2,2)$-mode phase at the empirical time nodes. This output vector may then be reinterpolated onto the full uniform time grid using the empirical interpolant, and inverse \minmax scaled to produce the full coprecessing mode phase $\phi_{22}$.}
    \label{fig:ANNSchematic}
\end{figure}

We now describe the architecture, training and optimization of our neural networks for the fitting coefficients of the coprecessing modes each decomposed into amplitude and phase and Euler angles, and discuss the achieved accuracy for each of component separately. We build the neural network for each model component using \tensorflow\cite{tensorflow2015-whitepaper} and \keras\cite{chollet2015keras}. Specifically, we use the \texttt{Sequential} model with fully-connected \texttt{Dense} layers.
A summary of the final neural network architectures for each coprecessing mode and the Euler angles is given in Table III. As an example, a graphical representation of the neural network architecture for the coprecessing $(2,2)$-mode phase is shown in Fig.~\ref{fig:ANNSchematic}. The neural network is shown by the red and teal rectangles, where the red ones represent the four fully-connected hidden layers, each with 320 neurons for this component and a Softplus activation function, and the teal ones show the input and output layers: $4$ neurons for the intrinsic parameters \X, and $29$ for the output layer as this is the number of empirical time nodes $T_i$ for this component. The output can then be reinterpolated onto the full uniform time grid, and inverse scaled to produce the full coprecessing $(2,2)$-mode phase $\phi_{22}$. For this particular component we have applied scaling to the \X and \Y data, as shown by the blue rectangles. For all ANNs we use 4 input neurons, but the detailed architecture is adapted for each component. 

The size of neural network differs between the coprecessing modes and Euler angles, as shown in Table III. Additionally, the coprecessing amplitudes will not undergo inverse \minmax scaling as we did not scale the amplitude training data in our model construction, and for the Euler angles the \X data is not scaled. Lastly, we note that the size of the neural network output layer will vary, as it is equal to the number of empirical time nodes for each model component.

\begin{table}[t!]
    \centering
    \begin{tabular}{c|cc}
    \hline
    \hline
     & $(\ell,m)$- & Euler \\
     & Amplitude \& Phase & Angles \\
    \hline
    X-data conditioning & \standard & None \\
    & & \\
    \multirow{2}{*}{Y-data conditioning} & None (Amplitude) & \minmax $(\alpha, \gamma)$  \\
    & \minmax (Phase) & None $(\beta)$ \\
    \hline
    Number of input neurons & $4$ & $4$ \\
    Number of layers & $4$ & $9$ \\
    Neurons per layer & $320$ & $128$ \\
    Optimiser & Adam & AdaMax \\
    Activation function & Softplus & Softplus \\
    Mini-batch size & $64$ & $512$ \\
    Number of training epochs & $10,000$ & $5000$ \\
    \hline
    \hline
    \end{tabular}
    \label{tab:ANNs}
    \caption{Details of the final neural network architecture for each component.}
\end{table}

\subsubsection{Coprecessing Modes}
\label{subsec:CoprecessingModes}
\begin{figure*}
    \centering
    \includegraphics[width=\textwidth]{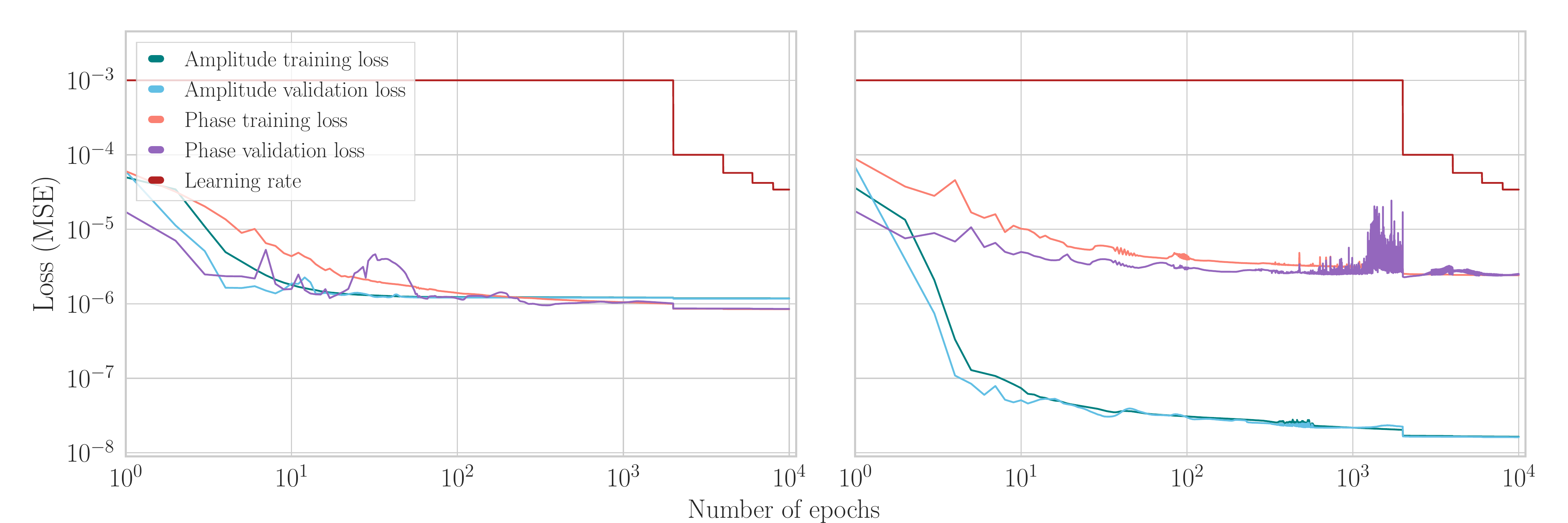}
    \caption{Training and validation losses for the $(2,2)$- (left) and $(2,1)$-mode (right), for both amplitude and phase. The loss shown is the mean squared error (MSE) as a function of training epochs. Also shown in both panels is the learning rate (red), which changes as a function of epoch as defined by Eq.~\eqref{eq:LearningRate}.}
    \label{fig:LossPlots}
\end{figure*}

When training our coprecessing mode neural networks, we use a mean squared error (MSE) loss function. This quantity can be computed over either the training dataset to monitor training progress, or the validation dataset as a control to check for over- or under-fitting. It is defined as 
\begin{equation}
    \textrm{MSE} = \frac{1}{N} \sum^{N}_{i=1} \left|\vec{y}^{~\textit{true}}_i - \vec{y}^{~\textit{pred}}_i\right|^2,
    \label{eq:MSE}
\end{equation}
where $\vec{y}_{i}^{~\textit{pred}}$ is the output from the neural network, $\vec{y}_{i}^{~\textit{true}}$ is the true \SEOB{} data, scaled accordingly if required and at the appropriate empirical time nodes, and $N$ is the number of points in either the training or validation set.

The final neural network architecture for the coprecessing modes is determined through optimisation via the hyperparameter sampling package \hyperopt \cite{Hyperopt:2013byc}. We parse choices for neural network hyperparameters, as well as a maximum number of neural network evaluations. The package then creates and trains neural networks with hyperparameters drawn from these choices, and returns the best performing hyperparameters based on a metric, which we specify to be the validation loss upon completion of training. The choices for optimisable hyperparameters are as follows: activation function (Relu \cite{fukushima1969visual,glorot2011deep}, Elu \cite{Clevert2015:elu}, Tanh, Softmax, Softplus, Softsign \cite{Elliott93abetter}); optimiser (Adam \cite{kingma2014adam}, Nadam \cite{dozat2016incorporating}, Adamax \cite{kingma2014adam}, Adadelta \cite{Zeiler:2012ad}), number of training epochs ($1000$, $2000$, $10000$); and mini-batch size ($32$, $64$, $128$). We refer the reader to \cite{Lederer2021:act} for a systematic overview of activation functions and \cite{Ruder2016:gdo} for an overview of gradient descent optimization algorithms. 
We also compare three sizes of neural network: 4 layers with 320 neurons per layer; 9 layers with 128 neurons per layer; and 4 layers where the number of neurons per layer is the next largest power of two from the reduced basis size. We find slightly improved performance with shallower, wider architectures, and so use the 4 layer, 320 neurons per layer architecture for the final networks. We also do not use dropout in our final configurations as we find this can create a lack of stability in training leading to higher mismatches. Our final optimal neural network architecture is detailed in Table III. 

Additionally, we use an adaptive learning rate as in Ref.~\cite{Khan:2020fso} in order to achieve faster convergence and prevent overshooting of the optimal trained weights. Our learning rate takes the form
\begin{equation}
    \tau_{i} = (\tau_{\text{init}} - \tau_{\text{final}}) / (1 + R \lfloor i / \Delta i \rfloor) + \tau_{\text{final}},
    \label{eq:LearningRate}
\end{equation}
where $\tau_{i}$ is the learning rate at epoch $i$, the initial learning rate $\tau_{\text{init}}=10^{-3}$, the final learning rate $\tau_{\text{init}}=10^{-5}$, the decay rate $R=10$, and our training epoch interval $\Delta i = 2000$. Thus our learning rate exhibits step-wise changes, decreasing every $2,000$ epochs. We use a mean squared error (MSE) as the loss metric, losses on both the training and validation datasets for the $(2,2)$ and $(2,1)$-mode amplitudes and phases are shown in Fig.~\ref{fig:LossPlots}, as well as the variable learning rate. We see that for the $(2,2)$-mode (left panel), both the amplitude and phase loss plateau around $10^{-6}$ after $\sim 100$ epochs of training, and for the $(2,1)$-mode (right panel) the phase reaches a similar loss plateau as the $(2,2)$-mode, however, the amplitude continues to improve to a loss value of $\sim 10^{-8}$. We also note that for all components, the training and validation losses are very comparable -- a sign that we are neither over- nor under-fitting in our training procedure. The training and validation losses for the $(3,3)$- and $(4,4)$-modes are shown in Fig.~\ref{fig:LossPlots3344} in Appendix~\ref{sec:app}.

\subsubsection{Euler Angles}
\label{subsec:EulerAngles}
In contrast to the architecture used for the coprecessing modes, for the Euler angles we use a network that is narrower and deeper consisting of 9 layers with 128 neurons per layer. We found that the Softplus activation function coupled with the Adamax optimizer produced robust results at the desired level of accuracy, though we did not perform the more exhaustive hyperparameter optimization used in the construction of the networks for the coprecessing modes. The networks are trained for $5000$ epochs using a mean squared error loss function, as defined in Eq.~\eqref{eq:MSE}. For the learning rate, we use an initial value of $10^{-2}$ and use an adaptive scheme that reduces the learning rate when the loss has stopped improving, as implemented by the \texttt{ReduceLROnPlateau} callback in \keras. We found no significant improvement when exploring the use of dropout regularization or $L^p$ regularizers\footnote{The $L^p$ norm is defined by $\| L \|_p = \sum_n \left( | x_n |^p \right)^{1/p}$ and we applied the regularization penalty to both the kernel and bias using the \texttt{L1L2 Class} in \keras.}, so do not include them in the final model.

In addition to the default network above, we also constructed a neural network for the residuals between the input empirical interpolation coefficients and the default neural network predictions (see also \cite{Fragkouli:2022lpt}), $\tilde{y}_k = y^{\rm true}_k - y^{\rm pred}_k$. This allows us to reconstruct the empirical interpolation coefficients using a two step procedure: we first evaluate the default neural network then we correct for any residual errors using the second network. However, we found this gave no noticeable improvement in accuracy. Due to the additional computational cost associated to the network evaluation, we opt not to use the residuals approach in the final model. 

\subsection{Complete Surrogate Model}
\label{subsec:CompleteSurrogateModel}
\begin{figure*}
    \centering
    \includegraphics[width=\textwidth]{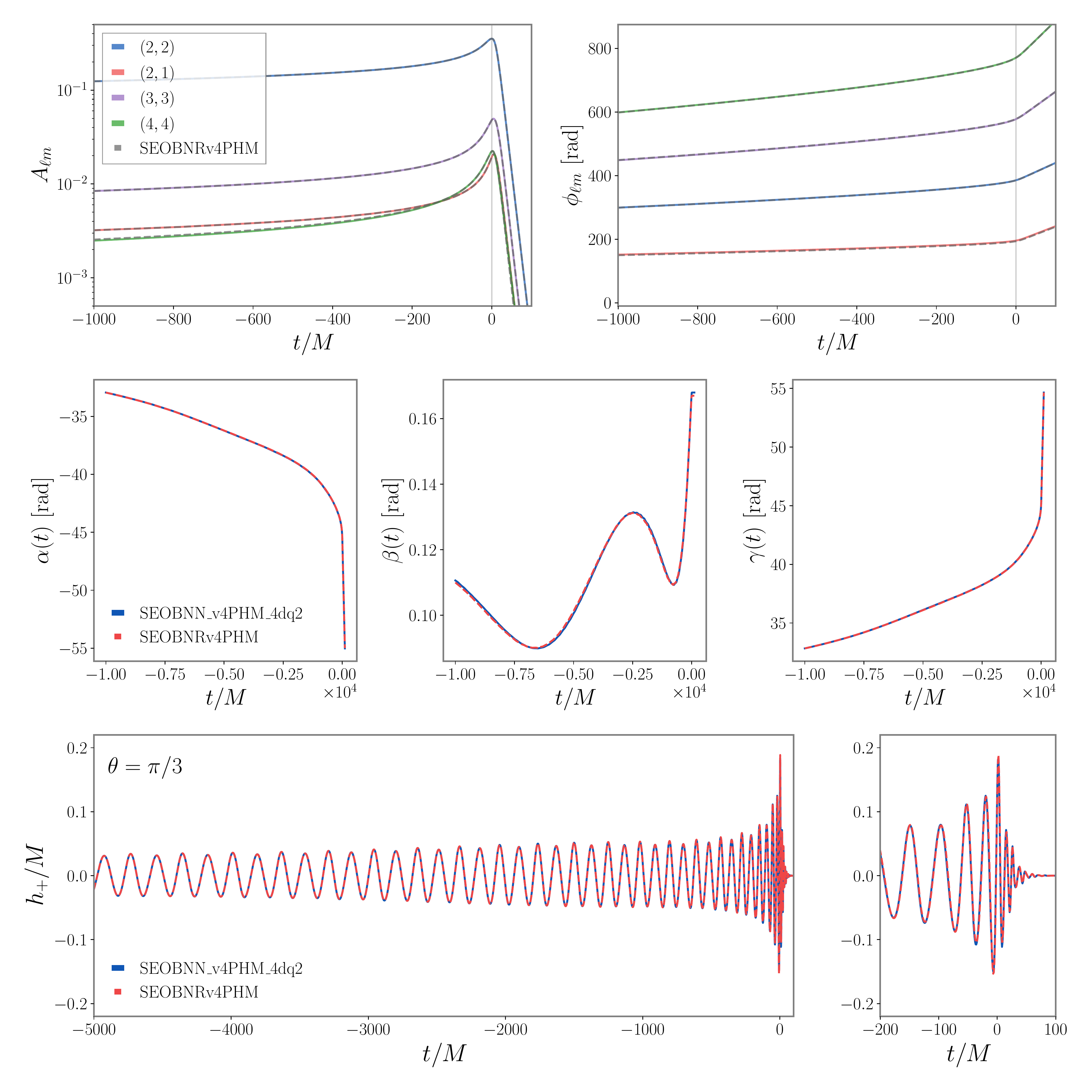}
    \caption{Top panel: Comparison of the coprecessing mode amplitudes (left) and phases (right) predicted by the surrogates (solid coloured lines) against the \SEOB{} data (dashed grey)for a fiducial binary with parameters $\lbrace q,\chi_{1x},\chi_{1y},\chi_{1z} \rbrace = \lbrace 1.86, 0.045, -0.283, 0.274 \rbrace$, where the Cartesian spin parameters are specified at a $(2,2)$-mode reference frequency of 4 Hz. The merger at $t=0M$ is indicated by the grey vertical line. Middle panel: Comparison of the Euler angles predicted by the neural network (blue) against the \SEOB{} data (red) for the fiducial binary. Bottom panel: The time-domain strain in the $J$-frame for our fiducial binary at an inclination of $\theta = \pi / 3$. We include all modes up to $\ell \leq 4$.}
    \label{fig:CoprecModesNNvsEOB}
\end{figure*}

Once the reduced bases and empirical interpolants are built and the neural networks have been trained, we have a total of 11 surrogate models for the different components that constitute the complete precessing model, \SEOBNN: The four coprecessing modes split into amplitudes and phases, and the three Euler angles. 
In Fig.~\ref{fig:CoprecModesNNvsEOB} we show an example for a fiducial binary with parameters $\vec{\lambda} = \lbrace q,\chi_{1x},\chi_{1y},\chi_{1z} \rbrace = \lbrace 1.86, 0.045, -0.283, 0.274 \rbrace$, i.e. a moderately precessing binary with a moderate unequal mass ratio. We note that this particular binary was not in our training or validation datasets. The top left panel shows the mode amplitudes as predicted by the surrogate for each coprecessing mode, the top right panel the corresponding phases. The \SEOB{} data are shown by the dashed curves in all panels. The middle panel shows the final surrogate models for the Euler angles. We note the excellent agreement between the true data and predictions, including around merger at $t=0 M$. In the bottom panel we show the time-domain strain (Eq.~\eqref{eq:strain}) obtained by combining the surrogate models (plus the conjugate modes) following the description in Eq.~\eqref{eq:modes}.  
We note, however, that we do not explicitly model the relative phase offsets between the coprecessing modes, which were incorporated manually from the true \SEOB{} data in the construction of the precessing strain. We leave the modelling of these relative phase offsets to future work.

Having seen the excellent agreement between prediction and true \SEOB{} data for a single fiducial binary, we now quantify the accuracy the surrogate models for each component across the parameter space.

\begin{figure}
    \centering
    \includegraphics[width=\columnwidth]{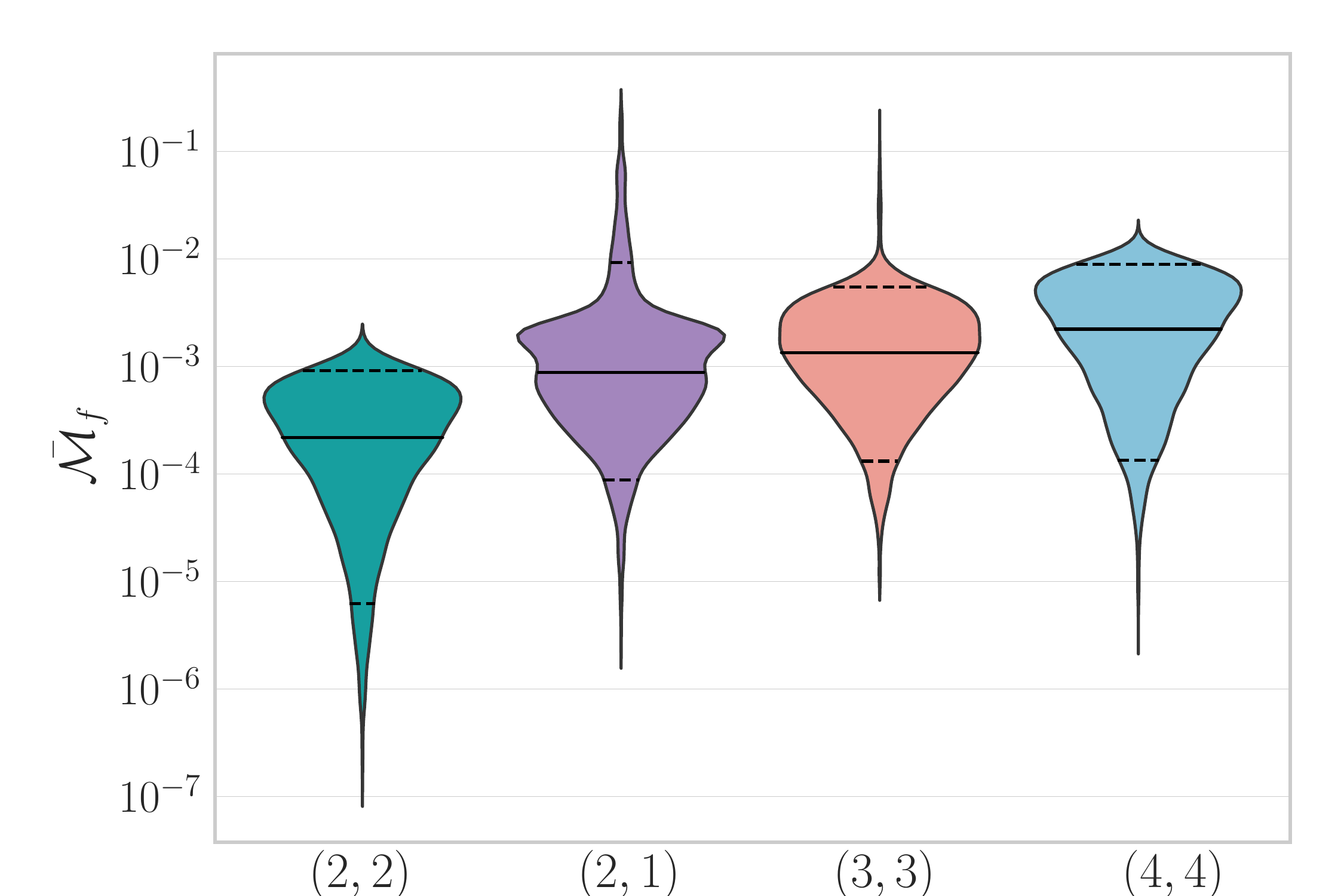}
    \caption{White noise mismatches between the \SEOB-generated coprecessing frame mode data and and the neural network-predicted coprecessing mode, for each of the four modes across the $10,000$ binary test set. Mismatch calculations start from an $f_{\text{min}} = 20$Hz for a total mass of $44 M_{\odot}$. Also shown for each coprecessing mode are the median mismatch (black) and 90\% intervals (black dashed).}
    \label{fig:CoprecMismatches}
\end{figure}

For each coprecessing mode we compute white noise frequency-domain mismatches $\bar{\mathcal{M}}_{f}$ between the true \SEOB{} coprecessing waveform modes and the surrogate predictions for the test dataset, which consists of $10^4$ waveforms that were not part of our training space $\mathcal{T}_M$ (see Sec.~\ref{subsec:TrainingAndValidationData}).  
We limit our mismatch integration to start at $f_\text{min}=20$ Hz and fix the total mass to $44 M_\odot$, which completely covers also the longest waveforms in our test set. The mismatch result for each of the four coprecessing modes is shown in in Fig.~\ref{fig:CoprecMismatches}. For each of the four coprecessing modes we find that the bulk of mismatches is less than $10^{-2}$ or $1\%$, with $4.6\%$ greater than this value for the $(2,1)$-mode, $0.8\%$ for the $(3,3)$-mode, and $2.6\%$ for the $(4,4)$-mode. For the $(2,2)$-mode  we find that it is less than $10^{-3}$ with only $3.3\%$ of mismatches greater than this, with a median mismatch of $\sim 3 \times 10^{-4}$. We find comparable performance for each of the three higher modes considered, with a median mismatch of $\sim 10^{-3}$, however we do note that there are tails of higher mismatches in the odd $m$-modes. Histograms of the mismatches for the coprecessing $(2,2)$ and $(2,1)$-modes at different total masses can be found in Fig.~\ref{fig:CoprecMismatchParamsMasses_22_21} in App.~\ref{sec:app}. We find almost identical results for the $(2,2)$-mode, and find small improvement for the $(2,1)$-mode as the total mass is increased. Therefore, the results in Fig.~\ref{fig:CoprecMismatches} represents the worst case scenario.

\begin{figure}
    \centering
    \includegraphics[width=\columnwidth]{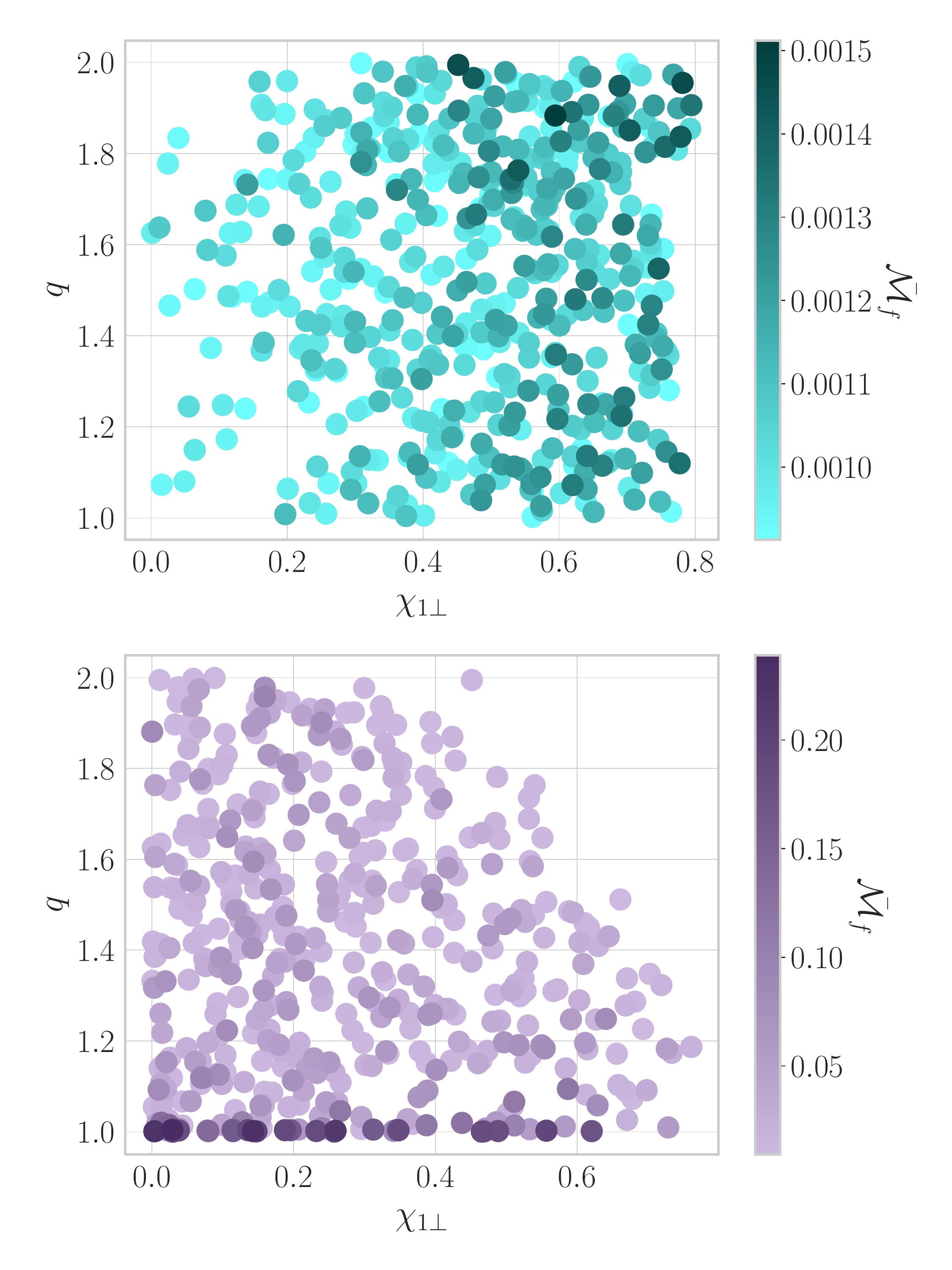}
    \caption{Worst $5\%$ of test dataset mismatches for the coprecessing $(2,2)$- (top), $(2,1)$-modes (bottom), shown in parameter space of mass ratio $q$ against in-plane spin magnitude $\left|\chi_{1\perp}\right|$. We note that the highest mismatches for the $(2,2)$-mode are scattered across much of this space, although with the worst mismatches at larger mass ratio and spin magnitude. In contrast, the worst mismatches for the $(2,1)$-mode lie in the region close to equal-mass where there is less asymmetry in the system and so this particular mode is heavily suppressed.}
    \label{fig:CoprecMismatchParams_22_21}
\end{figure}

To see where in parameter space the worst mismatches lie, particularly the high mismatch tails in the odd $m$-modes, we take the worst $5\%$ for each coprecessing mode and plot them in the space of mass ratio $q$ against $\chi_{1\perp}$, with the $(2,2)$- and $(2,1)$-modes shown in Fig.~\ref{fig:CoprecMismatchParams_22_21}, and the $(3,3)$- and $(4,4)$-modes in Fig.~\ref{fig:CoprecMismatchParams_33_44} in App.~\ref{sec:app}. We see that for the $(2,2)$- and $(4,4)$-modes, the highest mismatches lie broadly evenly across the parameter space, although with fewer high mismatches at low in-plane spin values. For the odd $m$-modes, however, the worst mismatches lie close to equal mass and at low in-plane spin values. In this region of parameter space, we expect the odd $m$-modes to be heavily suppressed, and so training data may be considerably more noisy, therefore leading to worse mismatches. It also means that when combining the modes into a full precessing strain, the contribution of these modes to the full signal is diminished and so will not have as much impact on the accuracy of the full waveform.

\begin{figure}[th!]
    \centering
    \includegraphics[width=\columnwidth]{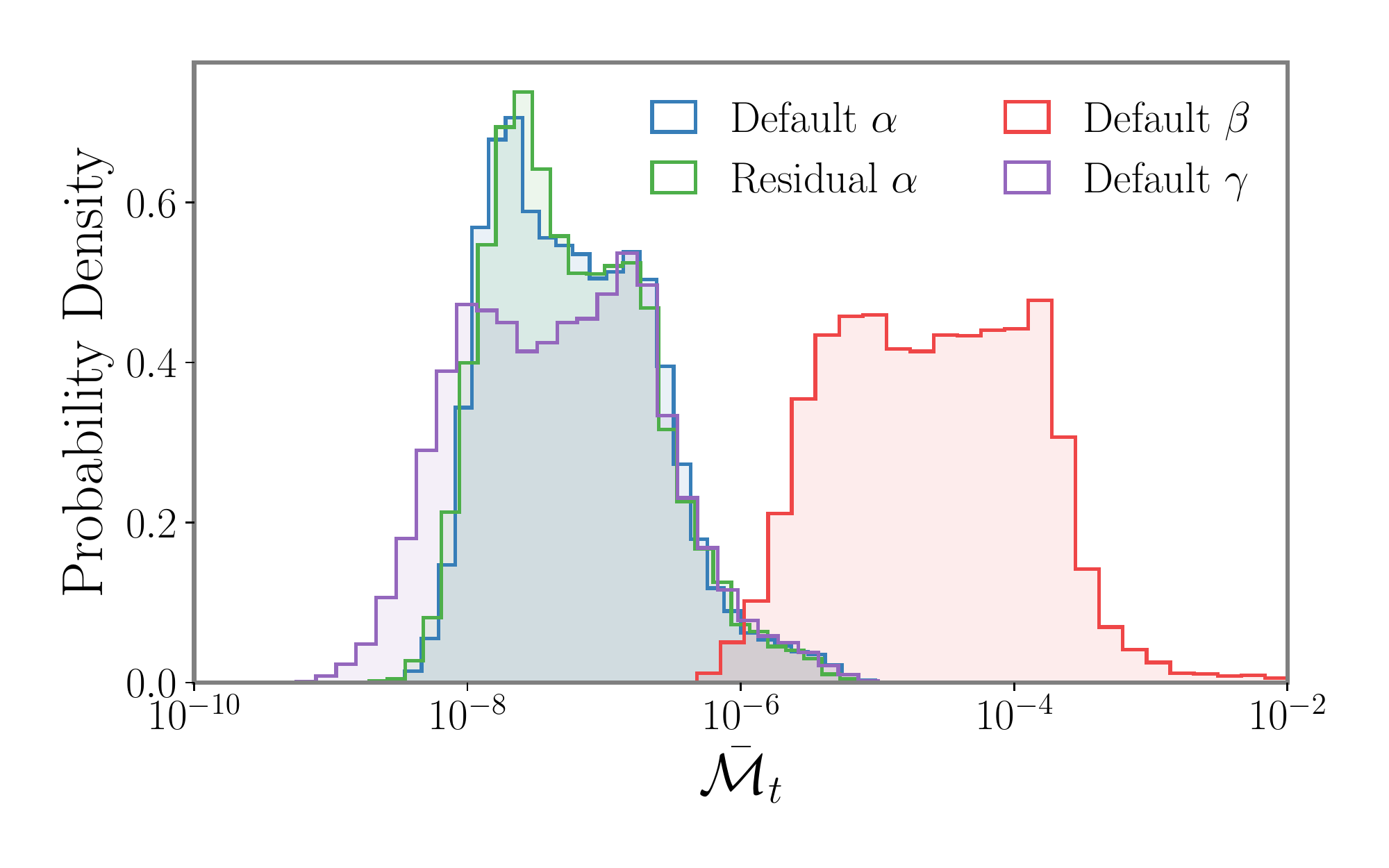}
    \caption{Time-domain mismatches for the surrogate model for the Euler angles against the training data. For $\alpha$, we show mismatches for the surrogate when using empirical coefficients predicted by the default network and when predicted by a two stage network that includes a fit to the residuals from the default network. We find no notable improvement in fitting the residuals.}
    \label{fig:EulerAngleAlpha_Residual}
\end{figure}

For the Euler angles, we use time-domain mismatches, see Eq.~\eqref{eq:MismatchTD}, as the main metric to quantify the accuracy of the surrogate prediction. We show the mismatches between the \SEOB{} data and the surrogate models for the Euler angles in Fig.~\ref{fig:EulerAngleAlpha_Residual}. We also demonstrate that the accuracy of the residual surrogate model outlined in Sec.~\ref{subsec:ANNs} offers no noticeable benefit with mismatches in broad agreement with our default model. 

Finally, whilst we find it convenient to work with the $SO(3)$ representation of the Euler angles, an appealing alternative approach is to parameterize the rotation group by a set of unit quaternions \cite{Boyle:2011gg,Boyle:2013nka}. Fundamentally, the quaternions still describe the time-dependent rotation of the frame but are endowed with a number of beneficial mathematical properties, such as the singularities that can occur in the Euler angle formalism. For the reduced parameter space considered here, we found no noticeable benefit to adopting the quaternion framework and opted to work with Euler angles out of simplicity. We leave a more detailed investigation of the wider parameter space to future work. 

\section{Model Evaluation}
\label{sec:ModelEvaluation}
\subsection{Waveform Accuracy}
\label{subsec:WaveformAccuracy}
\begin{figure}[th!]
    \centering
    \includegraphics[width=\columnwidth]{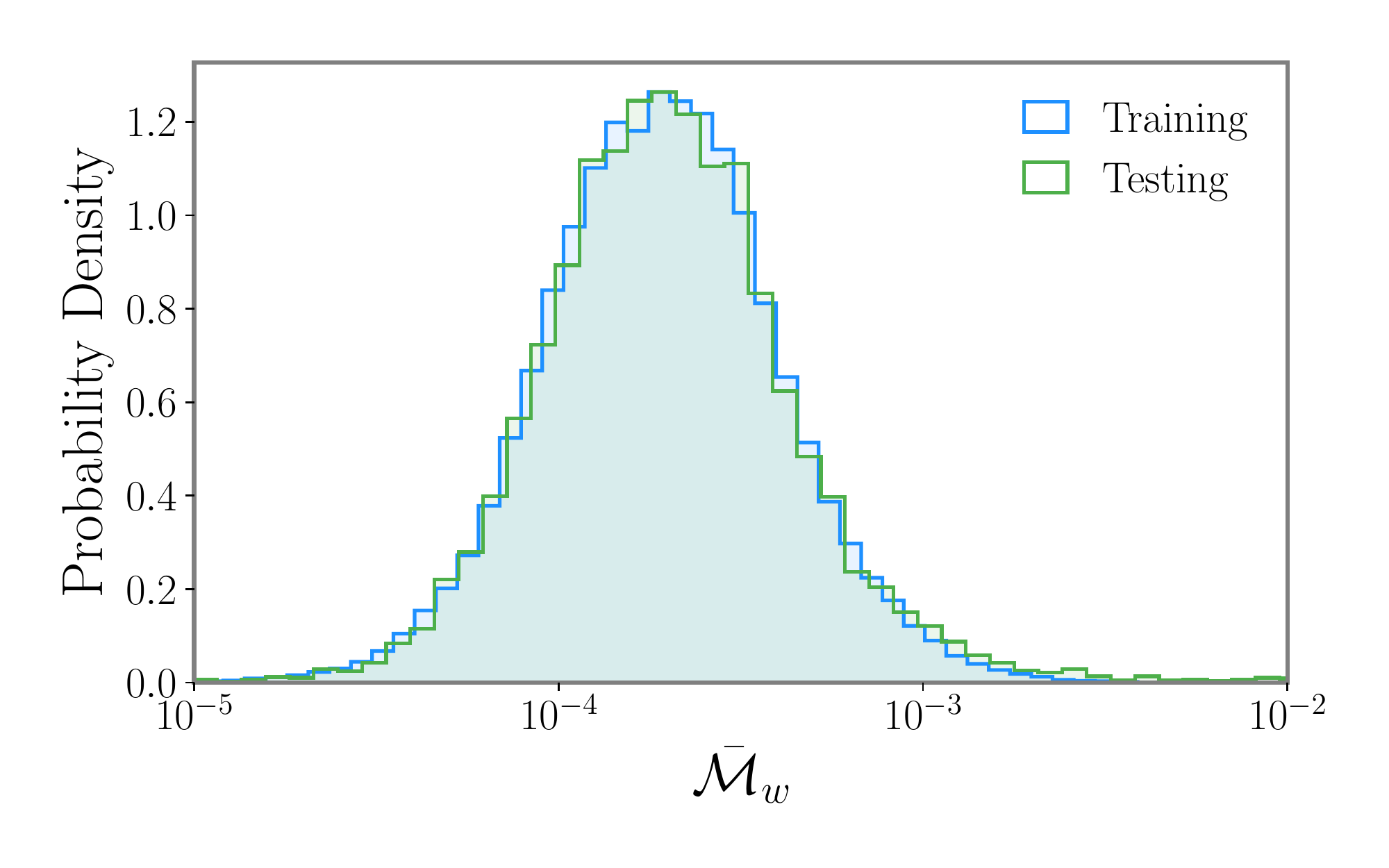}
    \caption{Orientation averaged mismatches for \SEOB{} against \SEOBNN{} for all $\leq 4$ modes in the $J$-frame. We show $9 \times 10^4$ binaries randomly drawn from the training (blue) dataset and $1 \times 10^4$ binaries randomly drawn from the independent testing (green) dataset, which the neural network has never been exposed to. We find excellent agreement irrespective of the dataset. 
    }
    \label{fig:ModelMismatches}
\end{figure}

The observed GW signal from single-spin precessing binary black holes depends on 12 parameters: the component masses $m_i$, the dimensionless spin $\vec{\chi}_1 (t)$, the direction from the source frame to the observer $(\iota,\phi_0)$, the polarization $\psi_0$, time of arrival $t$, the luminosity distance $d_L$ and the sky location $(\theta,\phi)$. Here we neglect the sky location and write the real-valued detector response $h_r(t)$ as 
\begin{align}
    h_r (t) = h_+ (t) \cos( 2 \psi) + h_{\times} (t) \sin (2 \psi).
\end{align} 
\newline 
where $h(t) = h_+(t) - i h_{\times}(t)$. We are now interested in validating the accuracy of our surrogate model, \SEOBNN, against the slow waveform model \SEOB{}. To do so, we calculate strain mismatches optimized over $\lbrace \psi, \varphi, t \rbrace$, as these quantities are not astrophysically relevant. We follow the approaches detailed in \cite{Schmidt:2014iyl,Harry:2016ijz,Pratten:2020ceb,Ossokine:2020kjp} and numerically optimize over the phase $\phi$ and analytically maximize over the template polarization $\psi$ and relative time shift $t$,
\begin{align}
    \mathcal{M}_{\kappa} (\phi_0^s,\psi^s_0) = \max_{t^h_0,\phi^h_0,\psi^h_0} \langle \hat{h}_r, \hat{s}_r  (\phi_0^s,\psi^s_0)  \rangle_f, 
\end{align}
\newline 
where $h_r$ denotes the template waveform, generated by our \SEOBNN{} surrogate, and $s_r$ is the signal waveform, taken to be \SEOB{}. We use the index $\kappa$ to distinguish the match optimised over the polarisation angle from Eq.~\eqref{eq:Match}.
Finally, we average the match by weighting each waveform (indexed by $i$) by its optimal signal-to-noise ratio $\rho$ to account for the likelihood that the signal would have been detected. This allows us to define an orientation-averaged match as \cite{Harry:2016ijz}
\begin{align}
    \mathcal{M}_{w} = \left( \frac{\sum_i \mathcal{M}_{\kappa,i}^3 \, \rho^3_i }{\sum_i \rho^3_i} \right)^{1/3},
\end{align}
\newline 
and the concomitant orientation-averaged mismatch $\bar{\mathcal{M}}_w = 1 - \mathcal{M}_w$. For the match calculation, we assume a lower cutoff frequency of $20 \rm{Hz}$ and use the projected PSD for Advanced LIGO in the upcoming fourth observing run (O4) \cite{KAGRA:2013rdx}, consisting of the Advanced LIGO and Virgo detectors as well as KAGRA. We take the masses to be uniformly distributed between $50 M_{\odot}$ and $200 M_{\odot}$ and the orientation angles to be isotropic on the unit sphere. The mass ratio, spin magnitude and spin orientation are as described in Table I. We reiterate that to construct the full precessing strain from our ANN waveform model, here we use the true \SEOB{} relative phase offsets between the coprecessing modes. The resulting mismatches are shown in Fig.~\ref{fig:ModelMismatches} using all $\ell \leq 4$ modes in the inertial $J$-frame as per Eq.~\eqref{eq:modes}. We show mismatches against the training dataset, used to construct our ANN waveform model, and the independent testing dataset to which the model has never been exposed. For both datasets we find excellent agreement and find a median mismatch of $1.9 \times 10^{-4}$. The $5$th and $95$th percentiles for the mismatches against the training dataset are $5.8 \times 10^{-5}$ and $6.5 \times 10^{-4}$ respectively. The mismatch errors here are approximately an order of magnitude below the anticipated error of \SEOB{} against precessing numerical relativity simulations \cite{Ossokine:2020kjp}. We find that the error of our model against the input data is competitive with the accuracy provided by other surrogate models, e.g. \cite{Blackman:2015pia,Varma:2019csw}. 

\subsection{Timing}
\label{subsec:Timing}
In order to test the efficiency of our surrogate model, we developed two interfaces. The first interface is built exclusively within the \numpy framework. The second interface uses the \tensorflow framework to provide GPU acceleration. When run on a single CPU, we find broad parity between the computational efficiency of the two implementations. However, when run on a GPU, the implementation in \tensorflow allows for significant computational speedup, as discussed below. 

The typical evaluation time for a single Euler angle surrogate model is on the order of $250 \mu$s. As a reminder, this includes the computational cost of producing a single prediction for the empirical interpolation coefficients from the neural network as well as the multiplication by the empirical interpolation matrix. The amplitude and phase surrogate models are marginally slower such that each waveform mode $h_{\ell m} = A_{\ell m} e^{-i \phi_{\ell m}}$ takes $\sim 925 \mu$s to generate. 

In required model components are constructed from 11 individual surrogate models: 3 Euler angles and 4 waveform modes plus their conjugates. To evaluate all 3 Euler angles takes $\sim 750 \mu$s and to evaluate the 4 waveform mode surrogates takes $\sim 3.7$ms. Array conjugation is a significantly cheaper operation requiring only $\sim 10 \mu$s per array. Evaluation of the waveform modes is the single most expensive operation in our model. 

Next we need to evaluate the Wigner-D matrices, $D^{\ell}_{m m^{\prime}} (\alpha,\beta,\gamma)$, in order to perform the time-dependent rotations. This is the second most expensive operation in \SEOBNN. In order to mitigate against the computational cost, we can perform a series of optimizations, such as pre-caching of numerical coefficients. This allows us to significantly reduce the cost of evaluating the Wigner-D matrices to $\sim 5.5$ms. Further optimization could be achieved through the use of interpolating non-uniform grids or pre-compilation in C. We leave such optimizations to the future. Performing the time-dependent rotations of the waveform modes from the $L$-frame to the $J$-frame is relatively efficient, requiring only $\sim 2$ms.

\begin{figure}
    \centering
    \includegraphics[width=0.99\columnwidth]{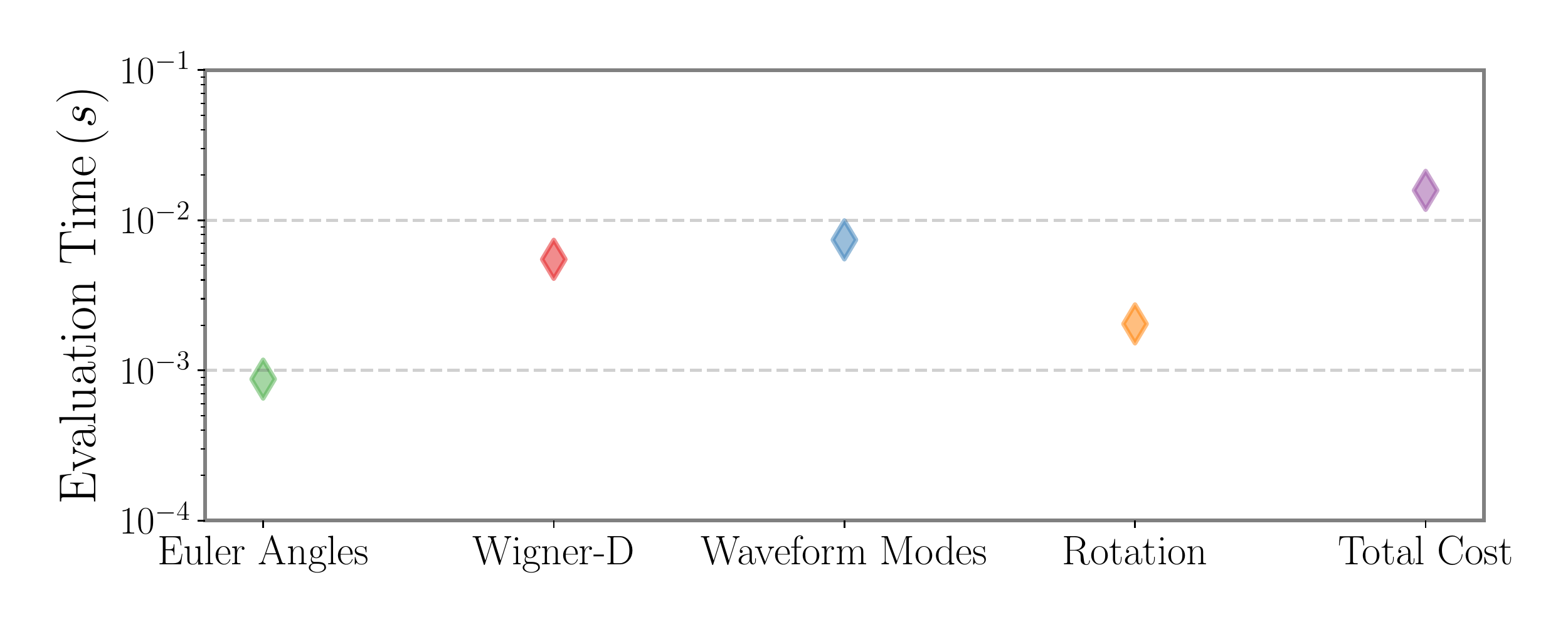}
    \caption{Computational cost for each step in the waveform construction. A notable bottleneck is the computation of the Wigner-D matrices $D^{\ell}_{m m^{\prime}}(\alpha,\beta,\gamma)$ over the full $10^4 M$ time grid.}
    \label{fig:timings}
\end{figure}

Altogether, we find that the typical waveform generation cost for a signal covering the surrogate length of $10,000M$ is on the order of $18$ms on a single CPU with \SEOBNN. This is on average $\mathcal{O}(10^2)$ times faster than the underlying \SEOB{} model, which takes $\sim 3000$ms. In addition, it is also almost three times as fast as the surrogate model presented in \cite{Gadre:2022sed}, though the surrogate model presented here is twice as long in duration spanning $10^4 M$ compared to $5 \times 10^3 M$ in \cite{Gadre:2022sed}. A notable caveat is that the surrogate model presented in \cite{Gadre:2022sed} covers a significantly larger domain of the parameter space making any direct comparison difficult. Nevertheless, the preliminary model presented here suggests that reduced order models for precessing multipolar waveform models powered by neural networks are highly competitive relative to alternative strategies, even on a CPU. We show the typical timings for each element and for the entire waveform in Fig.~\ref{fig:timings}. All CPU timings were generated using an Intel(R) Core(TM) i7-9750H CPU $@$ 2.60GHz using the \numpy interface. 

\begin{figure}
    \centering
    \includegraphics[width=\columnwidth]{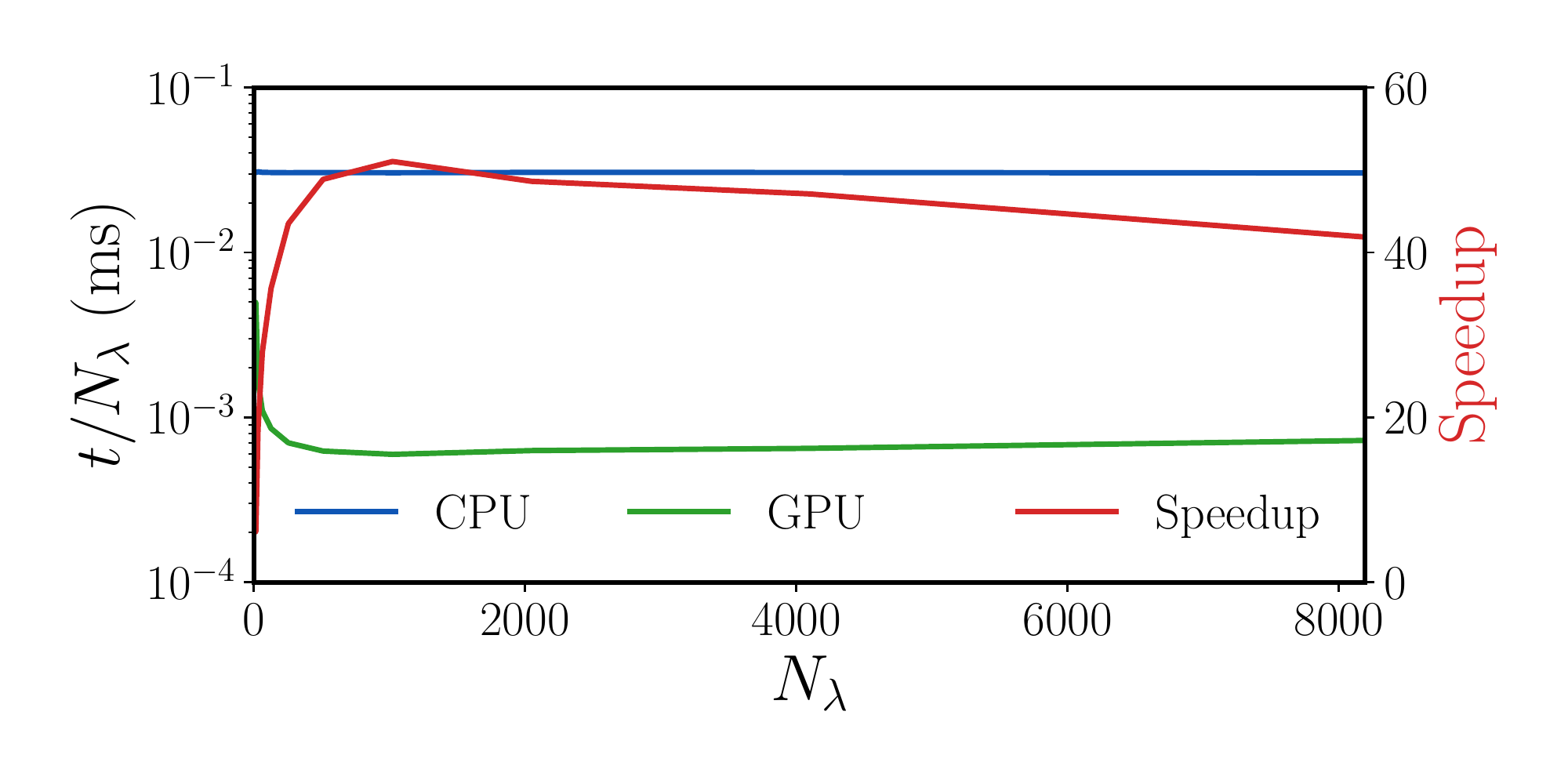}
    \caption{Computational cost per binary for evaluating the $22$-mode surrogate model over batches of $N_{\lambda}$ binaries. We show both CPU and GPU timings as well as the overall speedup enabled by GPU acceleration. }
    \label{fig:gpus}
\end{figure}

However, a significant benefit of reduced order models powered by the \tensorflow architecture is that they provide a convenient platform for GPU acceleration. In particular, GPU acceleration is most beneficial when generating batches of surrogate models, mitigating any overhead in the transfer of data between the CPU and the GPU. Evaluating the surrogate model for the $22$-mode over a varying number of binaries, we find that GPU acceleration leads to a factor $\sim \mathcal{O}(30)$ speedup in surrogate generation cost relative to CPUs. For $8192$ binary configurations, we find that on a CPU each surrogate model takes $\sim 30$ms compared to $\sim 0.7$ms on a GPU. For the CPU-GPU benchmarking, CPU timings were performed using an Intel(R) Xeon(R) CPU @ 2.30GHz and GPU timings were performed using an NVIDIA \textrm{Tesla P100-PCIE-16GB}. We show the comparative CPU and GPU timings in Fig.~\ref{fig:gpus} along with the relative speedup provided by GPU acceleration. 
 
\section{Discussion}
\label{sec:Discussion}
As the number of observations of GW signals from BBH mergers is only set to increase with improving detector sensitivity, the availability of accurate, highly computationally efficient theoretical models is critical for future GW data analysis. Faster surrogate models of very accurate but slower underlying waveform models will prove beneficial, motivating a need for exploration of novel surrogate-building methods. In this paper, we have constructed a proof-of-concept time-domain surrogate model of \SEOB{}, which makes use of neural networks to perform parameter space fits. We follow the techniques used in Refs.~\cite{Chua:2019wwt, Khan:2020fso}, extending them to precessing multipolar waveforms for the full inspiral-merger-ringdown signal. We decompose our surrogate model into two sections: coprecessing waveform modes defined in the non-inertial coprecessing $L$-frame which tracks the precessing motion of the binary, and the three Euler angles which represent the rotation between this frame and the $J$-frame. We consider four coprecessing modes: the dominant quadrupolar $(2,2)$-mode, and three higher multipoles $(2,1)$, $(3,3)$ and $(4,4)$, and each of these modes is then decomposed further into amplitude and phase. Therefore we model a total of $11$ components. For each component, we construct a reduced basis and empirical interpolant, before performing parameter space fits using artificial neural networks.

We demonstrate that the performance of our surrogate \SEOBNN{} is highly competitive in comparison to alternative surrogate modelling strategies, producing waveforms with precessing strain mismatches $\sim \mathcal{O}\left(10^{-3} - 10^{-4}\right)$ against the true \SEOB{} data. We also show that this model is computationally efficient, producing waveforms on a CPU two orders of magnitude faster than the underlying \SEOB{} model, and almost three times as fast as the recently developed surrogate model \cite{Gadre:2022sed} in the restricted intrinsic parameter space covered by our ANN model. We also note that our output waveforms are around two times longer than this surrogate, and that unlike the underlying \SEOB{} model, the evaluation time is independent upon the binary parameters. Additionally, we have shown that our surrogate model allows for an even more significant speed up in evaluation time when evaluating batches of waveforms simultaneously on GPUs. 

As a proof of concept for neural network surrogates of precessing multipolar waveforms, our model is built on a restricted parameter range of mass ratios $q\in\left[1,2\right]$ and single precessing spins $\left|\chi_{1}\right|\leq 0.8$, $\left|\chi_{2}\right|= 0$. This multidimensional portion of the precessing BBH parameter space is a starting point for surrogates which utilise neural networks, though we do not envisage any imminent roadblocks to incorporating additional information in order to extend towards the full 7D intrinsic parameter space of double precessing spins, with more unequal mass ratios. We note, however, that the size of training dataset would need to be significantly larger to accurately represents the full range of waveforms in this larger parameter space. Additionally, any higher dimensional training dataset would need to be thoroughly checked for data quality across the parameter space, as we noted that even in our restricted parameter we faced issues of pathologies in the underlying waveform model, where the coprecessing mode phases became discontinuous in the inspiral, possibly due to inaccurate next-to-quasicircular corrections in \SEOB{}. 

To explore how accurately our model can extrapolate outside the training range, we tested each coprecessing mode surrogate on $1,500$ single-spin binaries with mass ratio in $q\in\left(2,4\right]$ or with primary spin magnitudes $0.8<\left|\chi_{1}\right|\leq 0.99$ and computed mismatches against the true \SEOB{} data. 
We found that the extrapolation in spin magnitude is relatively smooth as long as the mass ratio is constrained to values that were in the original training space (i.e. $q\leq 2$), resulting in mismatches for each coprecessing mode approximately one order of magnitude worse than shown in Fig.~\ref{fig:CoprecMismatches}. However, for binaries with $q>2$, irrespective of the in-plane spin magnitude, each mode surrogate performs poorly. The same trends were observed for the Euler angles.

Additionally, we investigated whether our model is able to capture the behaviour of binaries with two spinning black holes by using the previously developed dimensional reduction mapping of \cite{Thomas:2020uqj}. To do so, we constructed $1,000$ double-spin binaries with parameters inside the training space, ensuring that the mapped spin magnitude was $\leq 0.8$. We found that the coprecessing $(2,2)$-mode is replicated with a mismatch accuracy of $\mathcal{O}(10^2 - 10^3)$, but that higher modes are less well reproduced.

When building our surrogate model, we explored several options to improve the accuracy of the coprecessing modes neural network fit. Before training the artificial neural networks for the coprecessing modes, we tried using principal component analysis on the reduced basis coefficient phase training data, to identify trend directions in the data which may be easier for the neural network to fit. Whilst this provided a small improvement in resulting mismatches for the $(2,2)$-mode, it led to marginally worse results for the $(2,1)$-mode and no noticeable difference in the $(3,3)$ and $(4,4)$-modes. We also attempted to improve the mismatches of our coprecessing modes by training the neural networks for longer than $10,000$ epochs. However, between $10,000$ and $100,000$ epochs, almost no improvements were seen in the loss values for both amplitudes and phases for all modes. Furthermore we tried training on the residual coprecessing phase, where the geometric mean has been subtracted to de-trend the phase data. We find this had no impact on either the reduced basis sizes or the accuracy with which we were able to train our artificial neural networks. For the Euler angles, we explored the possibility of training an additional neural network to model the residual error on the predicted $\alpha$, but found no noticeable improvement.

In addition, we also explored the effect of different sizes of training data sets upon the accuracy of the coprecessing mode fits. We found that the reduced basis size and projection errors were insensitive to smaller training set sizes for sets above $100$ waveforms, and similarly that the coprecessing mode mismatches for the $(2,1)$-mode shown in Fig.~\ref{fig:CoprecMismatches} were virtually identical when the $(2,1)$-phase was reconstructed on a random training subset of $10,000$. This suggests that our choice of training set size may have been conservative, and future models over this parameter space could attain similar accuracies with smaller training set sizes.

We have demonstrated the feasibility and efficacy of using neural networks as part of precessing multipolar IMR waveform surrogate models, and leave the extension to the full 7D precessing parameter space as well as the modelling of the spin evolution to further work.
We suggest that with even further consideration given to neural network optimisation and data de-trending over the full 7D parameter space of generically precessing BBHs, this could prove a promising pathway towards accurate, efficient gravitational waveform surrogate model building.

\section*{Acknowledgments}
The authors thank Alberto Vecchio for useful discussions and Vijay Varma for comments on the manuscript.
L.M.T. is supported by STFC, the School of Physics and Astronomy at the University of Birmingham and the Birmingham Institute for Gravitational Wave Astronomy. GP is grateful for support from a Royal Society University Research Fellowship URF{\textbackslash}R1{\textbackslash}221500 and STFC grant ST/V005677/1. P.S. acknowledges support from STFC grant ST/V005677/1 and the Dutch Research Council (NWO) Veni Grant No. 680-47-460. GP gratefully acknowledges support from an NVIDIA Academic Hardware Grant. Computations were performed using the University of Birmingham's BlueBEAR HPC service, which provides a High Performance Computing service to the University's research community, as well as on resources provided by Supercomputing Wales, funded by STFC grants ST/I006285/1 and ST/V001167/1 supporting the UK Involvement in the Operation of Advanced LIGO. Some computations were also performed using Google Colaboratory. Part of this research was performed while L.M.T., G.P. and P.S. were visiting the Institute for Pure and Applied Mathematics (IPAM), which is supported by the National Science Foundation (Grant No. DMS-1925919).
This manuscript has the LIGO document number P2200161.

\appendix
\section*{Appendix}
\label{sec:app}
\begin{figure}[h!]
    \centering
    \includegraphics[width=\textwidth]{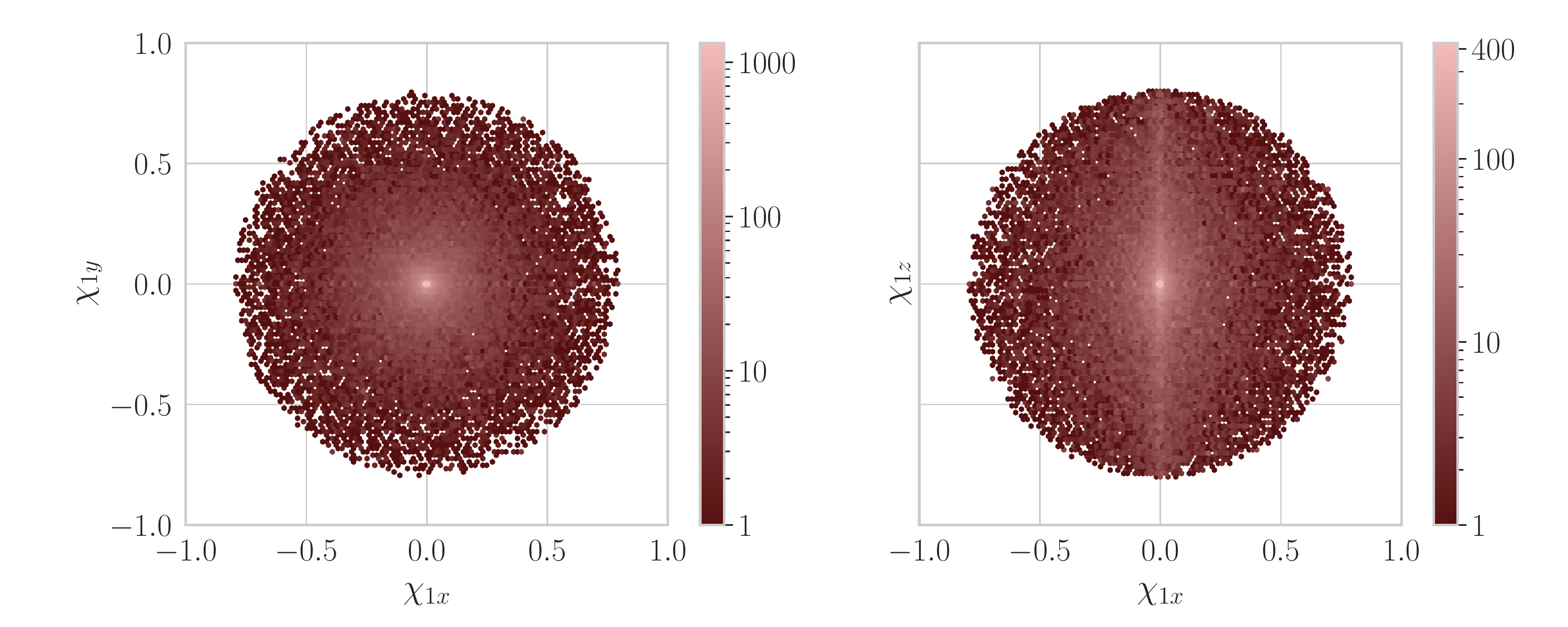}
    \caption{Visualisation of the spin parameters of the entire training dataset coloured by number density.}
    \label{fig:spins}
\end{figure}

Here we first show in Fig.~\ref{fig:spins} the distribution of the $2\times 10^5$ waveform training dataset, plotted in the space of primary spin components and coloured by number density. This dataset is made up of a systematically sampled subset of $774$ points (whose parameters are specified in Table I) in order to effectively sample the boundaries, as well as $199,226$ randomly sampled binaries, uniform in mass ratio, spin magnitude, azimuthal and tilt angles.

\begin{figure}
    \centering
    \includegraphics[width=\textwidth]{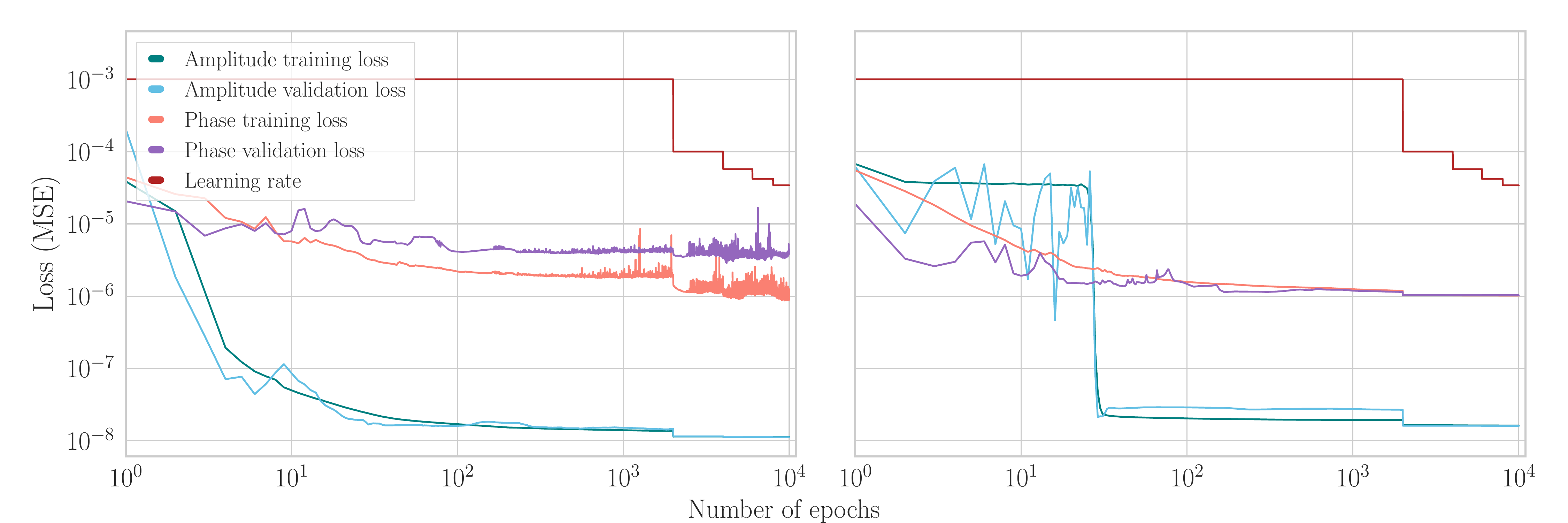}
    \caption{Training and validation losses for the $(3,3)$ (left) and $(4,4)$ (right) -modes, both amplitude and phase. The loss shown is the mean squared error (MSE) as a function of epochs trained. Also shown in both panels is the learning rate, which changes as a function of epoch as defined by Eq.~\eqref{eq:LearningRate}.}
    \label{fig:LossPlots3344}
\end{figure}

Next we show in Fig.~\ref{fig:LossPlots3344} the training and validation losses as defined by Eq.~\eqref{eq:MSE}, for both the amplitude and phase of the coprecessing $(3,3)$- (left) and $(4,4)$ (right) -modes, over the course of the neural network training. We also plot the adaptive learning rate, defined by Eq.~\eqref{eq:LearningRate}. We note that similarly to the $(2,1)$-mode amplitude as seen in Fig.~\ref{fig:LossPlots}, the amplitude losses for the $(3,3)$- and $(4,4)$-modes evolve to a minimum of around $10^{-8}$ at the end of training, and the phases to around $10^{-6}$. Additionally, we see that the training and validation losses remain similar in magnitude throughout the training process, suggesting we are neither over- nor under-fitting.

\begin{figure}
    \centering
    \includegraphics[width=\textwidth]{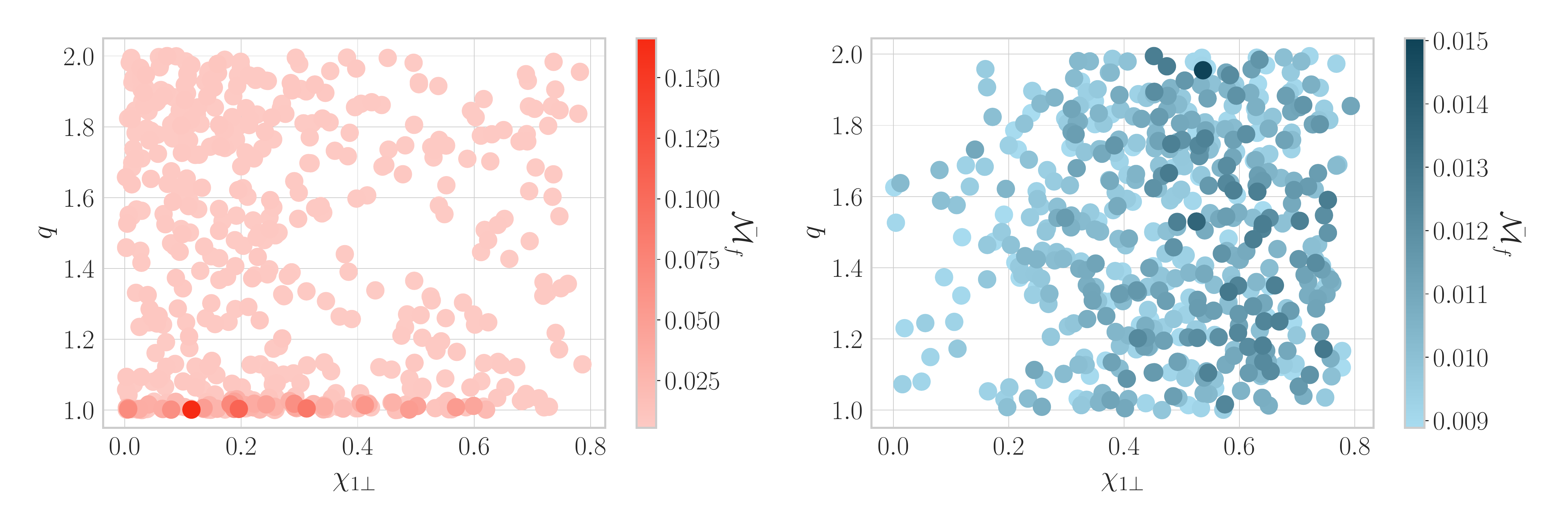}
    \caption{Worst $5\%$ of test dataset mismatches $\bar{\mathcal{M}}_f$ for the $(3,3)$- (left) and $(4,4)$ (right), modes, shown in parameter space of mass ratio and in-plane spin magnitude, and coloured by mismatch.}
    \label{fig:CoprecMismatchParams_33_44}
\end{figure}

We show in Fig.~\ref{fig:CoprecMismatchParams_33_44} the worst 5\% of mismatches $\bar{\mathcal{M}}_f$ over the test dataset between the surrogate-predicted coprecessing waveform modes and the true \SEOB{} data, for the $(3,3)$- (left) and $(4,4)$- (right) modes, plotted over mass ratio $q$ and in-plane spin magnitude $\left|\chi_{1\perp}\right|$ and coloured by mismatch. We see that for the $(3,3)$-mode, similarly to the $(2,1)$-mode in Fig.~\ref{fig:CoprecMismatchParams_22_21}, the worst mismatches appear around equal mass and less in-plane spin where there is less asymmetry in the system and so these modes are heavily suppressed in the full precessing strain. In contrast, and similarly to the $(2,2)$-mode in Fig.~\ref{fig:CoprecMismatchParams_22_21}, the $(4,4)$-mode exhibits lower mismatches overall, and more evenly spread across the parameter space, although the worst mismatches tend to be at more unequal mass ratios and larger in-plane spins.

\begin{figure}
    \centering
    \includegraphics[width=\textwidth]{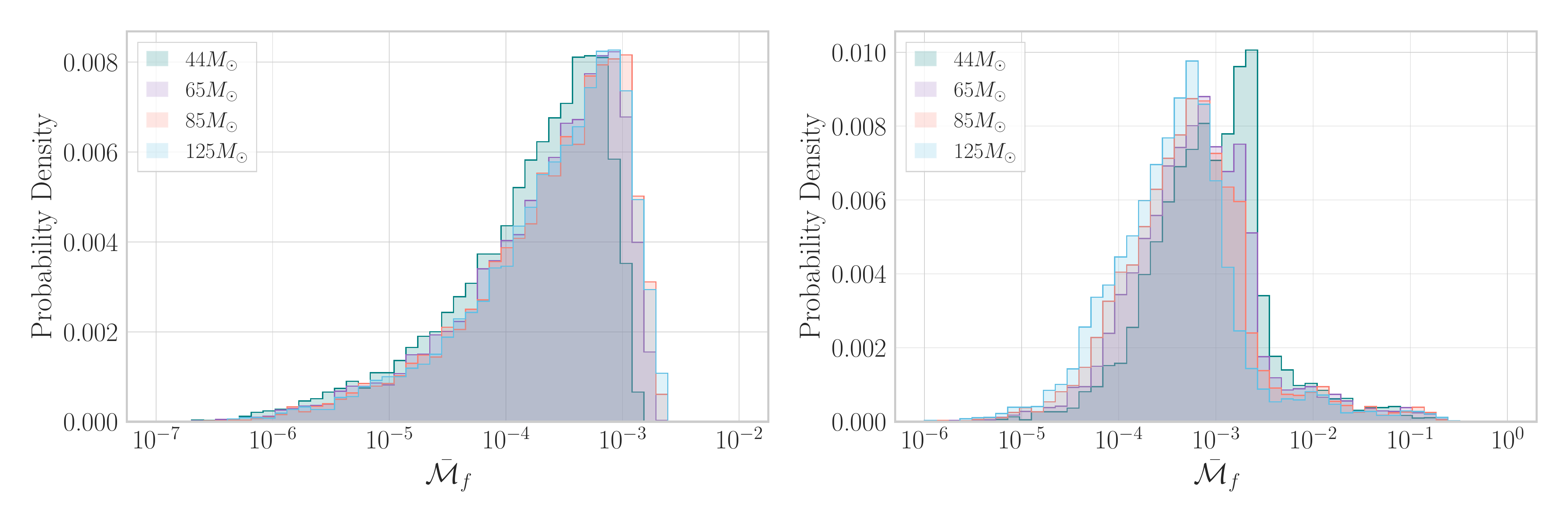}
    \caption{White noise mismatches between the \SEOB{}-generated coprecessing frame mode data and and the neural network-predicted coprecessing mode, for the $(2,2)$ (left) and the $(2,1)$-mode (right), as computed in Fig.~\ref{fig:CoprecMismatches} but for a range of different total masses $M_{\rm{tot}}=\left[44, 65, 85, 125\right] M_{\odot}$. Mismatch calculations start from an $f_{\rm{min}} = 20$Hz and are computed across the $10,000$ binary test set.}
    \label{fig:CoprecMismatchParamsMasses_22_21}
\end{figure}

Finally, we show in Fig.~\ref{fig:CoprecMismatchParamsMasses_22_21} the effect of changing the total mass of the binary upon the coprecessing mode mismatches shown in Fig.~\ref{fig:CoprecMismatches}. We choose a representative sample of four total masses $M_{\rm{tot}}\in \lbrace 44, 65, 85, 125\rbrace M_{\odot}$ and recompute mismatches in the same way as shown in Fig.~\ref{fig:CoprecMismatches}, from a low frequency cutoff of $20$ Hz each time, over the $10,000$ binary test set, for each of the $(2,2)$- (left) and $(2,1)$-modes. We find that the change in total mass makes little difference in both cases, and that in fact a higher total mass than $44 M_{\odot}$ slightly improves the $(2,1)$-mode mismatches, as may be expected since the higher total mass is effectively a decrease in waveform length.

\onecolumngrid

\clearpage
\twocolumngrid
\bibliography{References}

\begin{thebibliography}{74}%
\makeatletter
\providecommand \@ifxundefined [1]{%
 \@ifx{#1\undefined}
}%
\providecommand \@ifnum [1]{%
 \ifnum #1\expandafter \@firstoftwo
 \else \expandafter \@secondoftwo
 \fi
}%
\providecommand \@ifx [1]{%
 \ifx #1\expandafter \@firstoftwo
 \else \expandafter \@secondoftwo
 \fi
}%
\providecommand \natexlab [1]{#1}%
\providecommand \enquote  [1]{``#1''}%
\providecommand \bibnamefont  [1]{#1}%
\providecommand \bibfnamefont [1]{#1}%
\providecommand \citenamefont [1]{#1}%
\providecommand \href@noop [0]{\@secondoftwo}%
\providecommand \href [0]{\begingroup \@sanitize@url \@href}%
\providecommand \@href[1]{\@@startlink{#1}\@@href}%
\providecommand \@@href[1]{\endgroup#1\@@endlink}%
\providecommand \@sanitize@url [0]{\catcode `\\12\catcode `\$12\catcode
  `\&12\catcode `\#12\catcode `\^12\catcode `\_12\catcode `\%12\relax}%
\providecommand \@@startlink[1]{}%
\providecommand \@@endlink[0]{}%
\providecommand \url  [0]{\begingroup\@sanitize@url \@url }%
\providecommand \@url [1]{\endgroup\@href {#1}{\urlprefix }}%
\providecommand \urlprefix  [0]{URL }%
\providecommand \Eprint [0]{\href }%
\providecommand \doibase [0]{http://dx.doi.org/}%
\providecommand \selectlanguage [0]{\@gobble}%
\providecommand \bibinfo  [0]{\@secondoftwo}%
\providecommand \bibfield  [0]{\@secondoftwo}%
\providecommand \translation [1]{[#1]}%
\providecommand \BibitemOpen [0]{}%
\providecommand \bibitemStop [0]{}%
\providecommand \bibitemNoStop [0]{.\EOS\space}%
\providecommand \EOS [0]{\spacefactor3000\relax}%
\providecommand \BibitemShut  [1]{\csname bibitem#1\endcsname}%
\let\auto@bib@innerbib\@empty
\bibitem [{\citenamefont {Aasi}\ \emph {et~al.}(2015)\citenamefont {Aasi} \emph
  {et~al.}}]{LIGOScientific:2014pky}%
  \BibitemOpen
  \bibfield  {author} {\bibinfo {author} {\bibfnamefont {J.}~\bibnamefont
  {Aasi}} \emph {et~al.} (\bibinfo {collaboration} {LIGO Scientific}),\
  }\bibfield  {title} {\enquote {\bibinfo {title} {{Advanced LIGO}},}\ }\href
  {\doibase 10.1088/0264-9381/32/7/074001} {\bibfield  {journal} {\bibinfo
  {journal} {Class. Quant. Grav.}\ }\textbf {\bibinfo {volume} {32}},\ \bibinfo
  {pages} {074001} (\bibinfo {year} {2015})},\ \Eprint
  {http://arxiv.org/abs/1411.4547} {arXiv:1411.4547 [gr-qc]} \BibitemShut
  {NoStop}%
\bibitem [{\citenamefont {Acernese}\ \emph {et~al.}(2015)\citenamefont
  {Acernese} \emph {et~al.}}]{VIRGO:2014yos}%
  \BibitemOpen
  \bibfield  {author} {\bibinfo {author} {\bibfnamefont {F.}~\bibnamefont
  {Acernese}} \emph {et~al.} (\bibinfo {collaboration} {VIRGO}),\ }\bibfield
  {title} {\enquote {\bibinfo {title} {{Advanced Virgo: a second-generation
  interferometric gravitational wave detector}},}\ }\href {\doibase
  10.1088/0264-9381/32/2/024001} {\bibfield  {journal} {\bibinfo  {journal}
  {Class. Quant. Grav.}\ }\textbf {\bibinfo {volume} {32}},\ \bibinfo {pages}
  {024001} (\bibinfo {year} {2015})},\ \Eprint {http://arxiv.org/abs/1408.3978}
  {arXiv:1408.3978 [gr-qc]} \BibitemShut {NoStop}%
\bibitem [{\citenamefont {Akutsu}\ \emph {et~al.}(2021)\citenamefont {Akutsu}
  \emph {et~al.}}]{KAGRA:2020tym}%
  \BibitemOpen
  \bibfield  {author} {\bibinfo {author} {\bibfnamefont {T.}~\bibnamefont
  {Akutsu}} \emph {et~al.} (\bibinfo {collaboration} {KAGRA}),\ }\bibfield
  {title} {\enquote {\bibinfo {title} {{Overview of KAGRA: Detector design and
  construction history}},}\ }\href {\doibase 10.1093/ptep/ptaa125} {\bibfield
  {journal} {\bibinfo  {journal} {PTEP}\ }\textbf {\bibinfo {volume} {2021}},\
  \bibinfo {pages} {05A101} (\bibinfo {year} {2021})},\ \Eprint
  {http://arxiv.org/abs/2005.05574} {arXiv:2005.05574 [physics.ins-det]}
  \BibitemShut {NoStop}%
\bibitem [{\citenamefont {Abbott}\ \emph {et~al.}(2019)\citenamefont {Abbott}
  \emph {et~al.}}]{LIGOScientific:2018mvr}%
  \BibitemOpen
  \bibfield  {author} {\bibinfo {author} {\bibfnamefont {B.~P.}\ \bibnamefont
  {Abbott}} \emph {et~al.} (\bibinfo {collaboration} {LIGO Scientific,
  Virgo}),\ }\bibfield  {title} {\enquote {\bibinfo {title} {{GWTC-1: A
  Gravitational-Wave Transient Catalog of Compact Binary Mergers Observed by
  LIGO and Virgo during the First and Second Observing Runs}},}\ }\href
  {\doibase 10.1103/PhysRevX.9.031040} {\bibfield  {journal} {\bibinfo
  {journal} {Phys. Rev. X}\ }\textbf {\bibinfo {volume} {9}},\ \bibinfo {pages}
  {031040} (\bibinfo {year} {2019})},\ \Eprint
  {http://arxiv.org/abs/1811.12907} {arXiv:1811.12907 [astro-ph.HE]}
  \BibitemShut {NoStop}%
\bibitem [{\citenamefont {Abbott}\ \emph
  {et~al.}(2021{\natexlab{a}})\citenamefont {Abbott} \emph
  {et~al.}}]{LIGOScientific:2020ibl}%
  \BibitemOpen
  \bibfield  {author} {\bibinfo {author} {\bibfnamefont {R.}~\bibnamefont
  {Abbott}} \emph {et~al.} (\bibinfo {collaboration} {LIGO Scientific,
  Virgo}),\ }\bibfield  {title} {\enquote {\bibinfo {title} {{GWTC-2: Compact
  Binary Coalescences Observed by LIGO and Virgo During the First Half of the
  Third Observing Run}},}\ }\href {\doibase 10.1103/PhysRevX.11.021053}
  {\bibfield  {journal} {\bibinfo  {journal} {Phys. Rev. X}\ }\textbf {\bibinfo
  {volume} {11}},\ \bibinfo {pages} {021053} (\bibinfo {year}
  {2021}{\natexlab{a}})},\ \Eprint {http://arxiv.org/abs/2010.14527}
  {arXiv:2010.14527 [gr-qc]} \BibitemShut {NoStop}%
\bibitem [{\citenamefont {Abbott}\ \emph
  {et~al.}(2021{\natexlab{b}})\citenamefont {Abbott} \emph
  {et~al.}}]{LIGOScientific:2021usb}%
  \BibitemOpen
  \bibfield  {author} {\bibinfo {author} {\bibfnamefont {R.}~\bibnamefont
  {Abbott}} \emph {et~al.} (\bibinfo {collaboration} {LIGO Scientific,
  VIRGO}),\ }\bibfield  {title} {\enquote {\bibinfo {title} {{GWTC-2.1: Deep
  Extended Catalog of Compact Binary Coalescences Observed by LIGO and Virgo
  During the First Half of the Third Observing Run}},}\ }\href@noop {} {\
  (\bibinfo {year} {2021}{\natexlab{b}})},\ \Eprint
  {http://arxiv.org/abs/2108.01045} {arXiv:2108.01045 [gr-qc]} \BibitemShut
  {NoStop}%
\bibitem [{\citenamefont {Abbott}\ \emph
  {et~al.}(2021{\natexlab{c}})\citenamefont {Abbott} \emph
  {et~al.}}]{LIGOScientific:2021djp}%
  \BibitemOpen
  \bibfield  {author} {\bibinfo {author} {\bibfnamefont {R.}~\bibnamefont
  {Abbott}} \emph {et~al.} (\bibinfo {collaboration} {LIGO Scientific, VIRGO,
  KAGRA}),\ }\bibfield  {title} {\enquote {\bibinfo {title} {{GWTC-3: Compact
  Binary Coalescences Observed by LIGO and Virgo During the Second Part of the
  Third Observing Run}},}\ }\href@noop {} {\  (\bibinfo {year}
  {2021}{\natexlab{c}})},\ \Eprint {http://arxiv.org/abs/2111.03606}
  {arXiv:2111.03606 [gr-qc]} \BibitemShut {NoStop}%
\bibitem [{\citenamefont {Abbott}\ \emph
  {et~al.}(2021{\natexlab{d}})\citenamefont {Abbott} \emph
  {et~al.}}]{LIGOScientific:2021psn}%
  \BibitemOpen
  \bibfield  {author} {\bibinfo {author} {\bibfnamefont {R.}~\bibnamefont
  {Abbott}} \emph {et~al.} (\bibinfo {collaboration} {LIGO Scientific, VIRGO,
  KAGRA}),\ }\bibfield  {title} {\enquote {\bibinfo {title} {{The population of
  merging compact binaries inferred using gravitational waves through
  GWTC-3}},}\ }\href@noop {} {\  (\bibinfo {year} {2021}{\natexlab{d}})},\
  \Eprint {http://arxiv.org/abs/2111.03634} {arXiv:2111.03634 [astro-ph.HE]}
  \BibitemShut {NoStop}%
\bibitem [{\citenamefont {Pratten}\ \emph {et~al.}(2020)\citenamefont
  {Pratten}, \citenamefont {Husa}, \citenamefont {Garcia-Quiros}, \citenamefont
  {Colleoni}, \citenamefont {Ramos-Buades}, \citenamefont {Estelles},\ and\
  \citenamefont {Jaume}}]{Pratten:2020fqn}%
  \BibitemOpen
  \bibfield  {author} {\bibinfo {author} {\bibfnamefont {Geraint}\ \bibnamefont
  {Pratten}}, \bibinfo {author} {\bibfnamefont {Sascha}\ \bibnamefont {Husa}},
  \bibinfo {author} {\bibfnamefont {Cecilio}\ \bibnamefont {Garcia-Quiros}},
  \bibinfo {author} {\bibfnamefont {Marta}\ \bibnamefont {Colleoni}}, \bibinfo
  {author} {\bibfnamefont {Antoni}\ \bibnamefont {Ramos-Buades}}, \bibinfo
  {author} {\bibfnamefont {Hector}\ \bibnamefont {Estelles}}, \ and\ \bibinfo
  {author} {\bibfnamefont {Rafel}\ \bibnamefont {Jaume}},\ }\bibfield  {title}
  {\enquote {\bibinfo {title} {{Setting the cornerstone for a family of models
  for gravitational waves from compact binaries: The dominant harmonic for
  nonprecessing quasicircular black holes}},}\ }\href {\doibase
  10.1103/PhysRevD.102.064001} {\bibfield  {journal} {\bibinfo  {journal}
  {Phys. Rev. D}\ }\textbf {\bibinfo {volume} {102}},\ \bibinfo {pages}
  {064001} (\bibinfo {year} {2020})},\ \Eprint
  {http://arxiv.org/abs/2001.11412} {arXiv:2001.11412 [gr-qc]} \BibitemShut
  {NoStop}%
\bibitem [{\citenamefont {Field}\ \emph {et~al.}(2011)\citenamefont {Field},
  \citenamefont {Galley}, \citenamefont {Herrmann}, \citenamefont {Hesthaven},
  \citenamefont {Ochsner},\ and\ \citenamefont {Tiglio}}]{Field:2011mf}%
  \BibitemOpen
  \bibfield  {author} {\bibinfo {author} {\bibfnamefont {Scott~E.}\
  \bibnamefont {Field}}, \bibinfo {author} {\bibfnamefont {Chad~R.}\
  \bibnamefont {Galley}}, \bibinfo {author} {\bibfnamefont {Frank}\
  \bibnamefont {Herrmann}}, \bibinfo {author} {\bibfnamefont {Jan~S.}\
  \bibnamefont {Hesthaven}}, \bibinfo {author} {\bibfnamefont {Evan}\
  \bibnamefont {Ochsner}}, \ and\ \bibinfo {author} {\bibfnamefont {Manuel}\
  \bibnamefont {Tiglio}},\ }\bibfield  {title} {\enquote {\bibinfo {title}
  {{Reduced basis catalogs for gravitational wave templates}},}\ }\href
  {\doibase 10.1103/PhysRevLett.106.221102} {\bibfield  {journal} {\bibinfo
  {journal} {Phys. Rev. Lett.}\ }\textbf {\bibinfo {volume} {106}},\ \bibinfo
  {pages} {221102} (\bibinfo {year} {2011})},\ \Eprint
  {http://arxiv.org/abs/1101.3765} {arXiv:1101.3765 [gr-qc]} \BibitemShut
  {NoStop}%
\bibitem [{\citenamefont {Field}\ \emph {et~al.}(2014)\citenamefont {Field},
  \citenamefont {Galley}, \citenamefont {Hesthaven}, \citenamefont {Kaye},\
  and\ \citenamefont {Tiglio}}]{Field:2013cfa}%
  \BibitemOpen
  \bibfield  {author} {\bibinfo {author} {\bibfnamefont {Scott~E.}\
  \bibnamefont {Field}}, \bibinfo {author} {\bibfnamefont {Chad~R.}\
  \bibnamefont {Galley}}, \bibinfo {author} {\bibfnamefont {Jan~S.}\
  \bibnamefont {Hesthaven}}, \bibinfo {author} {\bibfnamefont {Jason}\
  \bibnamefont {Kaye}}, \ and\ \bibinfo {author} {\bibfnamefont {Manuel}\
  \bibnamefont {Tiglio}},\ }\bibfield  {title} {\enquote {\bibinfo {title}
  {{Fast prediction and evaluation of gravitational waveforms using surrogate
  models}},}\ }\href {\doibase 10.1103/PhysRevX.4.031006} {\bibfield  {journal}
  {\bibinfo  {journal} {Phys. Rev. X}\ }\textbf {\bibinfo {volume} {4}},\
  \bibinfo {pages} {031006} (\bibinfo {year} {2014})},\ \Eprint
  {http://arxiv.org/abs/1308.3565} {arXiv:1308.3565 [gr-qc]} \BibitemShut
  {NoStop}%
\bibitem [{\citenamefont {P\"urrer}(2014)}]{Purrer:2014fza}%
  \BibitemOpen
  \bibfield  {author} {\bibinfo {author} {\bibfnamefont {Michael}\ \bibnamefont
  {P\"urrer}},\ }\bibfield  {title} {\enquote {\bibinfo {title} {{Frequency
  domain reduced order models for gravitational waves from aligned-spin compact
  binaries}},}\ }\href {\doibase 10.1088/0264-9381/31/19/195010} {\bibfield
  {journal} {\bibinfo  {journal} {Class. Quant. Grav.}\ }\textbf {\bibinfo
  {volume} {31}},\ \bibinfo {pages} {195010} (\bibinfo {year} {2014})},\
  \Eprint {http://arxiv.org/abs/1402.4146} {arXiv:1402.4146 [gr-qc]}
  \BibitemShut {NoStop}%
\bibitem [{\citenamefont {Blackman}\ \emph {et~al.}(2015)\citenamefont
  {Blackman}, \citenamefont {Field}, \citenamefont {Galley}, \citenamefont
  {Szil\'agyi}, \citenamefont {Scheel}, \citenamefont {Tiglio},\ and\
  \citenamefont {Hemberger}}]{Blackman:2015pia}%
  \BibitemOpen
  \bibfield  {author} {\bibinfo {author} {\bibfnamefont {Jonathan}\
  \bibnamefont {Blackman}}, \bibinfo {author} {\bibfnamefont {Scott~E.}\
  \bibnamefont {Field}}, \bibinfo {author} {\bibfnamefont {Chad~R.}\
  \bibnamefont {Galley}}, \bibinfo {author} {\bibfnamefont {B\'ela}\
  \bibnamefont {Szil\'agyi}}, \bibinfo {author} {\bibfnamefont {Mark~A.}\
  \bibnamefont {Scheel}}, \bibinfo {author} {\bibfnamefont {Manuel}\
  \bibnamefont {Tiglio}}, \ and\ \bibinfo {author} {\bibfnamefont {Daniel~A.}\
  \bibnamefont {Hemberger}},\ }\bibfield  {title} {\enquote {\bibinfo {title}
  {{Fast and Accurate Prediction of Numerical Relativity Waveforms from Binary
  Black Hole Coalescences Using Surrogate Models}},}\ }\href {\doibase
  10.1103/PhysRevLett.115.121102} {\bibfield  {journal} {\bibinfo  {journal}
  {Phys. Rev. Lett.}\ }\textbf {\bibinfo {volume} {115}},\ \bibinfo {pages}
  {121102} (\bibinfo {year} {2015})},\ \Eprint
  {http://arxiv.org/abs/1502.07758} {arXiv:1502.07758 [gr-qc]} \BibitemShut
  {NoStop}%
\bibitem [{\citenamefont {Blackman}\ \emph {et~al.}(2017)\citenamefont
  {Blackman}, \citenamefont {Field}, \citenamefont {Scheel}, \citenamefont
  {Galley}, \citenamefont {Hemberger}, \citenamefont {Schmidt},\ and\
  \citenamefont {Smith}}]{Blackman:2017dfb}%
  \BibitemOpen
  \bibfield  {author} {\bibinfo {author} {\bibfnamefont {Jonathan}\
  \bibnamefont {Blackman}}, \bibinfo {author} {\bibfnamefont {Scott~E.}\
  \bibnamefont {Field}}, \bibinfo {author} {\bibfnamefont {Mark~A.}\
  \bibnamefont {Scheel}}, \bibinfo {author} {\bibfnamefont {Chad~R.}\
  \bibnamefont {Galley}}, \bibinfo {author} {\bibfnamefont {Daniel~A.}\
  \bibnamefont {Hemberger}}, \bibinfo {author} {\bibfnamefont {Patricia}\
  \bibnamefont {Schmidt}}, \ and\ \bibinfo {author} {\bibfnamefont {Rory}\
  \bibnamefont {Smith}},\ }\bibfield  {title} {\enquote {\bibinfo {title} {{A
  Surrogate Model of Gravitational Waveforms from Numerical Relativity
  Simulations of Precessing Binary Black Hole Mergers}},}\ }\href {\doibase
  10.1103/PhysRevD.95.104023} {\bibfield  {journal} {\bibinfo  {journal} {Phys.
  Rev. D}\ }\textbf {\bibinfo {volume} {95}},\ \bibinfo {pages} {104023}
  (\bibinfo {year} {2017})},\ \Eprint {http://arxiv.org/abs/1701.00550}
  {arXiv:1701.00550 [gr-qc]} \BibitemShut {NoStop}%
\bibitem [{\citenamefont {Galley}\ and\ \citenamefont
  {Schmidt}(2016{\natexlab{a}})}]{Galley:2016mvy}%
  \BibitemOpen
  \bibfield  {author} {\bibinfo {author} {\bibfnamefont {Chad~R.}\ \bibnamefont
  {Galley}}\ and\ \bibinfo {author} {\bibfnamefont {Patricia}\ \bibnamefont
  {Schmidt}},\ }\bibfield  {title} {\enquote {\bibinfo {title} {{Fast and
  efficient evaluation of gravitational waveforms via reduced-order spline
  interpolation}},}\ }\href@noop {} {\  (\bibinfo {year}
  {2016}{\natexlab{a}})},\ \Eprint {http://arxiv.org/abs/1611.07529}
  {arXiv:1611.07529 [gr-qc]} \BibitemShut {NoStop}%
\bibitem [{\citenamefont {Varma}\ \emph
  {et~al.}(2019{\natexlab{a}})\citenamefont {Varma}, \citenamefont {Field},
  \citenamefont {Scheel}, \citenamefont {Blackman}, \citenamefont {Gerosa},
  \citenamefont {Stein}, \citenamefont {Kidder},\ and\ \citenamefont
  {Pfeiffer}}]{Varma:2019csw}%
  \BibitemOpen
  \bibfield  {author} {\bibinfo {author} {\bibfnamefont {Vijay}\ \bibnamefont
  {Varma}}, \bibinfo {author} {\bibfnamefont {Scott~E.}\ \bibnamefont {Field}},
  \bibinfo {author} {\bibfnamefont {Mark~A.}\ \bibnamefont {Scheel}}, \bibinfo
  {author} {\bibfnamefont {Jonathan}\ \bibnamefont {Blackman}}, \bibinfo
  {author} {\bibfnamefont {Davide}\ \bibnamefont {Gerosa}}, \bibinfo {author}
  {\bibfnamefont {Leo~C.}\ \bibnamefont {Stein}}, \bibinfo {author}
  {\bibfnamefont {Lawrence~E.}\ \bibnamefont {Kidder}}, \ and\ \bibinfo
  {author} {\bibfnamefont {Harald~P.}\ \bibnamefont {Pfeiffer}},\ }\bibfield
  {title} {\enquote {\bibinfo {title} {{Surrogate models for precessing binary
  black hole simulations with unequal masses}},}\ }\href {\doibase
  10.1103/PhysRevResearch.1.033015} {\bibfield  {journal} {\bibinfo  {journal}
  {Phys. Rev. Research.}\ }\textbf {\bibinfo {volume} {1}},\ \bibinfo {pages}
  {033015} (\bibinfo {year} {2019}{\natexlab{a}})},\ \Eprint
  {http://arxiv.org/abs/1905.09300} {arXiv:1905.09300 [gr-qc]} \BibitemShut
  {NoStop}%
\bibitem [{\citenamefont {Schmidt}\ \emph {et~al.}(2015)\citenamefont
  {Schmidt}, \citenamefont {Ohme},\ and\ \citenamefont
  {Hannam}}]{Schmidt:2014iyl}%
  \BibitemOpen
  \bibfield  {author} {\bibinfo {author} {\bibfnamefont {Patricia}\
  \bibnamefont {Schmidt}}, \bibinfo {author} {\bibfnamefont {Frank}\
  \bibnamefont {Ohme}}, \ and\ \bibinfo {author} {\bibfnamefont {Mark}\
  \bibnamefont {Hannam}},\ }\bibfield  {title} {\enquote {\bibinfo {title}
  {{Towards models of gravitational waveforms from generic binaries II:
  Modelling precession effects with a single effective precession
  parameter}},}\ }\href {\doibase 10.1103/PhysRevD.91.024043} {\bibfield
  {journal} {\bibinfo  {journal} {Phys. Rev. D}\ }\textbf {\bibinfo {volume}
  {91}},\ \bibinfo {pages} {024043} (\bibinfo {year} {2015})},\ \Eprint
  {http://arxiv.org/abs/1408.1810} {arXiv:1408.1810 [gr-qc]} \BibitemShut
  {NoStop}%
\bibitem [{\citenamefont {Thomas}\ \emph {et~al.}(2020)\citenamefont {Thomas},
  \citenamefont {Schmidt},\ and\ \citenamefont {Pratten}}]{Thomas:2020uqj}%
  \BibitemOpen
  \bibfield  {author} {\bibinfo {author} {\bibfnamefont {Lucy~M.}\ \bibnamefont
  {Thomas}}, \bibinfo {author} {\bibfnamefont {Patricia}\ \bibnamefont
  {Schmidt}}, \ and\ \bibinfo {author} {\bibfnamefont {Geraint}\ \bibnamefont
  {Pratten}},\ }\bibfield  {title} {\enquote {\bibinfo {title} {{A new
  effective precession spin for modelling multi-modal gravitational waveforms
  in the strong-field regime}},}\ }\href@noop {} {\  (\bibinfo {year}
  {2020})},\ \Eprint {http://arxiv.org/abs/2012.02209} {arXiv:2012.02209
  [gr-qc]} \BibitemShut {NoStop}%
\bibitem [{\citenamefont {Khan}\ and\ \citenamefont
  {Green}(2021)}]{Khan:2020fso}%
  \BibitemOpen
  \bibfield  {author} {\bibinfo {author} {\bibfnamefont {Sebastian}\
  \bibnamefont {Khan}}\ and\ \bibinfo {author} {\bibfnamefont {Rhys}\
  \bibnamefont {Green}},\ }\bibfield  {title} {\enquote {\bibinfo {title}
  {{Gravitational-wave surrogate models powered by artificial neural
  networks}},}\ }\href {\doibase 10.1103/PhysRevD.103.064015} {\bibfield
  {journal} {\bibinfo  {journal} {Phys. Rev. D}\ }\textbf {\bibinfo {volume}
  {103}},\ \bibinfo {pages} {064015} (\bibinfo {year} {2021})},\ \Eprint
  {http://arxiv.org/abs/2008.12932} {arXiv:2008.12932 [gr-qc]} \BibitemShut
  {NoStop}%
\bibitem [{\citenamefont {Cornish}(2010)}]{Cornish:2010kf}%
  \BibitemOpen
  \bibfield  {author} {\bibinfo {author} {\bibfnamefont {Neil~J.}\ \bibnamefont
  {Cornish}},\ }\bibfield  {title} {\enquote {\bibinfo {title} {{Fast Fisher
  Matrices and Lazy Likelihoods}},}\ }\href@noop {} {\  (\bibinfo {year}
  {2010})},\ \Eprint {http://arxiv.org/abs/1007.4820} {arXiv:1007.4820 [gr-qc]}
  \BibitemShut {NoStop}%
\bibitem [{\citenamefont {Canizares}\ \emph {et~al.}(2013)\citenamefont
  {Canizares}, \citenamefont {Field}, \citenamefont {Gair},\ and\ \citenamefont
  {Tiglio}}]{Canizares:2013ywa}%
  \BibitemOpen
  \bibfield  {author} {\bibinfo {author} {\bibfnamefont {Priscilla}\
  \bibnamefont {Canizares}}, \bibinfo {author} {\bibfnamefont {Scott~E.}\
  \bibnamefont {Field}}, \bibinfo {author} {\bibfnamefont {Jonathan~R.}\
  \bibnamefont {Gair}}, \ and\ \bibinfo {author} {\bibfnamefont {Manuel}\
  \bibnamefont {Tiglio}},\ }\bibfield  {title} {\enquote {\bibinfo {title}
  {{Gravitational wave parameter estimation with compressed likelihood
  evaluations}},}\ }\href {\doibase 10.1103/PhysRevD.87.124005} {\bibfield
  {journal} {\bibinfo  {journal} {Phys. Rev. D}\ }\textbf {\bibinfo {volume}
  {87}},\ \bibinfo {pages} {124005} (\bibinfo {year} {2013})},\ \Eprint
  {http://arxiv.org/abs/1304.0462} {arXiv:1304.0462 [gr-qc]} \BibitemShut
  {NoStop}%
\bibitem [{\citenamefont {Canizares}\ \emph {et~al.}(2015)\citenamefont
  {Canizares}, \citenamefont {Field}, \citenamefont {Gair}, \citenamefont
  {Raymond}, \citenamefont {Smith},\ and\ \citenamefont
  {Tiglio}}]{Canizares:2014fya}%
  \BibitemOpen
  \bibfield  {author} {\bibinfo {author} {\bibfnamefont {Priscilla}\
  \bibnamefont {Canizares}}, \bibinfo {author} {\bibfnamefont {Scott~E.}\
  \bibnamefont {Field}}, \bibinfo {author} {\bibfnamefont {Jonathan}\
  \bibnamefont {Gair}}, \bibinfo {author} {\bibfnamefont {Vivien}\ \bibnamefont
  {Raymond}}, \bibinfo {author} {\bibfnamefont {Rory}\ \bibnamefont {Smith}}, \
  and\ \bibinfo {author} {\bibfnamefont {Manuel}\ \bibnamefont {Tiglio}},\
  }\bibfield  {title} {\enquote {\bibinfo {title} {{Accelerated
  gravitational-wave parameter estimation with reduced order modeling}},}\
  }\href {\doibase 10.1103/PhysRevLett.114.071104} {\bibfield  {journal}
  {\bibinfo  {journal} {Phys. Rev. Lett.}\ }\textbf {\bibinfo {volume} {114}},\
  \bibinfo {pages} {071104} (\bibinfo {year} {2015})},\ \Eprint
  {http://arxiv.org/abs/1404.6284} {arXiv:1404.6284 [gr-qc]} \BibitemShut
  {NoStop}%
\bibitem [{\citenamefont {Smith}\ \emph {et~al.}(2016)\citenamefont {Smith},
  \citenamefont {Field}, \citenamefont {Blackburn}, \citenamefont {Haster},
  \citenamefont {P\"urrer}, \citenamefont {Raymond},\ and\ \citenamefont
  {Schmidt}}]{Smith:2016qas}%
  \BibitemOpen
  \bibfield  {author} {\bibinfo {author} {\bibfnamefont {Rory}\ \bibnamefont
  {Smith}}, \bibinfo {author} {\bibfnamefont {Scott~E.}\ \bibnamefont {Field}},
  \bibinfo {author} {\bibfnamefont {Kent}\ \bibnamefont {Blackburn}}, \bibinfo
  {author} {\bibfnamefont {Carl-Johan}\ \bibnamefont {Haster}}, \bibinfo
  {author} {\bibfnamefont {Michael}\ \bibnamefont {P\"urrer}}, \bibinfo
  {author} {\bibfnamefont {Vivien}\ \bibnamefont {Raymond}}, \ and\ \bibinfo
  {author} {\bibfnamefont {Patricia}\ \bibnamefont {Schmidt}},\ }\bibfield
  {title} {\enquote {\bibinfo {title} {{Fast and accurate inference on
  gravitational waves from precessing compact binaries}},}\ }\href {\doibase
  10.1103/PhysRevD.94.044031} {\bibfield  {journal} {\bibinfo  {journal} {Phys.
  Rev. D}\ }\textbf {\bibinfo {volume} {94}},\ \bibinfo {pages} {044031}
  (\bibinfo {year} {2016})},\ \Eprint {http://arxiv.org/abs/1604.08253}
  {arXiv:1604.08253 [gr-qc]} \BibitemShut {NoStop}%
\bibitem [{\citenamefont {Zackay}\ \emph {et~al.}(2018)\citenamefont {Zackay},
  \citenamefont {Dai},\ and\ \citenamefont {Venumadhav}}]{Zackay:2018qdy}%
  \BibitemOpen
  \bibfield  {author} {\bibinfo {author} {\bibfnamefont {Barak}\ \bibnamefont
  {Zackay}}, \bibinfo {author} {\bibfnamefont {Liang}\ \bibnamefont {Dai}}, \
  and\ \bibinfo {author} {\bibfnamefont {Tejaswi}\ \bibnamefont {Venumadhav}},\
  }\bibfield  {title} {\enquote {\bibinfo {title} {{Relative Binning and Fast
  Likelihood Evaluation for Gravitational Wave Parameter Estimation}},}\
  }\href@noop {} {\  (\bibinfo {year} {2018})},\ \Eprint
  {http://arxiv.org/abs/1806.08792} {arXiv:1806.08792 [astro-ph.IM]}
  \BibitemShut {NoStop}%
\bibitem [{\citenamefont {Cornish}(2021)}]{Cornish:2021lje}%
  \BibitemOpen
  \bibfield  {author} {\bibinfo {author} {\bibfnamefont {Neil~J.}\ \bibnamefont
  {Cornish}},\ }\bibfield  {title} {\enquote {\bibinfo {title} {{Heterodyned
  likelihood for rapid gravitational wave parameter inference}},}\ }\href
  {\doibase 10.1103/PhysRevD.104.104054} {\bibfield  {journal} {\bibinfo
  {journal} {Phys. Rev. D}\ }\textbf {\bibinfo {volume} {104}},\ \bibinfo
  {pages} {104054} (\bibinfo {year} {2021})},\ \Eprint
  {http://arxiv.org/abs/2109.02728} {arXiv:2109.02728 [gr-qc]} \BibitemShut
  {NoStop}%
\bibitem [{\citenamefont {Leslie}\ \emph {et~al.}(2021)\citenamefont {Leslie},
  \citenamefont {Dai},\ and\ \citenamefont {Pratten}}]{Leslie:2021ssu}%
  \BibitemOpen
  \bibfield  {author} {\bibinfo {author} {\bibfnamefont {Nathaniel}\
  \bibnamefont {Leslie}}, \bibinfo {author} {\bibfnamefont {Liang}\
  \bibnamefont {Dai}}, \ and\ \bibinfo {author} {\bibfnamefont {Geraint}\
  \bibnamefont {Pratten}},\ }\bibfield  {title} {\enquote {\bibinfo {title}
  {{Mode-by-mode relative binning: Fast likelihood estimation for gravitational
  waveforms with spin-orbit precession and multiple harmonics}},}\ }\href
  {\doibase 10.1103/PhysRevD.104.123030} {\bibfield  {journal} {\bibinfo
  {journal} {Phys. Rev. D}\ }\textbf {\bibinfo {volume} {104}},\ \bibinfo
  {pages} {123030} (\bibinfo {year} {2021})},\ \Eprint
  {http://arxiv.org/abs/2109.09872} {arXiv:2109.09872 [astro-ph.IM]}
  \BibitemShut {NoStop}%
\bibitem [{\citenamefont {Morisaki}(2021)}]{Morisaki:2021ngj}%
  \BibitemOpen
  \bibfield  {author} {\bibinfo {author} {\bibfnamefont {Soichiro}\
  \bibnamefont {Morisaki}},\ }\bibfield  {title} {\enquote {\bibinfo {title}
  {{Accelerating parameter estimation of gravitational waves from compact
  binary coalescence using adaptive frequency resolutions}},}\ }\href {\doibase
  10.1103/PhysRevD.104.044062} {\bibfield  {journal} {\bibinfo  {journal}
  {Phys. Rev. D}\ }\textbf {\bibinfo {volume} {104}},\ \bibinfo {pages}
  {044062} (\bibinfo {year} {2021})},\ \Eprint
  {http://arxiv.org/abs/2104.07813} {arXiv:2104.07813 [gr-qc]} \BibitemShut
  {NoStop}%
\bibitem [{\citenamefont {Chua}\ and\ \citenamefont
  {Vallisneri}(2020)}]{Chua:2019wwt}%
  \BibitemOpen
  \bibfield  {author} {\bibinfo {author} {\bibfnamefont {Alvin J.~K.}\
  \bibnamefont {Chua}}\ and\ \bibinfo {author} {\bibfnamefont {Michele}\
  \bibnamefont {Vallisneri}},\ }\bibfield  {title} {\enquote {\bibinfo {title}
  {{Learning Bayesian posteriors with neural networks for gravitational-wave
  inference}},}\ }\href {\doibase 10.1103/PhysRevLett.124.041102} {\bibfield
  {journal} {\bibinfo  {journal} {Phys. Rev. Lett.}\ }\textbf {\bibinfo
  {volume} {124}},\ \bibinfo {pages} {041102} (\bibinfo {year} {2020})},\
  \Eprint {http://arxiv.org/abs/1909.05966} {arXiv:1909.05966 [gr-qc]}
  \BibitemShut {NoStop}%
\bibitem [{\citenamefont {Dax}\ \emph {et~al.}(2021)\citenamefont {Dax},
  \citenamefont {Green}, \citenamefont {Gair}, \citenamefont {Macke},
  \citenamefont {Buonanno},\ and\ \citenamefont {Sch\"olkopf}}]{Dax:2021tsq}%
  \BibitemOpen
  \bibfield  {author} {\bibinfo {author} {\bibfnamefont {Maximilian}\
  \bibnamefont {Dax}}, \bibinfo {author} {\bibfnamefont {Stephen~R.}\
  \bibnamefont {Green}}, \bibinfo {author} {\bibfnamefont {Jonathan}\
  \bibnamefont {Gair}}, \bibinfo {author} {\bibfnamefont {Jakob~H.}\
  \bibnamefont {Macke}}, \bibinfo {author} {\bibfnamefont {Alessandra}\
  \bibnamefont {Buonanno}}, \ and\ \bibinfo {author} {\bibfnamefont {Bernhard}\
  \bibnamefont {Sch\"olkopf}},\ }\bibfield  {title} {\enquote {\bibinfo {title}
  {{Real-Time Gravitational Wave Science with Neural Posterior Estimation}},}\
  }\href {\doibase 10.1103/PhysRevLett.127.241103} {\bibfield  {journal}
  {\bibinfo  {journal} {Phys. Rev. Lett.}\ }\textbf {\bibinfo {volume} {127}},\
  \bibinfo {pages} {241103} (\bibinfo {year} {2021})},\ \Eprint
  {http://arxiv.org/abs/2106.12594} {arXiv:2106.12594 [gr-qc]} \BibitemShut
  {NoStop}%
\bibitem [{\citenamefont {Ossokine}\ \emph {et~al.}(2020)\citenamefont
  {Ossokine} \emph {et~al.}}]{Ossokine:2020kjp}%
  \BibitemOpen
  \bibfield  {author} {\bibinfo {author} {\bibfnamefont {Serguei}\ \bibnamefont
  {Ossokine}} \emph {et~al.},\ }\bibfield  {title} {\enquote {\bibinfo {title}
  {{Multipolar Effective-One-Body Waveforms for Precessing Binary Black Holes:
  Construction and Validation}},}\ }\href {\doibase
  10.1103/PhysRevD.102.044055} {\bibfield  {journal} {\bibinfo  {journal}
  {Phys. Rev. D}\ }\textbf {\bibinfo {volume} {102}},\ \bibinfo {pages}
  {044055} (\bibinfo {year} {2020})},\ \Eprint
  {http://arxiv.org/abs/2004.09442} {arXiv:2004.09442 [gr-qc]} \BibitemShut
  {NoStop}%
\bibitem [{\citenamefont {Apostolatos}\ \emph {et~al.}(1994)\citenamefont
  {Apostolatos}, \citenamefont {Cutler}, \citenamefont {Sussman},\ and\
  \citenamefont {Thorne}}]{Apostolatos:1994mx}%
  \BibitemOpen
  \bibfield  {author} {\bibinfo {author} {\bibfnamefont {Theocharis~A.}\
  \bibnamefont {Apostolatos}}, \bibinfo {author} {\bibfnamefont {Curt}\
  \bibnamefont {Cutler}}, \bibinfo {author} {\bibfnamefont {Gerald~J.}\
  \bibnamefont {Sussman}}, \ and\ \bibinfo {author} {\bibfnamefont {Kip~S.}\
  \bibnamefont {Thorne}},\ }\bibfield  {title} {\enquote {\bibinfo {title}
  {{Spin induced orbital precession and its modulation of the gravitational
  wave forms from merging binaries}},}\ }\href {\doibase
  10.1103/PhysRevD.49.6274} {\bibfield  {journal} {\bibinfo  {journal} {Phys.
  Rev. D}\ }\textbf {\bibinfo {volume} {49}},\ \bibinfo {pages} {6274--6297}
  (\bibinfo {year} {1994})}\BibitemShut {NoStop}%
\bibitem [{\citenamefont {Fragkouli}\ \emph {et~al.}(2022)\citenamefont
  {Fragkouli}, \citenamefont {Nousi}, \citenamefont {Passalis}, \citenamefont
  {Iosif}, \citenamefont {Stergioulas},\ and\ \citenamefont
  {Tefas}}]{Fragkouli:2022lpt}%
  \BibitemOpen
  \bibfield  {author} {\bibinfo {author} {\bibfnamefont {Styliani-Christina}\
  \bibnamefont {Fragkouli}}, \bibinfo {author} {\bibfnamefont {Paraskevi}\
  \bibnamefont {Nousi}}, \bibinfo {author} {\bibfnamefont {Nikolaos}\
  \bibnamefont {Passalis}}, \bibinfo {author} {\bibfnamefont {Panagiotis}\
  \bibnamefont {Iosif}}, \bibinfo {author} {\bibfnamefont {Nikolaos}\
  \bibnamefont {Stergioulas}}, \ and\ \bibinfo {author} {\bibfnamefont
  {Anastasios}\ \bibnamefont {Tefas}},\ }\bibfield  {title} {\enquote {\bibinfo
  {title} {{Deep Residual Error and Bag-of-Tricks Learning for Gravitational
  Wave Surrogate Modeling}},}\ }\href@noop {} {\  (\bibinfo {year} {2022})},\
  \Eprint {http://arxiv.org/abs/2203.08434} {arXiv:2203.08434 [astro-ph.IM]}
  \BibitemShut {NoStop}%
\bibitem [{\citenamefont {Varma}\ \emph
  {et~al.}(2019{\natexlab{b}})\citenamefont {Varma}, \citenamefont {Field},
  \citenamefont {Scheel}, \citenamefont {Blackman}, \citenamefont {Kidder},\
  and\ \citenamefont {Pfeiffer}}]{Varma:2018mmi}%
  \BibitemOpen
  \bibfield  {author} {\bibinfo {author} {\bibfnamefont {Vijay}\ \bibnamefont
  {Varma}}, \bibinfo {author} {\bibfnamefont {Scott~E.}\ \bibnamefont {Field}},
  \bibinfo {author} {\bibfnamefont {Mark~A.}\ \bibnamefont {Scheel}}, \bibinfo
  {author} {\bibfnamefont {Jonathan}\ \bibnamefont {Blackman}}, \bibinfo
  {author} {\bibfnamefont {Lawrence~E.}\ \bibnamefont {Kidder}}, \ and\
  \bibinfo {author} {\bibfnamefont {Harald~P.}\ \bibnamefont {Pfeiffer}},\
  }\bibfield  {title} {\enquote {\bibinfo {title} {{Surrogate model of
  hybridized numerical relativity binary black hole waveforms}},}\ }\href
  {\doibase 10.1103/PhysRevD.99.064045} {\bibfield  {journal} {\bibinfo
  {journal} {Phys. Rev. D}\ }\textbf {\bibinfo {volume} {99}},\ \bibinfo
  {pages} {064045} (\bibinfo {year} {2019}{\natexlab{b}})},\ \Eprint
  {http://arxiv.org/abs/1812.07865} {arXiv:1812.07865 [gr-qc]} \BibitemShut
  {NoStop}%
\bibitem [{\citenamefont {Islam}\ \emph {et~al.}(2021)\citenamefont {Islam},
  \citenamefont {Varma}, \citenamefont {Lodman}, \citenamefont {Field},
  \citenamefont {Khanna}, \citenamefont {Scheel}, \citenamefont {Pfeiffer},
  \citenamefont {Gerosa},\ and\ \citenamefont {Kidder}}]{Islam:2021mha}%
  \BibitemOpen
  \bibfield  {author} {\bibinfo {author} {\bibfnamefont {Tousif}\ \bibnamefont
  {Islam}}, \bibinfo {author} {\bibfnamefont {Vijay}\ \bibnamefont {Varma}},
  \bibinfo {author} {\bibfnamefont {Jackie}\ \bibnamefont {Lodman}}, \bibinfo
  {author} {\bibfnamefont {Scott~E.}\ \bibnamefont {Field}}, \bibinfo {author}
  {\bibfnamefont {Gaurav}\ \bibnamefont {Khanna}}, \bibinfo {author}
  {\bibfnamefont {Mark~A.}\ \bibnamefont {Scheel}}, \bibinfo {author}
  {\bibfnamefont {Harald~P.}\ \bibnamefont {Pfeiffer}}, \bibinfo {author}
  {\bibfnamefont {Davide}\ \bibnamefont {Gerosa}}, \ and\ \bibinfo {author}
  {\bibfnamefont {Lawrence~E.}\ \bibnamefont {Kidder}},\ }\bibfield  {title}
  {\enquote {\bibinfo {title} {{Eccentric binary black hole surrogate models
  for the gravitational waveform and remnant properties: comparable mass,
  nonspinning case}},}\ }\href {\doibase 10.1103/PhysRevD.103.064022}
  {\bibfield  {journal} {\bibinfo  {journal} {Phys. Rev. D}\ }\textbf {\bibinfo
  {volume} {103}},\ \bibinfo {pages} {064022} (\bibinfo {year} {2021})},\
  \Eprint {http://arxiv.org/abs/2101.11798} {arXiv:2101.11798 [gr-qc]}
  \BibitemShut {NoStop}%
\bibitem [{\citenamefont {Islam}\ \emph {et~al.}(2022)\citenamefont {Islam},
  \citenamefont {Field}, \citenamefont {Hughes}, \citenamefont {Khanna},
  \citenamefont {Varma}, \citenamefont {Giesler}, \citenamefont {Scheel},
  \citenamefont {Kidder},\ and\ \citenamefont {Pfeiffer}}]{Islam:2022laz}%
  \BibitemOpen
  \bibfield  {author} {\bibinfo {author} {\bibfnamefont {Tousif}\ \bibnamefont
  {Islam}}, \bibinfo {author} {\bibfnamefont {Scott~E.}\ \bibnamefont {Field}},
  \bibinfo {author} {\bibfnamefont {Scott~A.}\ \bibnamefont {Hughes}}, \bibinfo
  {author} {\bibfnamefont {Gaurav}\ \bibnamefont {Khanna}}, \bibinfo {author}
  {\bibfnamefont {Vijay}\ \bibnamefont {Varma}}, \bibinfo {author}
  {\bibfnamefont {Matthew}\ \bibnamefont {Giesler}}, \bibinfo {author}
  {\bibfnamefont {Mark~A.}\ \bibnamefont {Scheel}}, \bibinfo {author}
  {\bibfnamefont {Lawrence~E.}\ \bibnamefont {Kidder}}, \ and\ \bibinfo
  {author} {\bibfnamefont {Harald~P.}\ \bibnamefont {Pfeiffer}},\ }\bibfield
  {title} {\enquote {\bibinfo {title} {{Surrogate model for gravitational wave
  signals from non-spinning, comparable- to large-mass-ratio black hole
  binaries built on black hole perturbation theory waveforms calibrated to
  numerical relativity}},}\ }\href@noop {} {\  (\bibinfo {year} {2022})},\
  \Eprint {http://arxiv.org/abs/2204.01972} {arXiv:2204.01972 [gr-qc]}
  \BibitemShut {NoStop}%
\bibitem [{\citenamefont {Boh\'e}\ \emph {et~al.}(2017)\citenamefont {Boh\'e}
  \emph {et~al.}}]{Bohe:2016gbl}%
  \BibitemOpen
  \bibfield  {author} {\bibinfo {author} {\bibfnamefont {Alejandro}\
  \bibnamefont {Boh\'e}} \emph {et~al.},\ }\bibfield  {title} {\enquote
  {\bibinfo {title} {{Improved effective-one-body model of spinning,
  nonprecessing binary black holes for the era of gravitational-wave
  astrophysics with advanced detectors}},}\ }\href {\doibase
  10.1103/PhysRevD.95.044028} {\bibfield  {journal} {\bibinfo  {journal} {Phys.
  Rev. D}\ }\textbf {\bibinfo {volume} {95}},\ \bibinfo {pages} {044028}
  (\bibinfo {year} {2017})},\ \Eprint {http://arxiv.org/abs/1611.03703}
  {arXiv:1611.03703 [gr-qc]} \BibitemShut {NoStop}%
\bibitem [{\citenamefont {Nagar}\ \emph {et~al.}(2018)\citenamefont {Nagar}
  \emph {et~al.}}]{Nagar:2018zoe}%
  \BibitemOpen
  \bibfield  {author} {\bibinfo {author} {\bibfnamefont {Alessandro}\
  \bibnamefont {Nagar}} \emph {et~al.},\ }\bibfield  {title} {\enquote
  {\bibinfo {title} {{Time-domain effective-one-body gravitational waveforms
  for coalescing compact binaries with nonprecessing spins, tides and self-spin
  effects}},}\ }\href {\doibase 10.1103/PhysRevD.98.104052} {\bibfield
  {journal} {\bibinfo  {journal} {Phys. Rev. D}\ }\textbf {\bibinfo {volume}
  {98}},\ \bibinfo {pages} {104052} (\bibinfo {year} {2018})},\ \Eprint
  {http://arxiv.org/abs/1806.01772} {arXiv:1806.01772 [gr-qc]} \BibitemShut
  {NoStop}%
\bibitem [{\citenamefont {Schmidt}\ \emph {et~al.}(2021)\citenamefont
  {Schmidt}, \citenamefont {Breschi}, \citenamefont {Gamba}, \citenamefont
  {Pagano}, \citenamefont {Rettegno}, \citenamefont {Riemenschneider},
  \citenamefont {Bernuzzi}, \citenamefont {Nagar},\ and\ \citenamefont
  {Del~Pozzo}}]{Schmidt:2020yuu}%
  \BibitemOpen
  \bibfield  {author} {\bibinfo {author} {\bibfnamefont {Stefano}\ \bibnamefont
  {Schmidt}}, \bibinfo {author} {\bibfnamefont {Matteo}\ \bibnamefont
  {Breschi}}, \bibinfo {author} {\bibfnamefont {Rossella}\ \bibnamefont
  {Gamba}}, \bibinfo {author} {\bibfnamefont {Giulia}\ \bibnamefont {Pagano}},
  \bibinfo {author} {\bibfnamefont {Piero}\ \bibnamefont {Rettegno}}, \bibinfo
  {author} {\bibfnamefont {Gunnar}\ \bibnamefont {Riemenschneider}}, \bibinfo
  {author} {\bibfnamefont {Sebastiano}\ \bibnamefont {Bernuzzi}}, \bibinfo
  {author} {\bibfnamefont {Alessandro}\ \bibnamefont {Nagar}}, \ and\ \bibinfo
  {author} {\bibfnamefont {Walter}\ \bibnamefont {Del~Pozzo}},\ }\bibfield
  {title} {\enquote {\bibinfo {title} {{Machine Learning Gravitational Waves
  from Binary Black Hole Mergers}},}\ }\href {\doibase
  10.1103/PhysRevD.103.043020} {\bibfield  {journal} {\bibinfo  {journal}
  {Phys. Rev. D}\ }\textbf {\bibinfo {volume} {103}},\ \bibinfo {pages}
  {043020} (\bibinfo {year} {2021})},\ \Eprint
  {http://arxiv.org/abs/2011.01958} {arXiv:2011.01958 [gr-qc]} \BibitemShut
  {NoStop}%
\bibitem [{\citenamefont {Barrault}\ \emph {et~al.}(2004)\citenamefont
  {Barrault}, \citenamefont {Maday}, \citenamefont {Nguyen},\ and\
  \citenamefont {Patera}}]{BARRAULT2004667}%
  \BibitemOpen
  \bibfield  {author} {\bibinfo {author} {\bibfnamefont {Maxime}\ \bibnamefont
  {Barrault}}, \bibinfo {author} {\bibfnamefont {Yvon}\ \bibnamefont {Maday}},
  \bibinfo {author} {\bibfnamefont {Ngoc~Cuong}\ \bibnamefont {Nguyen}}, \ and\
  \bibinfo {author} {\bibfnamefont {Anthony~T.}\ \bibnamefont {Patera}},\
  }\bibfield  {title} {\enquote {\bibinfo {title} {An ‘empirical
  interpolation’ method: application to efficient reduced-basis
  discretization of partial differential equations},}\ }\href {\doibase
  https://doi.org/10.1016/j.crma.2004.08.006} {\bibfield  {journal} {\bibinfo
  {journal} {Comptes Rendus Mathematique}\ }\textbf {\bibinfo {volume} {339}},\
  \bibinfo {pages} {667--672} (\bibinfo {year} {2004})}\BibitemShut {NoStop}%
\bibitem [{\citenamefont {Maday}\ \emph {et~al.}(2009)\citenamefont {Maday},
  \citenamefont {Nguyen}, \citenamefont {Patera},\ and\ \citenamefont
  {Pau}}]{Maday:2009}%
  \BibitemOpen
  \bibfield  {author} {\bibinfo {author} {\bibfnamefont {Yvon}\ \bibnamefont
  {Maday}}, \bibinfo {author} {\bibfnamefont {Ngoc~Cuong}\ \bibnamefont
  {Nguyen}}, \bibinfo {author} {\bibfnamefont {Anthony~T.}\ \bibnamefont
  {Patera}}, \ and\ \bibinfo {author} {\bibfnamefont {S.~H.}\ \bibnamefont
  {Pau}},\ }\bibfield  {title} {\enquote {\bibinfo {title} {A general
  multipurpose interpolation procedure: the magic points},}\ }\href@noop {}
  {\bibfield  {journal} {\bibinfo  {journal} {Communications on Pure and
  Applied Analysis}\ }\textbf {\bibinfo {volume} {8}},\ \bibinfo {pages}
  {383--404} (\bibinfo {year} {2009})}\BibitemShut {NoStop}%
\bibitem [{\citenamefont {Williams}\ \emph {et~al.}(2020)\citenamefont
  {Williams}, \citenamefont {Heng}, \citenamefont {Gair}, \citenamefont
  {Clark},\ and\ \citenamefont {Khamesra}}]{Williams:2019vub}%
  \BibitemOpen
  \bibfield  {author} {\bibinfo {author} {\bibfnamefont {Daniel}\ \bibnamefont
  {Williams}}, \bibinfo {author} {\bibfnamefont {Ik~Siong}\ \bibnamefont
  {Heng}}, \bibinfo {author} {\bibfnamefont {Jonathan}\ \bibnamefont {Gair}},
  \bibinfo {author} {\bibfnamefont {James~A.}\ \bibnamefont {Clark}}, \ and\
  \bibinfo {author} {\bibfnamefont {Bhavesh}\ \bibnamefont {Khamesra}},\
  }\bibfield  {title} {\enquote {\bibinfo {title} {{Precessing numerical
  relativity waveform surrogate model for binary black holes: A Gaussian
  process regression approach}},}\ }\href {\doibase
  10.1103/PhysRevD.101.063011} {\bibfield  {journal} {\bibinfo  {journal}
  {Phys. Rev. D}\ }\textbf {\bibinfo {volume} {101}},\ \bibinfo {pages}
  {063011} (\bibinfo {year} {2020})},\ \Eprint
  {http://arxiv.org/abs/1903.09204} {arXiv:1903.09204 [gr-qc]} \BibitemShut
  {NoStop}%
\bibitem [{\citenamefont {Yoo}\ \emph {et~al.}(2022)\citenamefont {Yoo},
  \citenamefont {Varma}, \citenamefont {Giesler}, \citenamefont {Scheel},
  \citenamefont {Haster}, \citenamefont {Pfeiffer}, \citenamefont {Kidder},\
  and\ \citenamefont {Boyle}}]{Yoo:2022erv}%
  \BibitemOpen
  \bibfield  {author} {\bibinfo {author} {\bibfnamefont {Jooheon}\ \bibnamefont
  {Yoo}}, \bibinfo {author} {\bibfnamefont {Vijay}\ \bibnamefont {Varma}},
  \bibinfo {author} {\bibfnamefont {Matthew}\ \bibnamefont {Giesler}}, \bibinfo
  {author} {\bibfnamefont {Mark~A.}\ \bibnamefont {Scheel}}, \bibinfo {author}
  {\bibfnamefont {Carl-Johan}\ \bibnamefont {Haster}}, \bibinfo {author}
  {\bibfnamefont {Harald~P.}\ \bibnamefont {Pfeiffer}}, \bibinfo {author}
  {\bibfnamefont {Lawrence~E.}\ \bibnamefont {Kidder}}, \ and\ \bibinfo
  {author} {\bibfnamefont {Michael}\ \bibnamefont {Boyle}},\ }\bibfield
  {title} {\enquote {\bibinfo {title} {{Targeted large mass ratio numerical
  relativity surrogate waveform model for GW190814}},}\ }\href@noop {} {\
  (\bibinfo {year} {2022})},\ \Eprint {http://arxiv.org/abs/2203.10109}
  {arXiv:2203.10109 [gr-qc]} \BibitemShut {NoStop}%
\bibitem [{\citenamefont {{Galley, C. R.}}()}]{rompy}%
  \BibitemOpen
  \bibfield  {author} {\bibinfo {author} {\bibnamefont {{Galley, C. R.}}},\
  }\href {\doibase https://bitbucket.org/chadgalley/rompy/src/master/}
  {\enquote {\bibinfo {title} {Rompy},}\ }\bibinfo {howpublished} {free
  software (GPL)}\BibitemShut {NoStop}%
\bibitem [{\citenamefont {Kidder}(1995)}]{Kidder:1995zr}%
  \BibitemOpen
  \bibfield  {author} {\bibinfo {author} {\bibfnamefont {Lawrence~E.}\
  \bibnamefont {Kidder}},\ }\bibfield  {title} {\enquote {\bibinfo {title}
  {{Coalescing binary systems of compact objects to postNewtonian 5/2 order. 5.
  Spin effects}},}\ }\href {\doibase 10.1103/PhysRevD.52.821} {\bibfield
  {journal} {\bibinfo  {journal} {Phys. Rev. D}\ }\textbf {\bibinfo {volume}
  {52}},\ \bibinfo {pages} {821--847} (\bibinfo {year} {1995})},\ \Eprint
  {http://arxiv.org/abs/gr-qc/9506022} {arXiv:gr-qc/9506022} \BibitemShut
  {NoStop}%
\bibitem [{\citenamefont {Buonanno}\ \emph {et~al.}(2003)\citenamefont
  {Buonanno}, \citenamefont {Chen},\ and\ \citenamefont
  {Vallisneri}}]{Buonanno:2002fy}%
  \BibitemOpen
  \bibfield  {author} {\bibinfo {author} {\bibfnamefont {Alessandra}\
  \bibnamefont {Buonanno}}, \bibinfo {author} {\bibfnamefont {Yan-bei}\
  \bibnamefont {Chen}}, \ and\ \bibinfo {author} {\bibfnamefont {Michele}\
  \bibnamefont {Vallisneri}},\ }\bibfield  {title} {\enquote {\bibinfo {title}
  {{Detecting gravitational waves from precessing binaries of spinning compact
  objects: Adiabatic limit}},}\ }\href {\doibase 10.1103/PhysRevD.67.104025}
  {\bibfield  {journal} {\bibinfo  {journal} {Phys. Rev. D}\ }\textbf {\bibinfo
  {volume} {67}},\ \bibinfo {pages} {104025} (\bibinfo {year} {2003})},\
  \bibinfo {note} {[Erratum: Phys.Rev.D 74, 029904 (2006)]},\ \Eprint
  {http://arxiv.org/abs/gr-qc/0211087} {arXiv:gr-qc/0211087} \BibitemShut
  {NoStop}%
\bibitem [{\citenamefont {Schmidt}\ \emph {et~al.}(2011)\citenamefont
  {Schmidt}, \citenamefont {Hannam}, \citenamefont {Husa},\ and\ \citenamefont
  {Ajith}}]{Schmidt:2010it}%
  \BibitemOpen
  \bibfield  {author} {\bibinfo {author} {\bibfnamefont {Patricia}\
  \bibnamefont {Schmidt}}, \bibinfo {author} {\bibfnamefont {Mark}\
  \bibnamefont {Hannam}}, \bibinfo {author} {\bibfnamefont {Sascha}\
  \bibnamefont {Husa}}, \ and\ \bibinfo {author} {\bibfnamefont
  {P.}~\bibnamefont {Ajith}},\ }\bibfield  {title} {\enquote {\bibinfo {title}
  {{Tracking the precession of compact binaries from their gravitational-wave
  signal}},}\ }\href {\doibase 10.1103/PhysRevD.84.024046} {\bibfield
  {journal} {\bibinfo  {journal} {Phys. Rev. D}\ }\textbf {\bibinfo {volume}
  {84}},\ \bibinfo {pages} {024046} (\bibinfo {year} {2011})},\ \Eprint
  {http://arxiv.org/abs/1012.2879} {arXiv:1012.2879 [gr-qc]} \BibitemShut
  {NoStop}%
\bibitem [{\citenamefont {Schmidt}\ \emph {et~al.}(2012)\citenamefont
  {Schmidt}, \citenamefont {Hannam},\ and\ \citenamefont
  {Husa}}]{Schmidt:2012rh}%
  \BibitemOpen
  \bibfield  {author} {\bibinfo {author} {\bibfnamefont {Patricia}\
  \bibnamefont {Schmidt}}, \bibinfo {author} {\bibfnamefont {Mark}\
  \bibnamefont {Hannam}}, \ and\ \bibinfo {author} {\bibfnamefont {Sascha}\
  \bibnamefont {Husa}},\ }\bibfield  {title} {\enquote {\bibinfo {title}
  {{Towards models of gravitational waveforms from generic binaries: A simple
  approximate mapping between precessing and non-precessing inspiral
  signals}},}\ }\href {\doibase 10.1103/PhysRevD.86.104063} {\bibfield
  {journal} {\bibinfo  {journal} {Phys. Rev. D}\ }\textbf {\bibinfo {volume}
  {86}},\ \bibinfo {pages} {104063} (\bibinfo {year} {2012})},\ \Eprint
  {http://arxiv.org/abs/1207.3088} {arXiv:1207.3088 [gr-qc]} \BibitemShut
  {NoStop}%
\bibitem [{\citenamefont {Hannam}\ \emph {et~al.}(2014)\citenamefont {Hannam},
  \citenamefont {Schmidt}, \citenamefont {Boh\'e}, \citenamefont {Haegel},
  \citenamefont {Husa}, \citenamefont {Ohme}, \citenamefont {Pratten},\ and\
  \citenamefont {P\"urrer}}]{Hannam:2013oca}%
  \BibitemOpen
  \bibfield  {author} {\bibinfo {author} {\bibfnamefont {Mark}\ \bibnamefont
  {Hannam}}, \bibinfo {author} {\bibfnamefont {Patricia}\ \bibnamefont
  {Schmidt}}, \bibinfo {author} {\bibfnamefont {Alejandro}\ \bibnamefont
  {Boh\'e}}, \bibinfo {author} {\bibfnamefont {Le\"\i{}la}\ \bibnamefont
  {Haegel}}, \bibinfo {author} {\bibfnamefont {Sascha}\ \bibnamefont {Husa}},
  \bibinfo {author} {\bibfnamefont {Frank}\ \bibnamefont {Ohme}}, \bibinfo
  {author} {\bibfnamefont {Geraint}\ \bibnamefont {Pratten}}, \ and\ \bibinfo
  {author} {\bibfnamefont {Michael}\ \bibnamefont {P\"urrer}},\ }\bibfield
  {title} {\enquote {\bibinfo {title} {{Simple Model of Complete Precessing
  Black-Hole-Binary Gravitational Waveforms}},}\ }\href {\doibase
  10.1103/PhysRevLett.113.151101} {\bibfield  {journal} {\bibinfo  {journal}
  {Phys. Rev. Lett.}\ }\textbf {\bibinfo {volume} {113}},\ \bibinfo {pages}
  {151101} (\bibinfo {year} {2014})},\ \Eprint {http://arxiv.org/abs/1308.3271}
  {arXiv:1308.3271 [gr-qc]} \BibitemShut {NoStop}%
\bibitem [{\citenamefont {Pratten}\ \emph {et~al.}(2021)\citenamefont {Pratten}
  \emph {et~al.}}]{Pratten:2020ceb}%
  \BibitemOpen
  \bibfield  {author} {\bibinfo {author} {\bibfnamefont {Geraint}\ \bibnamefont
  {Pratten}} \emph {et~al.},\ }\bibfield  {title} {\enquote {\bibinfo {title}
  {{Computationally efficient models for the dominant and subdominant harmonic
  modes of precessing binary black holes}},}\ }\href {\doibase
  10.1103/PhysRevD.103.104056} {\bibfield  {journal} {\bibinfo  {journal}
  {Phys. Rev. D}\ }\textbf {\bibinfo {volume} {103}},\ \bibinfo {pages}
  {104056} (\bibinfo {year} {2021})},\ \Eprint
  {http://arxiv.org/abs/2004.06503} {arXiv:2004.06503 [gr-qc]} \BibitemShut
  {NoStop}%
\bibitem [{\citenamefont {Ramos-Buades}\ \emph {et~al.}(2020)\citenamefont
  {Ramos-Buades}, \citenamefont {Schmidt}, \citenamefont {Pratten},\ and\
  \citenamefont {Husa}}]{Ramos-Buades:2020noq}%
  \BibitemOpen
  \bibfield  {author} {\bibinfo {author} {\bibfnamefont {Antoni}\ \bibnamefont
  {Ramos-Buades}}, \bibinfo {author} {\bibfnamefont {Patricia}\ \bibnamefont
  {Schmidt}}, \bibinfo {author} {\bibfnamefont {Geraint}\ \bibnamefont
  {Pratten}}, \ and\ \bibinfo {author} {\bibfnamefont {Sascha}\ \bibnamefont
  {Husa}},\ }\bibfield  {title} {\enquote {\bibinfo {title} {{Validity of
  common modeling approximations for precessing binary black holes with
  higher-order modes}},}\ }\href {\doibase 10.1103/PhysRevD.101.103014}
  {\bibfield  {journal} {\bibinfo  {journal} {Phys. Rev. D}\ }\textbf {\bibinfo
  {volume} {101}},\ \bibinfo {pages} {103014} (\bibinfo {year} {2020})},\
  \Eprint {http://arxiv.org/abs/2001.10936} {arXiv:2001.10936 [gr-qc]}
  \BibitemShut {NoStop}%
\bibitem [{\citenamefont {Boyle}\ \emph {et~al.}(2014)\citenamefont {Boyle},
  \citenamefont {Kidder}, \citenamefont {Ossokine},\ and\ \citenamefont
  {Pfeiffer}}]{Boyle:2014ioa}%
  \BibitemOpen
  \bibfield  {author} {\bibinfo {author} {\bibfnamefont {Michael}\ \bibnamefont
  {Boyle}}, \bibinfo {author} {\bibfnamefont {Lawrence~E.}\ \bibnamefont
  {Kidder}}, \bibinfo {author} {\bibfnamefont {Serguei}\ \bibnamefont
  {Ossokine}}, \ and\ \bibinfo {author} {\bibfnamefont {Harald~P.}\
  \bibnamefont {Pfeiffer}},\ }\bibfield  {title} {\enquote {\bibinfo {title}
  {{Gravitational-wave modes from precessing black-hole binaries}},}\
  }\href@noop {} {\  (\bibinfo {year} {2014})},\ \Eprint
  {http://arxiv.org/abs/1409.4431} {arXiv:1409.4431 [gr-qc]} \BibitemShut
  {NoStop}%
\bibitem [{\citenamefont {Cotesta}\ \emph {et~al.}(2018)\citenamefont
  {Cotesta}, \citenamefont {Buonanno}, \citenamefont {Boh\'e}, \citenamefont
  {Taracchini}, \citenamefont {Hinder},\ and\ \citenamefont
  {Ossokine}}]{Cotesta:2018fcv}%
  \BibitemOpen
  \bibfield  {author} {\bibinfo {author} {\bibfnamefont {Roberto}\ \bibnamefont
  {Cotesta}}, \bibinfo {author} {\bibfnamefont {Alessandra}\ \bibnamefont
  {Buonanno}}, \bibinfo {author} {\bibfnamefont {Alejandro}\ \bibnamefont
  {Boh\'e}}, \bibinfo {author} {\bibfnamefont {Andrea}\ \bibnamefont
  {Taracchini}}, \bibinfo {author} {\bibfnamefont {Ian}\ \bibnamefont
  {Hinder}}, \ and\ \bibinfo {author} {\bibfnamefont {Serguei}\ \bibnamefont
  {Ossokine}},\ }\bibfield  {title} {\enquote {\bibinfo {title} {{Enriching the
  Symphony of Gravitational Waves from Binary Black Holes by Tuning Higher
  Harmonics}},}\ }\href {\doibase 10.1103/PhysRevD.98.084028} {\bibfield
  {journal} {\bibinfo  {journal} {Phys. Rev. D}\ }\textbf {\bibinfo {volume}
  {98}},\ \bibinfo {pages} {084028} (\bibinfo {year} {2018})},\ \Eprint
  {http://arxiv.org/abs/1803.10701} {arXiv:1803.10701 [gr-qc]} \BibitemShut
  {NoStop}%
\bibitem [{\citenamefont {Boyle}(2013)}]{Boyle:2013nka}%
  \BibitemOpen
  \bibfield  {author} {\bibinfo {author} {\bibfnamefont {Michael}\ \bibnamefont
  {Boyle}},\ }\bibfield  {title} {\enquote {\bibinfo {title} {{Angular velocity
  of gravitational radiation from precessing binaries and the corotating
  frame}},}\ }\href {\doibase 10.1103/PhysRevD.87.104006} {\bibfield  {journal}
  {\bibinfo  {journal} {Phys. Rev. D}\ }\textbf {\bibinfo {volume} {87}},\
  \bibinfo {pages} {104006} (\bibinfo {year} {2013})},\ \Eprint
  {http://arxiv.org/abs/1302.2919} {arXiv:1302.2919 [gr-qc]} \BibitemShut
  {NoStop}%
\bibitem [{\citenamefont {{LIGO Scientific Collaboration}}(2018)}]{lalsuite}%
  \BibitemOpen
  \bibfield  {author} {\bibinfo {author} {\bibnamefont {{LIGO Scientific
  Collaboration}}},\ }\href {\doibase 10.7935/GT1W-FZ16} {\enquote {\bibinfo
  {title} {{LIGO} {A}lgorithm {L}ibrary - {LALS}uite},}\ }\bibinfo
  {howpublished} {free software (GPL)} (\bibinfo {year} {2018})\BibitemShut
  {NoStop}%
\bibitem [{\citenamefont {Racine}(2008)}]{Racine:2008qv}%
  \BibitemOpen
  \bibfield  {author} {\bibinfo {author} {\bibfnamefont {Etienne}\ \bibnamefont
  {Racine}},\ }\bibfield  {title} {\enquote {\bibinfo {title} {{Analysis of
  spin precession in binary black hole systems including quadrupole-monopole
  interaction}},}\ }\href {\doibase 10.1103/PhysRevD.78.044021} {\bibfield
  {journal} {\bibinfo  {journal} {Phys. Rev. D}\ }\textbf {\bibinfo {volume}
  {78}},\ \bibinfo {pages} {044021} (\bibinfo {year} {2008})},\ \Eprint
  {http://arxiv.org/abs/0803.1820} {arXiv:0803.1820 [gr-qc]} \BibitemShut
  {NoStop}%
\bibitem [{\citenamefont {Harry}\ \emph {et~al.}(2016)\citenamefont {Harry},
  \citenamefont {Privitera}, \citenamefont {Boh\'e},\ and\ \citenamefont
  {Buonanno}}]{Harry:2016ijz}%
  \BibitemOpen
  \bibfield  {author} {\bibinfo {author} {\bibfnamefont {Ian}\ \bibnamefont
  {Harry}}, \bibinfo {author} {\bibfnamefont {Stephen}\ \bibnamefont
  {Privitera}}, \bibinfo {author} {\bibfnamefont {Alejandro}\ \bibnamefont
  {Boh\'e}}, \ and\ \bibinfo {author} {\bibfnamefont {Alessandra}\ \bibnamefont
  {Buonanno}},\ }\bibfield  {title} {\enquote {\bibinfo {title} {{Searching for
  Gravitational Waves from Compact Binaries with Precessing Spins}},}\ }\href
  {\doibase 10.1103/PhysRevD.94.024012} {\bibfield  {journal} {\bibinfo
  {journal} {Phys. Rev. D}\ }\textbf {\bibinfo {volume} {94}},\ \bibinfo
  {pages} {024012} (\bibinfo {year} {2016})},\ \Eprint
  {http://arxiv.org/abs/1603.02444} {arXiv:1603.02444 [gr-qc]} \BibitemShut
  {NoStop}%
\bibitem [{\citenamefont {Johnson-McDaniel}\ \emph {et~al.}(2021)\citenamefont
  {Johnson-McDaniel}, \citenamefont {Kulkarni},\ and\ \citenamefont
  {Gupta}}]{Johnson-McDaniel:2021rvv}%
  \BibitemOpen
  \bibfield  {author} {\bibinfo {author} {\bibfnamefont {Nathan~K.}\
  \bibnamefont {Johnson-McDaniel}}, \bibinfo {author} {\bibfnamefont {Sumeet}\
  \bibnamefont {Kulkarni}}, \ and\ \bibinfo {author} {\bibfnamefont {Anuradha}\
  \bibnamefont {Gupta}},\ }\bibfield  {title} {\enquote {\bibinfo {title}
  {{Inferring spin tilts at formation from gravitational wave observations of
  binary black holes: Interfacing precession-averaged and orbit-averaged spin
  evolution}},}\ }\href@noop {} {\  (\bibinfo {year} {2021})},\ \Eprint
  {http://arxiv.org/abs/2107.11902} {arXiv:2107.11902 [astro-ph.HE]}
  \BibitemShut {NoStop}%
\bibitem [{\citenamefont {Pedregosa}\ \emph {et~al.}(2011)\citenamefont
  {Pedregosa}, \citenamefont {Varoquaux}, \citenamefont {Gramfort},
  \citenamefont {Michel}, \citenamefont {Thirion}, \citenamefont {Grisel},
  \citenamefont {Blondel}, \citenamefont {Prettenhofer}, \citenamefont {Weiss},
  \citenamefont {Dubourg}, \citenamefont {Vanderplas}, \citenamefont {Passos},
  \citenamefont {Cournapeau}, \citenamefont {Brucher}, \citenamefont {Perrot},\
  and\ \citenamefont {Duchesnay}}]{scikit-learn}%
  \BibitemOpen
  \bibfield  {author} {\bibinfo {author} {\bibfnamefont {F.}~\bibnamefont
  {Pedregosa}}, \bibinfo {author} {\bibfnamefont {G.}~\bibnamefont
  {Varoquaux}}, \bibinfo {author} {\bibfnamefont {A.}~\bibnamefont {Gramfort}},
  \bibinfo {author} {\bibfnamefont {V.}~\bibnamefont {Michel}}, \bibinfo
  {author} {\bibfnamefont {B.}~\bibnamefont {Thirion}}, \bibinfo {author}
  {\bibfnamefont {O.}~\bibnamefont {Grisel}}, \bibinfo {author} {\bibfnamefont
  {M.}~\bibnamefont {Blondel}}, \bibinfo {author} {\bibfnamefont
  {P.}~\bibnamefont {Prettenhofer}}, \bibinfo {author} {\bibfnamefont
  {R.}~\bibnamefont {Weiss}}, \bibinfo {author} {\bibfnamefont
  {V.}~\bibnamefont {Dubourg}}, \bibinfo {author} {\bibfnamefont
  {J.}~\bibnamefont {Vanderplas}}, \bibinfo {author} {\bibfnamefont
  {A.}~\bibnamefont {Passos}}, \bibinfo {author} {\bibfnamefont
  {D.}~\bibnamefont {Cournapeau}}, \bibinfo {author} {\bibfnamefont
  {M.}~\bibnamefont {Brucher}}, \bibinfo {author} {\bibfnamefont
  {M.}~\bibnamefont {Perrot}}, \ and\ \bibinfo {author} {\bibfnamefont
  {E.}~\bibnamefont {Duchesnay}},\ }\bibfield  {title} {\enquote {\bibinfo
  {title} {Scikit-learn: Machine learning in {P}ython},}\ }\href@noop {}
  {\bibfield  {journal} {\bibinfo  {journal} {Journal of Machine Learning
  Research}\ }\textbf {\bibinfo {volume} {12}},\ \bibinfo {pages} {2825--2830}
  (\bibinfo {year} {2011})}\BibitemShut {NoStop}%
\bibitem [{\citenamefont {Galley}\ and\ \citenamefont
  {Schmidt}(2016{\natexlab{b}})}]{Galley:2016}%
  \BibitemOpen
  \bibfield  {author} {\bibinfo {author} {\bibfnamefont {Chad.}\ \bibnamefont
  {Galley}}\ and\ \bibinfo {author} {\bibfnamefont {Patricia}\ \bibnamefont
  {Schmidt}},\ }\bibfield  {title} {\enquote {\bibinfo {title} {{Fast and
  efficient evaluation of gravitational waveforms via reduced-order spline
  interpolation}},}\ }\href@noop {} {\  (\bibinfo {year}
  {2016}{\natexlab{b}})},\ \Eprint {http://arxiv.org/abs/1611.07529}
  {arXiv:1611.07529 [gr-qc]} \BibitemShut {NoStop}%
\bibitem [{\citenamefont {Abadi}\ \emph {et~al.}(2015)\citenamefont {Abadi},
  \citenamefont {Agarwal}, \citenamefont {Barham}, \citenamefont {Brevdo},
  \citenamefont {Chen}, \citenamefont {Citro}, \citenamefont {Corrado},
  \citenamefont {Davis}, \citenamefont {Dean}, \citenamefont {Devin},
  \citenamefont {Ghemawat}, \citenamefont {Goodfellow}, \citenamefont {Harp},
  \citenamefont {Irving}, \citenamefont {Isard}, \citenamefont {Jia},
  \citenamefont {Jozefowicz}, \citenamefont {Kaiser}, \citenamefont {Kudlur},
  \citenamefont {Levenberg}, \citenamefont {Man\'{e}}, \citenamefont {Monga},
  \citenamefont {Moore}, \citenamefont {Murray}, \citenamefont {Olah},
  \citenamefont {Schuster}, \citenamefont {Shlens}, \citenamefont {Steiner},
  \citenamefont {Sutskever}, \citenamefont {Talwar}, \citenamefont {Tucker},
  \citenamefont {Vanhoucke}, \citenamefont {Vasudevan}, \citenamefont
  {Vi\'{e}gas}, \citenamefont {Vinyals}, \citenamefont {Warden}, \citenamefont
  {Wattenberg}, \citenamefont {Wicke}, \citenamefont {Yu},\ and\ \citenamefont
  {Zheng}}]{tensorflow2015-whitepaper}%
  \BibitemOpen
  \bibfield  {author} {\bibinfo {author} {\bibfnamefont {Mart\'{i}n}\
  \bibnamefont {Abadi}}, \bibinfo {author} {\bibfnamefont {Ashish}\
  \bibnamefont {Agarwal}}, \bibinfo {author} {\bibfnamefont {Paul}\
  \bibnamefont {Barham}}, \bibinfo {author} {\bibfnamefont {Eugene}\
  \bibnamefont {Brevdo}}, \bibinfo {author} {\bibfnamefont {Zhifeng}\
  \bibnamefont {Chen}}, \bibinfo {author} {\bibfnamefont {Craig}\ \bibnamefont
  {Citro}}, \bibinfo {author} {\bibfnamefont {Greg~S.}\ \bibnamefont
  {Corrado}}, \bibinfo {author} {\bibfnamefont {Andy}\ \bibnamefont {Davis}},
  \bibinfo {author} {\bibfnamefont {Jeffrey}\ \bibnamefont {Dean}}, \bibinfo
  {author} {\bibfnamefont {Matthieu}\ \bibnamefont {Devin}}, \bibinfo {author}
  {\bibfnamefont {Sanjay}\ \bibnamefont {Ghemawat}}, \bibinfo {author}
  {\bibfnamefont {Ian}\ \bibnamefont {Goodfellow}}, \bibinfo {author}
  {\bibfnamefont {Andrew}\ \bibnamefont {Harp}}, \bibinfo {author}
  {\bibfnamefont {Geoffrey}\ \bibnamefont {Irving}}, \bibinfo {author}
  {\bibfnamefont {Michael}\ \bibnamefont {Isard}}, \bibinfo {author}
  {\bibfnamefont {Yangqing}\ \bibnamefont {Jia}}, \bibinfo {author}
  {\bibfnamefont {Rafal}\ \bibnamefont {Jozefowicz}}, \bibinfo {author}
  {\bibfnamefont {Lukasz}\ \bibnamefont {Kaiser}}, \bibinfo {author}
  {\bibfnamefont {Manjunath}\ \bibnamefont {Kudlur}}, \bibinfo {author}
  {\bibfnamefont {Josh}\ \bibnamefont {Levenberg}}, \bibinfo {author}
  {\bibfnamefont {Dandelion}\ \bibnamefont {Man\'{e}}}, \bibinfo {author}
  {\bibfnamefont {Rajat}\ \bibnamefont {Monga}}, \bibinfo {author}
  {\bibfnamefont {Sherry}\ \bibnamefont {Moore}}, \bibinfo {author}
  {\bibfnamefont {Derek}\ \bibnamefont {Murray}}, \bibinfo {author}
  {\bibfnamefont {Chris}\ \bibnamefont {Olah}}, \bibinfo {author}
  {\bibfnamefont {Mike}\ \bibnamefont {Schuster}}, \bibinfo {author}
  {\bibfnamefont {Jonathon}\ \bibnamefont {Shlens}}, \bibinfo {author}
  {\bibfnamefont {Benoit}\ \bibnamefont {Steiner}}, \bibinfo {author}
  {\bibfnamefont {Ilya}\ \bibnamefont {Sutskever}}, \bibinfo {author}
  {\bibfnamefont {Kunal}\ \bibnamefont {Talwar}}, \bibinfo {author}
  {\bibfnamefont {Paul}\ \bibnamefont {Tucker}}, \bibinfo {author}
  {\bibfnamefont {Vincent}\ \bibnamefont {Vanhoucke}}, \bibinfo {author}
  {\bibfnamefont {Vijay}\ \bibnamefont {Vasudevan}}, \bibinfo {author}
  {\bibfnamefont {Fernanda}\ \bibnamefont {Vi\'{e}gas}}, \bibinfo {author}
  {\bibfnamefont {Oriol}\ \bibnamefont {Vinyals}}, \bibinfo {author}
  {\bibfnamefont {Pete}\ \bibnamefont {Warden}}, \bibinfo {author}
  {\bibfnamefont {Martin}\ \bibnamefont {Wattenberg}}, \bibinfo {author}
  {\bibfnamefont {Martin}\ \bibnamefont {Wicke}}, \bibinfo {author}
  {\bibfnamefont {Yuan}\ \bibnamefont {Yu}}, \ and\ \bibinfo {author}
  {\bibfnamefont {Xiaoqiang}\ \bibnamefont {Zheng}},\ }\href
  {https://www.tensorflow.org/} {\enquote {\bibinfo {title} {{TensorFlow}:
  Large-scale machine learning on heterogeneous systems},}\ } (\bibinfo {year}
  {2015}),\ \bibinfo {note} {software available from
  tensorflow.org}\BibitemShut {NoStop}%
\bibitem [{\citenamefont {Chollet}\ \emph {et~al.}(2015)\citenamefont {Chollet}
  \emph {et~al.}}]{chollet2015keras}%
  \BibitemOpen
  \bibfield  {author} {\bibinfo {author} {\bibfnamefont {Fran\c{c}ois}\
  \bibnamefont {Chollet}} \emph {et~al.},\ }\href@noop {} {\enquote {\bibinfo
  {title} {Keras},}\ }\bibinfo {howpublished} {\url{https://keras.io}}
  (\bibinfo {year} {2015})\BibitemShut {NoStop}%
\bibitem [{\citenamefont {Bergstra}\ \emph {et~al.}(2013)\citenamefont
  {Bergstra}, \citenamefont {Yamins},\ and\ \citenamefont
  {Cox}}]{Hyperopt:2013byc}%
  \BibitemOpen
  \bibfield  {author} {\bibinfo {author} {\bibfnamefont {James}\ \bibnamefont
  {Bergstra}}, \bibinfo {author} {\bibfnamefont {Daniel}\ \bibnamefont
  {Yamins}}, \ and\ \bibinfo {author} {\bibfnamefont {David}\ \bibnamefont
  {Cox}},\ }\bibfield  {title} {\enquote {\bibinfo {title} {Making a science of
  model search: Hyperparameter optimization in hundreds of dimensions for
  vision architectures},}\ }in\ \href@noop {} {\emph {\bibinfo {booktitle}
  {International conference on machine learning}}}\ (\bibinfo {organization}
  {PMLR},\ \bibinfo {year} {2013})\ pp.\ \bibinfo {pages}
  {115--123}\BibitemShut {NoStop}%
\bibitem [{\citenamefont {Fukushima}(1969)}]{fukushima1969visual}%
  \BibitemOpen
  \bibfield  {author} {\bibinfo {author} {\bibfnamefont {Kunihiko}\
  \bibnamefont {Fukushima}},\ }\bibfield  {title} {\enquote {\bibinfo {title}
  {Visual feature extraction by a multilayered network of analog threshold
  elements},}\ }\href@noop {} {\bibfield  {journal} {\bibinfo  {journal} {IEEE
  Transactions on Systems Science and Cybernetics}\ }\textbf {\bibinfo {volume}
  {5}},\ \bibinfo {pages} {322--333} (\bibinfo {year} {1969})}\BibitemShut
  {NoStop}%
\bibitem [{\citenamefont {Glorot}\ \emph {et~al.}(2011)\citenamefont {Glorot},
  \citenamefont {Bordes},\ and\ \citenamefont {Bengio}}]{glorot2011deep}%
  \BibitemOpen
  \bibfield  {author} {\bibinfo {author} {\bibfnamefont {Xavier}\ \bibnamefont
  {Glorot}}, \bibinfo {author} {\bibfnamefont {Antoine}\ \bibnamefont
  {Bordes}}, \ and\ \bibinfo {author} {\bibfnamefont {Yoshua}\ \bibnamefont
  {Bengio}},\ }\bibfield  {title} {\enquote {\bibinfo {title} {Deep sparse
  rectifier neural networks},}\ }in\ \href@noop {} {\emph {\bibinfo {booktitle}
  {Proceedings of the fourteenth international conference on artificial
  intelligence and statistics}}}\ (\bibinfo {organization} {JMLR Workshop and
  Conference Proceedings},\ \bibinfo {year} {2011})\ pp.\ \bibinfo {pages}
  {315--323}\BibitemShut {NoStop}%
\bibitem [{\citenamefont {Clevert}\ \emph {et~al.}(2015)\citenamefont
  {Clevert}, \citenamefont {Unterthiner},\ and\ \citenamefont
  {Hochreiter}}]{Clevert2015:elu}%
  \BibitemOpen
  \bibfield  {author} {\bibinfo {author} {\bibfnamefont {Djork-Arné}\
  \bibnamefont {Clevert}}, \bibinfo {author} {\bibfnamefont {Thomas}\
  \bibnamefont {Unterthiner}}, \ and\ \bibinfo {author} {\bibfnamefont {Sepp}\
  \bibnamefont {Hochreiter}},\ }\href {\doibase 10.48550/ARXIV.1511.07289}
  {\enquote {\bibinfo {title} {Fast and accurate deep network learning by
  exponential linear units (elus)},}\ } (\bibinfo {year} {2015})\BibitemShut
  {NoStop}%
\bibitem [{\citenamefont {Elliott}\ and\ \citenamefont
  {Elliott}(1993)}]{Elliott93abetter}%
  \BibitemOpen
  \bibfield  {author} {\bibinfo {author} {\bibfnamefont {D.L.}\ \bibnamefont
  {Elliott}}\ and\ \bibinfo {author} {\bibfnamefont {David~L.}\ \bibnamefont
  {Elliott}},\ }\href@noop {} {\enquote {\bibinfo {title} {A better activation
  function for artificial neural networks},}\ } (\bibinfo {year}
  {1993})\BibitemShut {NoStop}%
\bibitem [{\citenamefont {Kingma}\ and\ \citenamefont
  {Ba}(2014)}]{kingma2014adam}%
  \BibitemOpen
  \bibfield  {author} {\bibinfo {author} {\bibfnamefont {Diederik~P}\
  \bibnamefont {Kingma}}\ and\ \bibinfo {author} {\bibfnamefont {Jimmy}\
  \bibnamefont {Ba}},\ }\bibfield  {title} {\enquote {\bibinfo {title} {Adam: A
  method for stochastic optimization},}\ }\href@noop {} {\bibfield  {journal}
  {\bibinfo  {journal} {arXiv preprint arXiv:1412.6980}\ } (\bibinfo {year}
  {2014})}\BibitemShut {NoStop}%
\bibitem [{\citenamefont {Dozat}(2016)}]{dozat2016incorporating}%
  \BibitemOpen
  \bibfield  {author} {\bibinfo {author} {\bibfnamefont {Timothy}\ \bibnamefont
  {Dozat}},\ }\bibfield  {title} {\enquote {\bibinfo {title} {Incorporating
  nesterov momentum into adam},}\ }\href@noop {} {\bibfield  {journal}
  {\bibinfo  {journal} {ICLR Workshop}\ }\textbf {\bibinfo {volume}
  {1:2013-2016}} (\bibinfo {year} {2016})}\BibitemShut {NoStop}%
\bibitem [{\citenamefont {Zeiler}(2012)}]{Zeiler:2012ad}%
  \BibitemOpen
  \bibfield  {author} {\bibinfo {author} {\bibfnamefont {Matthew~D.}\
  \bibnamefont {Zeiler}},\ }\bibfield  {title} {\enquote {\bibinfo {title}
  {Adadelta: An adaptive learning rate method},}\ }\href {\doibase
  10.48550/ARXIV.1212.5701} {\  (\bibinfo {year} {2012}),\
  10.48550/ARXIV.1212.5701}\BibitemShut {NoStop}%
\bibitem [{\citenamefont {Lederer}(2021)}]{Lederer2021:act}%
  \BibitemOpen
  \bibfield  {author} {\bibinfo {author} {\bibfnamefont {Johannes}\
  \bibnamefont {Lederer}},\ }\href {\doibase 10.48550/ARXIV.2101.09957}
  {\enquote {\bibinfo {title} {Activation functions in artificial neural
  networks: A systematic overview},}\ } (\bibinfo {year} {2021})\BibitemShut
  {NoStop}%
\bibitem [{\citenamefont {Ruder}(2016)}]{Ruder2016:gdo}%
  \BibitemOpen
  \bibfield  {author} {\bibinfo {author} {\bibfnamefont {Sebastian}\
  \bibnamefont {Ruder}},\ }\href {\doibase 10.48550/ARXIV.1609.04747} {\enquote
  {\bibinfo {title} {An overview of gradient descent optimization
  algorithms},}\ } (\bibinfo {year} {2016})\BibitemShut {NoStop}%
\bibitem [{\citenamefont {Boyle}\ \emph {et~al.}(2011)\citenamefont {Boyle},
  \citenamefont {Owen},\ and\ \citenamefont {Pfeiffer}}]{Boyle:2011gg}%
  \BibitemOpen
  \bibfield  {author} {\bibinfo {author} {\bibfnamefont {Michael}\ \bibnamefont
  {Boyle}}, \bibinfo {author} {\bibfnamefont {Robert}\ \bibnamefont {Owen}}, \
  and\ \bibinfo {author} {\bibfnamefont {Harald~P.}\ \bibnamefont {Pfeiffer}},\
  }\bibfield  {title} {\enquote {\bibinfo {title} {{A geometric approach to the
  precession of compact binaries}},}\ }\href {\doibase
  10.1103/PhysRevD.84.124011} {\bibfield  {journal} {\bibinfo  {journal} {Phys.
  Rev. D}\ }\textbf {\bibinfo {volume} {84}},\ \bibinfo {pages} {124011}
  (\bibinfo {year} {2011})},\ \Eprint {http://arxiv.org/abs/1110.2965}
  {arXiv:1110.2965 [gr-qc]} \BibitemShut {NoStop}%
\bibitem [{\citenamefont {Abbott}\ \emph {et~al.}(2018)\citenamefont {Abbott}
  \emph {et~al.}}]{KAGRA:2013rdx}%
  \BibitemOpen
  \bibfield  {author} {\bibinfo {author} {\bibfnamefont {B.~P.}\ \bibnamefont
  {Abbott}} \emph {et~al.} (\bibinfo {collaboration} {KAGRA, LIGO Scientific,
  Virgo, VIRGO}),\ }\bibfield  {title} {\enquote {\bibinfo {title} {{Prospects
  for observing and localizing gravitational-wave transients with Advanced
  LIGO, Advanced Virgo and KAGRA}},}\ }\href {\doibase
  10.1007/s41114-020-00026-9} {\bibfield  {journal} {\bibinfo  {journal}
  {Living Rev. Rel.}\ }\textbf {\bibinfo {volume} {21}},\ \bibinfo {pages} {3}
  (\bibinfo {year} {2018})},\ \Eprint {http://arxiv.org/abs/1304.0670}
  {arXiv:1304.0670 [gr-qc]} \BibitemShut {NoStop}%
\bibitem [{\citenamefont {Gadre}\ \emph {et~al.}(2022)\citenamefont {Gadre},
  \citenamefont {P\"urrer}, \citenamefont {Field}, \citenamefont {Ossokine},\
  and\ \citenamefont {Varma}}]{Gadre:2022sed}%
  \BibitemOpen
  \bibfield  {author} {\bibinfo {author} {\bibfnamefont {Bhooshan}\
  \bibnamefont {Gadre}}, \bibinfo {author} {\bibfnamefont {Michael}\
  \bibnamefont {P\"urrer}}, \bibinfo {author} {\bibfnamefont {Scott~E.}\
  \bibnamefont {Field}}, \bibinfo {author} {\bibfnamefont {Serguei}\
  \bibnamefont {Ossokine}}, \ and\ \bibinfo {author} {\bibfnamefont {Vijay}\
  \bibnamefont {Varma}},\ }\bibfield  {title} {\enquote {\bibinfo {title} {{A
  fully precessing higher-mode surrogate model of effective-one-body
  waveforms}},}\ }\href@noop {} {\  (\bibinfo {year} {2022})},\ \Eprint
  {http://arxiv.org/abs/2203.00381} {arXiv:2203.00381 [gr-qc]} \BibitemShut
  {NoStop}%
\end{thebibliography}%

\end{document}